\documentclass{aa}  

\usepackage{graphicx}
\usepackage{txfonts}
\usepackage{CJKutf8}
\usepackage{rotating}
\begin{document} 
%\begin{CJK*}{UTF8}{gbsn}

   %\title{Kinematics of the CND in NGC 1068 using CO observations}
   \title{ALMA uncovers optically thin and multi-component CO gas in the outflowing circumnuclear disk of NGC1068}

   \author{
          Y. Zhang \begin{CJK*}{UTF8}{gbsn}(张宇泽)\end{CJK*}\inst{1}
          \and
          S. Viti \inst{1,2,3}
          \and
          S. Garc\'{i}a-Burillo \inst{4}
          \and
          K.-Y. Huang \inst{1}
          }

   \institute{Leiden Observatory, Leiden University, PO Box 9513, 2300 Leiden, RA, The Netherlands\\
              \email{yuzezhang@mail.strw.leidenuniv.nl}\\
              \and
              Transdisciplinary Research Area (TRA Matter) and Argelander Institut f\"{u}r Astronomie, Universit\"{a}t Bonn, Auf dem H\"{u}gel 71, D-53121 Bonn, Germany\\
              \email{viti@mail.strw.leidenuniv.nl}\\
              \and
              Department of Physics and Astronomy, UCL, Gower Place, London WC1E 6BT, UK\\
              \and
              Observatorio Astron\'{o}mico Nacional (OAN-IGN)-Observatorio de Madrid, Alfonso XII, 3, 28014 Madrid, Spain\\
              \email{s.gburillo@oan.es}\\
             }

   \date{Received ; accepted }

% \abstract{}{}{}{}{} 
% 5 {} token are mandatory
 
  \abstract
   {Active galactic nuclei (AGNs) influence host galaxies through winds and jets that generate molecular outflows, traceable with $^{12}$CO line emissions using the Atacama Large Millimeter Array (ALMA). Leveraging ALMA observations, recent studies proposed a 3D outflow geometry in the nearby Seyfert II galaxy NGC 1068---a primary testbed for AGN unification theories. Utilizing ALMA data of CO(2-1), CO(3-2), and CO(6-5) transitions at $\sim0.1''$ ($\sim7\,\rm pc$) resolution, we analyzed temperature, density, and kinematics within the circumnuclear disk (CND) of NGC 1068, focusing on molecular outflows. We selected regions across the CND based on a previously modeled AGN wind bicone. We performed local thermodynamic equilibrium (LTE) analysis to infer column densities and rotational temperatures, revealing optically thin gas with $X_{\mathrm{CO}}$ factors $4.8 \pm 0.4$–$9.6 \pm 0.9$ times smaller than the Milky Way value. Consequently, the molecular mass outflow rate across the CND is mostly below $5.5\,\rm M_{\odot}\,\rm yr^{-1}$, with the majority contributed from northeast of the AGN position ($\alpha_{2000} = 02^{\mathrm{h}}42^{\mathrm{m}}40.776^{\mathrm{s}}$, $\delta_{2000} = -00^\circ 00'47.714''$). After subtracting the rotation curve of the CND, we fitted averaged line profiles for each sampled region using single and weighted multi-component Gaussian models to investigate the kinematics of the non-rotating gas. The fitting results show that some line profiles close to or within the AGN wind bicone require multi-component Gaussian models, with each component exhibiting significant velocity departures from the galaxy's mean motion---a hallmark of a multi-component molecular outflow. We observe lateral variations of CO gas kinematics along the edge and center of the AGN wind bicone, as well as a misalignment of the orientation and spread between the molecular outflow and the ionized outflow. Overall, due to the optically thin condition, the dynamic impact of the ionized outflow to molecular gas inside the CND might not be as substantial as expected. Regardless, the outflowing molecular gas across the CND exhibits complex kinematics, highlighted by an asymmetry between the northeastern and southern CND, and our analyses do not eliminate the 3D outflow geometry as a possible outflow scenario within the CND of NGC 1068.
    }


   \titlerunning{CO gas in the CND of NGC 1068} 
   \authorrunning{Y. Zhang et al.}
   \maketitle
%
%-------------------------------------------------------------------

\section{Introduction}
\label{intro}

Active galactic nuclei (AGNs) are believed to play a crucial role in the co-evolution of supermassive black holes (SMBHs) and their host galaxies through a process known as AGN feedback (Marconi \& Hunt \citeyear{MH03}; Shankar et al. \citeyear{Shankar_etal_2009}). This relationship is evident from various scaling relations that link the mass of SMBHs ($M_{\rm{BH}}$) to the properties of the spheroidal components of their host galaxies, including the stellar velocity dispersion ($\sigma$; e.g., Gebhardt et al. \citeyear{Gebhardt_etal_2000}; Merritt \& Ferrarese \citeyear{MF01}; Ferrarese \& Ford \citeyear{FF05}; G\"{u}ltekin et al. \citeyear{GRG09}) and the bulge mass ($M_{\rm{bulge}}$; Kormendy \& Richstone \citeyear{KR95}; Magorrian et al. \citeyear{Magorrian_etal_98}; Marconi \& Hunt \citeyear{MH03}). AGNs are powered by the accretion of material onto SMBHs, releasing energy in the form of outflows that can be either mechanical (e.g.,  collimated jets or winds) or radiative (Best \& Heckman \citeyear{BH12}; Alonso-Herrero et al. \citeyear{AH19}). These outflows regulate AGN feedback in two primary ways: by disrupting the galactic gas reservoir, thus preventing cooling and expelling gas, or by triggering star formation through the compression of gas (Silk \& Rees \citeyear{SR98}; Gebhardt et al. \citeyear{Gebhardt_etal_2000}; Harrison \citeyear{Harrison_2017}; Maiolino et al. \citeyear{maiolino_etal_2017}; Gallagher et al. \citeyear{gallagher_etal_2019}). Whether through suppression or enhancement of star formation, AGN feedback is central to understanding the evolution of SMBHs and their host galaxies.

In the past, resolving the structures associated with AGN activity around the central regions of galaxies was challenging, even for nearby sources. Extended emissions around the central engines of nearby galaxies were initially observed using single-dish near- and mid-infrared (NIR and MIR) instruments (e.g., Alonso-Herrero et al. \citeyear{ah11}; Burtscher et al. \citeyear{burtscher_etal_2013}), followed by higher-spatial-resolution IR interferometric observations (e.g., Tristram et al. \citeyear{tristram_etal_2009}; Burtscher et al. \citeyear{burtscher_etal_2013}; H\"{o}nig et al. \citeyear{honig_etal_2013}; L\'{o}pez-Gonzaga et al. \citeyear{lg14}, \citeyear{lg16}). However, due to the limited $(u,v)$ coverage of the instruments, IR interferometric data often require reconstruction techniques in the $(u,v)$ plane to analyze complex structures around AGNs. This limitation can be overcome by submillimeter instruments like the Atacama Large Millimeter Array (ALMA). The high spatial resolution provided by ALMA has enabled detailed observations of the geometry of the outflows (Garc\'{i}a-Burillo et al. \citeyear{gb19}; Aalto et al. \citeyear{aalto_etal_2020}). 

One major component of these outflows is their molecular phase (Garc\'{i}a-Burillo et al. \citeyear{gb05}, \citeyear{gb14}, \citeyear{gb16}, \citeyear{gb17}, \citeyear{gb19}; Garc\'{i}a-Burillo \& Combes \citeyear{GC12}; Combes et al. \citeyear{CGC13}, \citeyear{CGC14}; Storchi-Bergmann \& Schnorr-Müller \citeyear{SS19}). The molecular phase is considered the most massive outflow component and plays a significant role in galaxy evolution (Fiore et al. \citeyear{fiore_etal_2017}). Observations of $^{12}$CO (referred to as CO hereafter) have been particularly useful for investigating molecular outflows and galactic kinematics, as CO effectively traces the bulk of molecular gas due to its high abundance and ease of detection (e.g., Combes et al. \citeyear{CGC13}, \citeyear{combes_etal_2019}; Cicone et al. \citeyear{cicone_etal_2014}; Morganti et al. \citeyear{morganti_etal_2015}; Dasyra et al. \citeyear{dasyra_etal_2016}; Alonso-Herrero et al. \citeyear{AH19}; Dom\'{i}nguez-Fern\'{a}ndez et al. \citeyear{df20}). 

NGC 1068, a nearby Seyfert II galaxy located at a distance of $D\sim14\,\rm{Mpc}$ (Bland-Hawthorn et al. \citeyear{bh1997}), serves as a prime example for studying AGN unifying schemes. The discovery of broad-line emissions and polarized continuum from this source (Miller \& Antonucci \citeyear{MA83}; Antonucci \& Miller \citeyear{AM85}) has made it a benchmark for AGN studies. NGC 1068 has been extensively observed across multiple wavelengths, including radio (Greenhill et al. \citeyear{greenhill_etal_96}), UV (Antonucci et al. \citeyear{antonucci_94}), and X-ray (Kinkhabwala et al. \citeyear{Kinkhabwala_02}), revealing significant structural details. In particular, molecular observations in millimeter and submillimeter wavelengths have identified a bar region of approximately $2.8\,\rm kpc$ in diameter containing a starburst (SB) ring of about $1.3\,\rm kpc$ in size via the CO(1-0) emission line. Additionally, a $300\,\rm pc$ circumnuclear disk (CND) surrounding the AGN has been detected through the CO(2-1) emission line (Schinnerer et al. \citeyear{schinnerer_etal_2000}; Garc\'{i}a-Burillo et al. \citeyear{gb14}). In addition, NIR and MIR interferometric observations have provided evidence of a multi-component dusty torus (e.g., Jaffe et al. \citeyear{jaffe_etal_2004}), while high-resolution ALMA observations have revealed a molecular torus 20--30 pc in size (Garc\'{i}a-Burillo et al. \citeyear{gb14}, \citeyear{gb16}, \citeyear{gb19}; Gallimore et al. \citeyear{gallimore_etal_2016}; Imanishi et al. \citeyear{imanishi_etal_18}, \citeyear{imanishi_etal_2020}; Impellizzeri et al. \citeyear{impellizzeri_etal_2019}). 

The CND of NGC 1068 exhibits an asymmetric ringed morphology, offset from the dusty molecular torus and enclosed within a $\sim130$ pc hollow region at its inner boundary, which is thought to be filled with ionized AGN wind (Garc\'{i}a-Burillo et al. \citeyear{gb14}, \citeyear{gb16}, \citeyear{gb19}). Recent molecular observations have detected line broadening and splitting in the spectra of CO line emissions, indicative of the presence of molecular outflows (e.g., Garc\'{i}a-Burillo et al. \citeyear{gb19}). Bright arcs of NIR polarized emission have also been observed within the CND, with the most prominent features in the southern region (Gratadour et al. \citeyear{gratadour_etal_15}), suggesting large-scale shocks at the working surface of the AGN wind. These features are consistent with previously developed AGN wind bicone models (Das et al. \citeyear{das_etal_2006}; Barbosa et al. \citeyear{barbosa_etal_14}). Additionally, Venturi et al. (\citeyear{venturi_etal_2021}), using [OIII] tracers, observed linewidth enhancements perpendicular to the path of the bent kpc-scale radio jet (Gallimore et al. \citeyear{gallimore_etal_96}, \citeyear{gallimore_etal_04}, \citeyear{gallimore_etal_06}), suggesting turbulence caused by jet-ISM interactions within the CND. These observations are further supported by simulations from Wagner et al. (\citeyear{wagner_etal_12}) and Mukherjee et al. (\citeyear{mukherjee_etal_16}).

In light of recent molecular observations, Garc\'{i}a-Burillo et al. (\citeyear{gb19}) proposed a model of the 3D outflow geometry around the AGN of NGC 1068, assisted by 	\textit{Kinemetry} and 3D-Barolo software (Krajnov\'{i}c et al. \citeyear{krajnovic_etal_06}; Di Teodoro \& Fraternali \citeyear{DiTF_15}). This model suggests that the wide-angle AGN wind ($FWHM_{\rm inner, outer} = 40^\circ,\, 80^\circ$ along an axis with $\rm PA\sim30^\circ$) launched by the central engine is sweeping through the surrounding molecular reservoir, initiating two constituents of the molecular outflow along the line of sight: a blueshifted and a redshifted outflow. Given the inclination of the molecular reservoir, which is roughly aligned with the galactic disk ($i = 41^\circ \pm 2^\circ$; Garc\'{i}a-Burillo et al. \citeyear{gb14}), only one constituent of the molecular outflow is visible within the CND. Specifically, the northern CND is dominated by the redshifted molecular outflow, while the southern CND contains the blueshifted molecular outflow. The linewidth enhancement observed in molecular tracers provides evidence of the working surface where the ionized AGN wind interacts with the molecular gas inside the northern CND, as viewed by observers.

In this study, we further analyze the high spatial resolution CO observations from Gallimore et al. (\citeyear{gallimore_etal_2016}) and Garc\'{i}a-Burillo et al. (\citeyear{gb16}, \citeyear{gb19}) across the CND of NGC 1068. Our goal is to better understand and map the temperature, density, and kinematics of this region, with a particular focus on the properties and behavior of the outflows. By leveraging these high-resolution datasets, we aim to provide a comprehensive characterization of the physical conditions within the CND and assess the impact of AGN-driven winds on the surrounding molecular gas.

The observations used in this paper are presented in Sect. \ref{obs}. In Sect. \ref{comaps}, we present velocity-integrated, velocity-width, and residual mean-velocity maps of the CO(2-1), (3-2), and (6-5) transitions. The investigation of gas physical properties, including line ratios and local thermodynamic equilibrium (LTE) analyses, is detailed in Sect. \ref{phys}. In Sect. \ref{kin}, we introduce line profile analyses to explore gas kinematics. In Sect. \ref{mdot}, we calculate the molecular mass outflow rate across the CND. Finally, we summarize our results in Sect. \ref{dis}.

%--------------------------------------------------------------------
\section{Observations}
\label{obs}

We made use of ALMA observations of three CO line transitions, CO(2-1), CO(3-2), and CO(6-5), at a spatial resolution of $\sim0.1''$, which were previously presented in Garc\'{i}a-Burillo et al. (\citeyear{gb16}, \citeyear{gb19}) and Gallimore et al. (\citeyear{gallimore_etal_2016}) and are summarised in Table \ref{tab1}. The CO(6-5) data was observed during Cycle 2, using the Band 9 receiver (project-ID: \#2013.1.00055.S), in combination with data used in Gallimore et al. (\citeyear{gallimore_etal_2016}), while the CO(2-1) and (3-2) data were observed during Cycle 4, using Band 6 and Band 7 receivers respectively (project-ID: \#2016.1.00232.S). The data in the three bands were calibrated by Garc\'{i}a-Burillo et al. (\citeyear{gb16}, \citeyear{gb19}) using the ALMA reduction package CASA. For all the observations, the phase tracking center was set to $\alpha_{2000} = \rm{02h42m40.771s}$, $\delta_{2000} = -00^{\circ}00'47.84''$, and the rest velocity is at $\nu_{o}(\rm{HEL}) = 1140\rm{kms^{-1}}$. A single pointing was used, with $\rm{FOV} = 25''$, $17''$, and $9''$ for band 6, 7 and 9 respectively. The absolute flux accuracy is estimated to be 10\% for all three bands. The observations do not contain short-spacing correction, and the flux is expected to be filtered out on scales $> 2''$. However, it was shown in Garc\'{i}a-Burillo et al. (\citeyear{gb19}) that almost no flux was lost due to the clumpy distribution of the gas. Assuming a distance to NGC 1068 of 14 Mpc (Bland-Hawthorn et al. \citeyear{bh97}), we use $1''$ = 70 pc for physical scales. The resolution for our observations at $\sim0.1''$, therefore, corresponds to roughly 7 pc on a linear scale.

\begin{table*}
\centering
\begin{tabular}{cccc}
\hline
\hline
Transition & CO(2-1) & CO(3-2) & CO(6-5) \\
\hline
ALMA Band & 6 & 7 & 9 \\
Frequency (GHz) & 230.54 & 345.79 & 691.47 \\
Resolution (angular scales) & $0.113'' \times 0.068''$ & $0.13'' \times 0.065''$ & $0.092'' \times 0.066''$ \\
Resolution (linear scales) & 7.91 $\times$ 4.76 pc & 9.1 $\times$ 4.55 pc & 6.44 $\times$ 4.62 pc \\
Sensitivity (mJy beam\(^{-1}\)) & 0.11 & 0.23 & 1.9 \\
LAS ($''$) & 1.3 & 1.8 & 2 \\
FOV ($''$) & 25.4 & 16.9 & 9 \\
\hline
\end{tabular}
\caption{Summary of observations presented in Garc\'{i}a-Burillo et al. (\citeyear{gb19}) for the CO(2-1) and CO(3-2) transitions, and a combined observation dataset from Garc\'{i}a-Burillo et al. (\citeyear{gb19}) and Gallimore et al. (\citeyear{gallimore_etal_2016}) for the CO(6-5) transition.}
\label{tab1}
\end{table*}

%--------------------------------------------------------------------
\section{CO maps}
\label{comaps}

Moment maps from the Common Astronomy Software Applications package (CASA; The CASA Team et al. \citeyear{casateam_etal_2022}), the primary data processing software for ALMA, are useful tools for visualizing basic physical properties from interferometric data before more detailed analyses. In this section, we present three different moment maps for each CO rotational transition, the velocity-integrated map (Moment 0 map), the velocity-width map (Moment 2 map), and the mean velocity map (Moment 1 map) to trace the total emission, the intensity weighted velocity dispersion, and the intensity weighted line of sight velocity, respectively.

\subsection{Velocity-integrated, velocity-width, and residual mean velocity maps}
\label{vivwrmv}

\begin{figure}
\resizebox{0.475\textwidth}{!}{\includegraphics{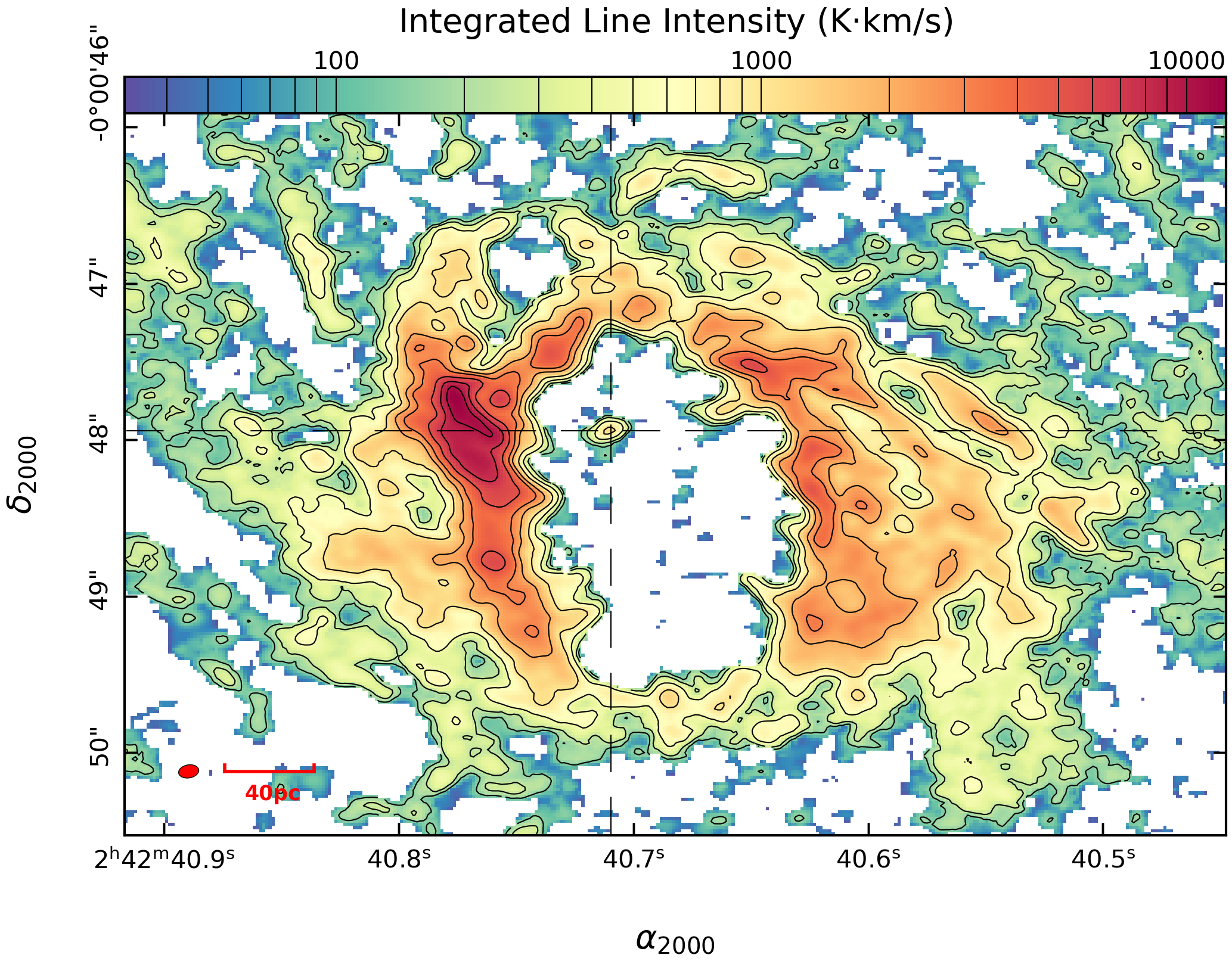}}
\caption{CO(2-1) velocity-integrated map in units of $\rm{K\,km\,s^{-1}}$ obtained with ALMA in the CND of NGC 1068. The color ranges from the minimum to the maximum value in the logarithmic scale, and the contour covers the same extent with levels 3$\sigma$, 5$\sigma$, 10$\sigma$, 20$\sigma$, 40$\sigma$, 60$\sigma$, 100$\sigma$--250$\sigma$ in steps of $50\sigma$, where $1\sigma = 50.6\,\rm{K\,km\,s^{-1}}$. Here, the $1\sigma$ value differs slightly from Garc\'{i}a-Burillo et al. (\citeyear{gb19}). We integrated over all channels instead of restricting to a certain range around $\rm v_{sys}$, which does not affect the velocity-integrated maps (this also applies to the CO(3-2) and CO(6-5) transitions, in comparison to Garc\'{i}a-Burillo et al.\ (\citeyear{gb16})). The smallest common beam size is indicated at the lower left corner of the figure, along with a linear scale indicator of 40 pc. The AGN position is also specified with two dashed lines.}

\label{fig:1}       
\end{figure}

We reproduced velocity-integrated intensity maps in K km $\rm{s^{-1}}$ for CO(2-1), CO(3-2), and CO(6-5) transitions from Garc\'{i}a-Burillo et al. (\citeyear{gb19}) and Garc\'{i}a-Burillo et al. (\citeyear{gb16}) respectively. Contrary to the original works, we first converted intensities in the original data from mJy $\rm{beam^{-1}}$ to K using
\begin{equation}\label{eqn1}
    \rm{T(K)}=\frac{1.22\times10^3}{\nu^2\theta_{\rm{maj}}\theta_{\rm{min}}}\rm{I}(\rm{mJy\ beam^{-1}})\,,
\end{equation}
where $\nu$ is in the unit of GHz, and $\theta_{\rm{maj}}$ and $\theta_{\rm{min}}$ are the major and minor axes of the beam in seconds of arc, respectively. We also adopted $3\sigma$ clipping to eliminate noise from the data. After following the same approach as Garc\'{i}a-Burillo et al. (\citeyear{gb19}) and Garc\'{i}a-Burillo et al. (\citeyear{gb16}), the resulting velocity-integrated maps were reduced to a common resolution of $0.13''\times0.833''$ with PA=$-80.3^\circ$, the smallest common beam of the three CO observations, through \verb|imsmooth| task in CASA using a Gaussian convolution kernel. The spatial coordinates of the smoothed maps were subsequently shifted based on the CO(3-2) transition map after applying \verb|imregrid| in CASA. Smoothing and regriding of the maps were conducted for more convenient comparisons among three CO observations. These velocity-integrated maps are displayed in Fig. \ref{fig:1} for the CO(2-1) transition as well as in Fig. \ref{fig:a1} \& \ref{fig:a2} in Appendix \ref{appa:addcom} for the CO(3-2) and CO(6-5) transitions respectively.

\begin{figure}
\resizebox{0.475\textwidth}{!}{\includegraphics{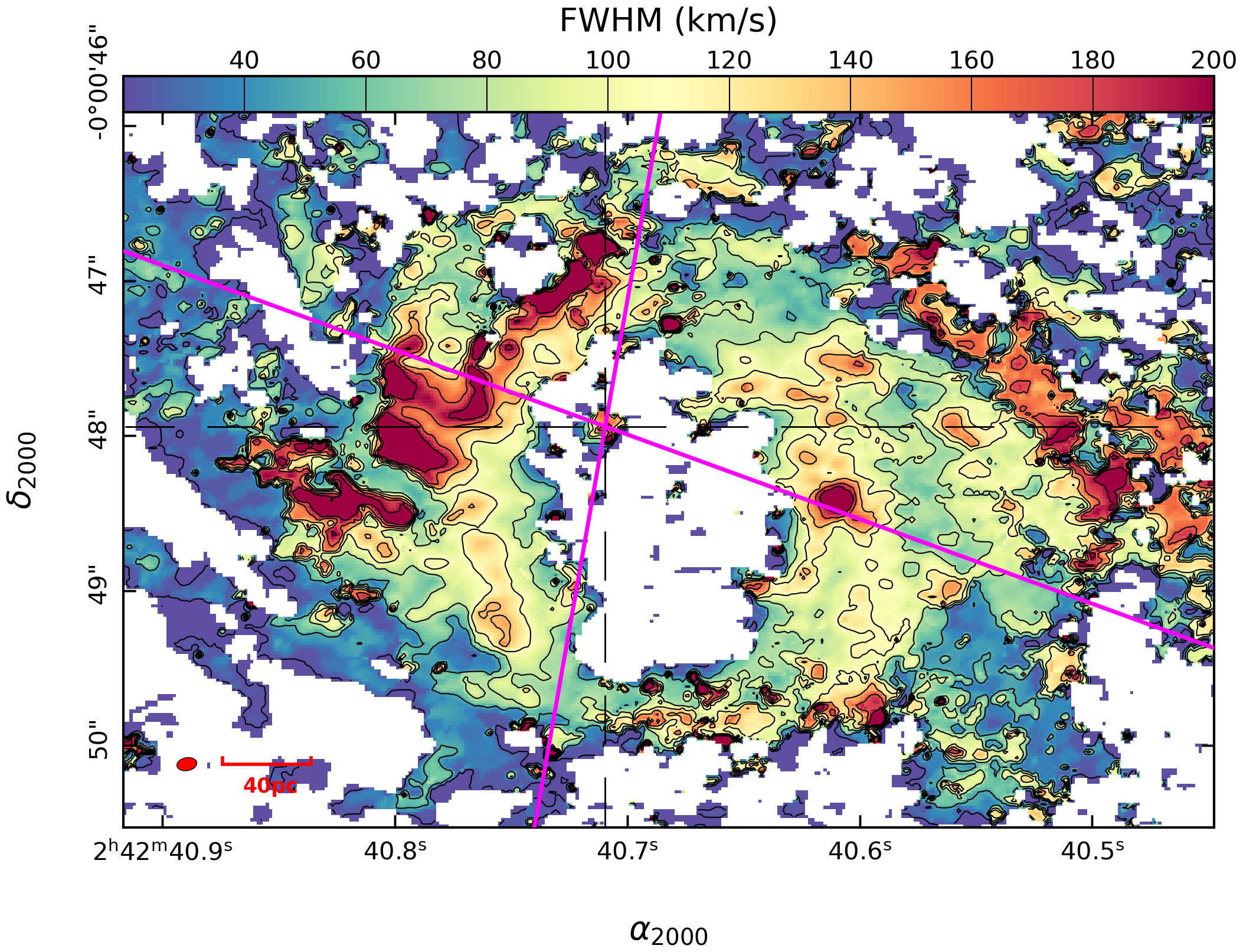}}
\caption{CO(2-1) velocity-width map of the CND in units of FWHM. The color scale spans the range [20, 200]$\rm{km\,s^{-1}}$, and the contours traverse the same range with a step size of 30 $\rm{km\,s^{-1}}$. The magenta lines trace the borders of the previously constructed AGN wind bicone (Das et al. \citeyear{das_etal_2006}). The remaining markers and symbols are the same as in Fig. \ref{fig:1}.}

\label{fig:2}      
\end{figure}

We also reproduced smoothed and regrided velocity-width maps in FWHM with the common resolution (Fig. \ref{fig:2} for the CO(2-1) transition as well as Fig. \ref{fig:a3} \& \ref{fig:a4} in Appendix \ref{appa:addcom} for the CO(3-2) and CO(6-5) transitions respectively) based on Garc\'{i}a-Burillo et al. (\citeyear{gb19}), using 3$\sigma$ clipping, to investigate molecular line widths throughout the CND.

\begin{figure}
\resizebox{0.475\textwidth}{!}{\includegraphics{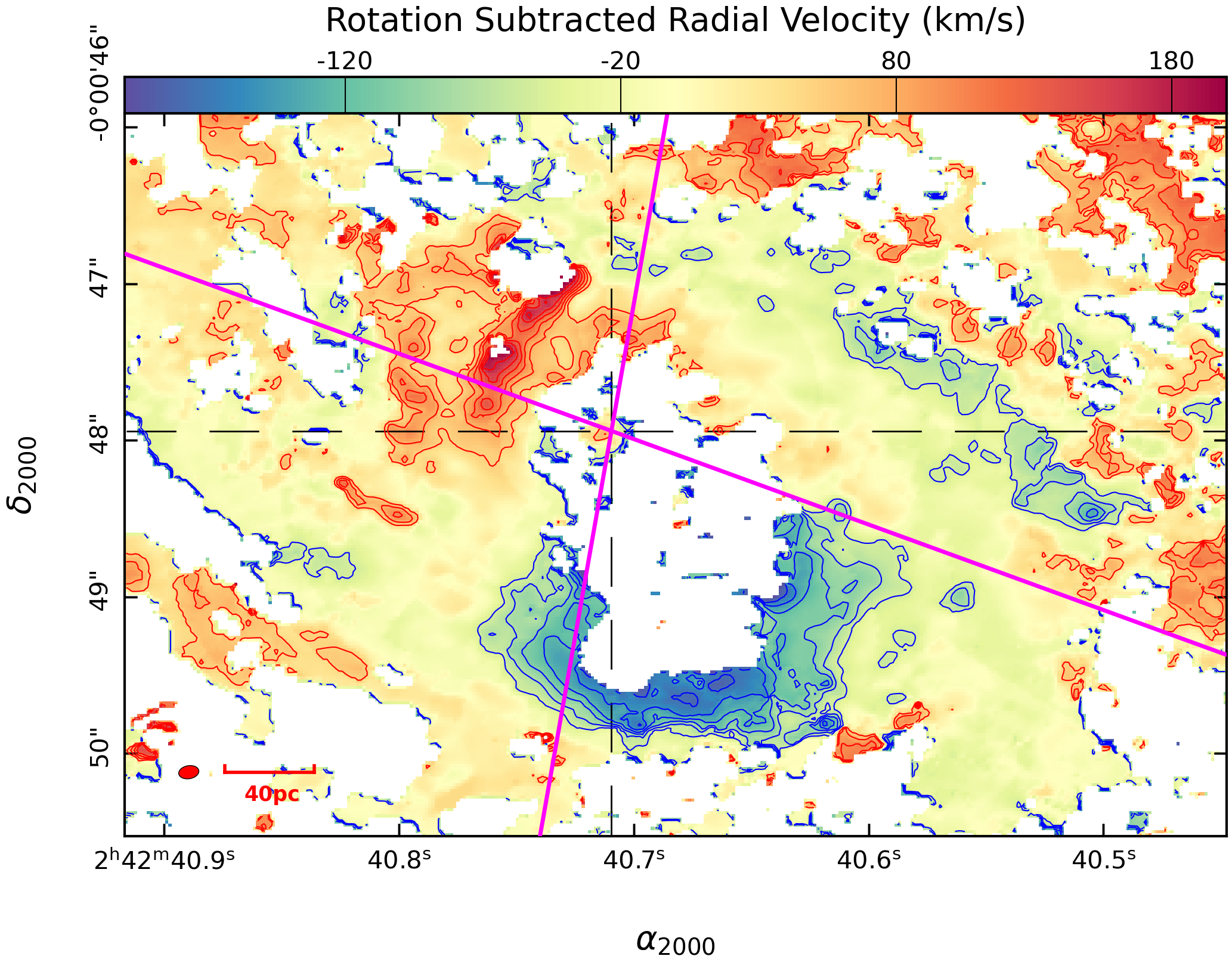}}
\caption{CO(2-1) residual mean-velocity field of the CND after the subtraction of projected CND rotation curve produced by Garc\'{i}a-Burillo et al. (\citeyear{gb19}). The color scale is within the range [-200, 200]$\rm{km\,s^{-1}}$ relative to the mean motion of the galaxy, $\rm{v_{sys}}=1120\rm{km\,s^{-1}}$, which is set to $0\,\rm{km\,s^{-1}}$. Blue (red) contours span from -200 (+50) to -50 (+200) $\rm{km\,s^{-1}}$ with a spacing of 25 $\rm{km\,s^{-1}}$ relative to $\rm{v_{sys}}$. The symbols and markers are the same as in Fig. \ref{fig:2}.}

\label{fig:3}       
\end{figure}

We likewise reproduced the smoothed and regrided residual mean velocity map from Garc\'{i}a-Burillo et al. (\citeyear{gb19})  with the common resolution for all CO transitions (Fig. \ref{fig:3} for the CO(2-1) transition as well as Fig. \ref{fig:a5} \& \ref{fig:a6} in Appendix \ref{appa:addcom} for the CO(3-2) and the CO(6-5) transitions respectively) after $3\sigma$ clipping.

\subsection{Region selections based on outflow goemetry}
\label{rsoutflow}

\begin{figure}
\resizebox{0.475\textwidth}{!}{\includegraphics{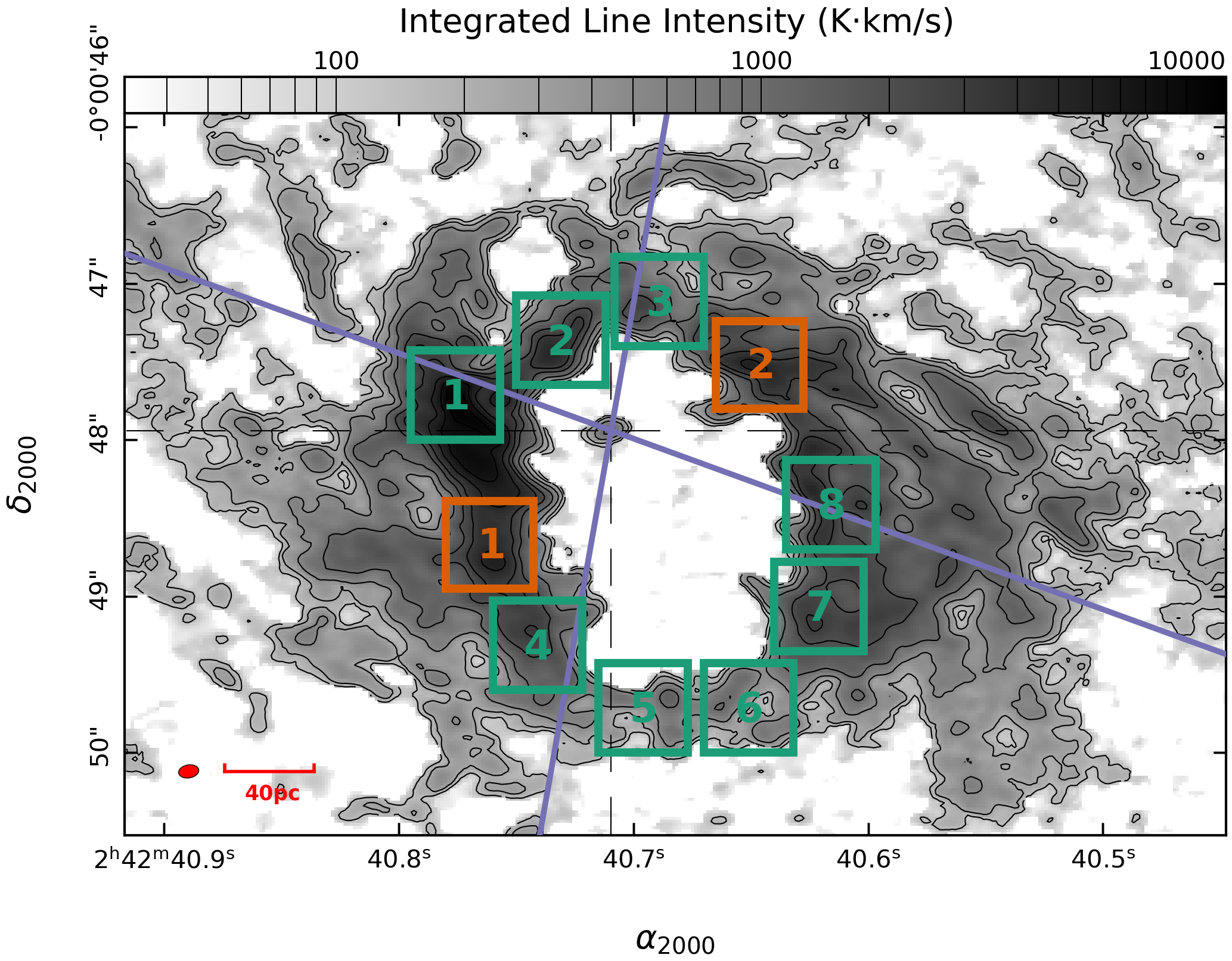}}
\caption{Region selection based on the AGN wind bicone indicated by the purple lines. The green regions are located either within the bicone or along the edge of the bicone, while the orange regions are outside the bicone. The numbers marked within each region match those in region codes displayed in Table \ref{tab2} and are defined in Sect. \ref{rsoutflow}. The background plot, including markers and symbols, is the de-colored CO(2-1) velocity-integrated map from Fig. \ref{fig:1}.}

\label{fig:4}      
\end{figure}

\begin{table*}[!htbp]
\centering
\begin{tabular}{c c c}
\hline
\hline
Region & Right Ascension & Declination \\ % Repeat this pattern 31 times
\hline
G1\_40 (N) & 02h42m40.776s & $\rm{-00^{\circ}00'47.714''}$ \\
G2\_40 (N) & 02h42m40.731s & $\rm{-00^{\circ}00'47.364''}$ \\
G3\_40 (N) & 02h42m40.689s & $\rm{-00^{\circ}00'47.114''}$ \\
G4\_40 (S) & 02h42m40.741s & $\rm{-00^{\circ}00'49.314''}$ \\
G5\_40 (S) & 02h42m40.696s & $\rm{-00^{\circ}00'49.714''}$ \\
G6\_40 (S) & 02h42m40.651s & $\rm{-00^{\circ}00'49.714''}$ \\
G7\_40 (S) & 02h42m40.621s & $\rm{-00^{\circ}00'49.064''}$ \\
G8\_40 (S) & 02h42m40.616s & $\rm{-00^{\circ}00'48.414''}$ \\
O1\_40 (E) & 02h42m40.761s & $\rm{-00^{\circ}00'48.514''}$ \\
O2\_40 (W) & 02h42m40.646s & $\rm{-00^{\circ}00'47.664''}$ \\
\hline
\end{tabular}
\caption{Coordinates in right ascension and declination at centers of regions sampled across the CND based on outflow. Definitions of the region numbers are stated in Sect. \ref{rsoutflow}. The location of each region within the CND is also provided in brackets (e.g., ``N'' stands for the northern CND).}
\label{tab2}
\end{table*}

In order to investigate the locations of molecular gas dynamically affected by the kinematics of the AGN wind and the radio jet in detail, we selected regions across the CND (Fig. \ref{fig:4}). The coordinates of these sampled regions are displayed in Table \ref{tab2}. Despite the high spatial resolution of our data, $\sim0.1''$ corresponding to a linear scale of 7pc, we used squared regions with a linear size of 40pc, matching an angular size of $\sim0.57''$, due to the high variance of physical behaviors across adjacent regions at smaller scales. We also defined these regions based on the AGN wind bicone (Das et al. \citeyear{das_etal_2006}), where the green-colored regions are situated either within the bicone or along the edge of the model, while the orange regions are completely outside the bicone. These regions were also sampled close to the inner edge of the CND, supposedly under more influence from the ionized outflow from the central engine compared to the CND outskirt. To prevent mixing up different regions defined across the CND, we adopted a naming scheme of the form ``inside/outside bicone (in color)''+``region number''+``\_size''+`` (position inside the CND)'' (e.g., G1\_40 (N) for green region 1 inside or along the edge of the AGN bicone in the northern CND and O2\_40 (W) for orange region 2 outside the bicone in the western CND). These regions (see Table 2) were used for the physical and kinematics interpretation of molecular gas inside the CND, as shown in the following sections.

\section{Physical properties of the CND}
\label{phys}

In this section, we construct a temperature and density map of the CO gas, assuming LTE conditions. Knowledge of the temperature and excitation conditions could help discern between gas swept by the outflow and gas with high temperature, which might indicate whether the outflowing gas in the northeastern part of the CND is being dynamically influenced by the ionized AGN wind.

\subsection{LTE analysis}
\label{LTE}

Assuming optically thin LTE conditions, we quantitatively estimated the total gas column densities with
\begin{equation}\label{eqn2}
    N=\frac{N_{\rm{u}}\textit{Z}}{g_{\rm{u}}e^{-E_{\rm{u}}/\rm{k}\textit{T}_k}}
\end{equation}
where $Z=\rm{k}\textit{T}_{k}/\rm{h}\textit{B}_{\rm{rot}}$ is the partition function ($B_{\rm{rot}}$ is the rotational constant of CO, and h is the Planck constant), k is the Boltzmann constant, $g_{\rm{u}}$ is the statistical weight of the upper-level u, $E_{\rm{u}}$ is the energy of the upper-level u, $T_{\rm{k}}=T_{\rm{rot}}$ is the kinetic temperature of the gas which is equal to the rotational temperature if all levels are thermalized, and
\begin{equation}\label{eqn3}
    N_{\rm{u}}=\frac{8\pi\rm{k}\nu^2\textit{I}}{\rm{hc^3}\textit{A}_{\rm{ul}}}
\end{equation}
is the column density of level u. Here, $\nu$ is the transition frequency, $I$ is the integrated line intensity in units of $\rm{K\,km\,s^{-1}}$, and $A_{\rm{ul}}$ is the Einstein A coefficient for transitions from upper level u to lower level l. When calculating $I$ for each of the 40 $\times$ 40 pc regions defined in Sect. \ref{rsoutflow} for each CO transition, we averaged only over available values within each region, ignoring pixels without detected emission (eliminated by 3$\sigma$ clipping). The effective area filling factor is thus the fraction of the region covered by remaining emissions\footnote{We do not treat the filling factor as a free parameter because it is difficult to constrain with only three CO lines.}. By calculating the total column densities, one can also estimate the rotational temperatures based on rotational diagrams of upper-level column densities from three CO transitions as an intermediate step, which relates the level column densities per statistical weight to the upper-level energy above ground state. If the LTE condition is satisfied, the gas level populations (i.e., $n_{\rm u}$ and $n_{\rm l}$ for the upper and lower-level number densities) follow that of the Boltzmann distribution,
\begin{equation}\label{eqnboltz}
    \frac{n_{\rm u}}{n_{\rm l}}=\frac{g_{\rm u}}{g_{\rm l}}\rm{e}^{-\frac{\mathit{E}_{\rm ul}}{\rm{k}\mathit{T_{\rm{kin}}}}}\,,
\end{equation}
which simplifies to a linear relation ($\rm y=\mathit{a}x$),
\begin{equation}\label{eqnboltz}
    \rm ln\left(\frac{\mathit{n}_u}{\mathit{g}_u}\right)-ln\left(\frac{\mathit{n}_l}{\mathit{g}_l}\right) = -\frac{\mathit{E}_{\rm ul}}{\rm{k}\mathit{T}_{\rm kin}}\,,
\end{equation}
where the left-hand side is the difference in y ($\delta \rm y$), and the right-hand side is the difference in x ($\delta \rm x = \mathit{E}_{\rm ul}/\rm k$) multiplied by the slop ($-1/T_{\rm kin}$). As a result, the rotational temperature is associated with the fitted slope, $a$, of a natural logarithmic rotational diagram through $T_{\rm{rot}}\sim T_{\rm kin}=-1/a$. While, in reality, it is unlikely that all the gas is in the LTE condition, total CO column densities and rotational temperatures derived following the formulations above can be instructive in providing lower limits on the actual column densities and actual kinetic temperatures of the gas.

\begin{figure*}
\centering
\resizebox{0.75\textwidth}{!}{\includegraphics{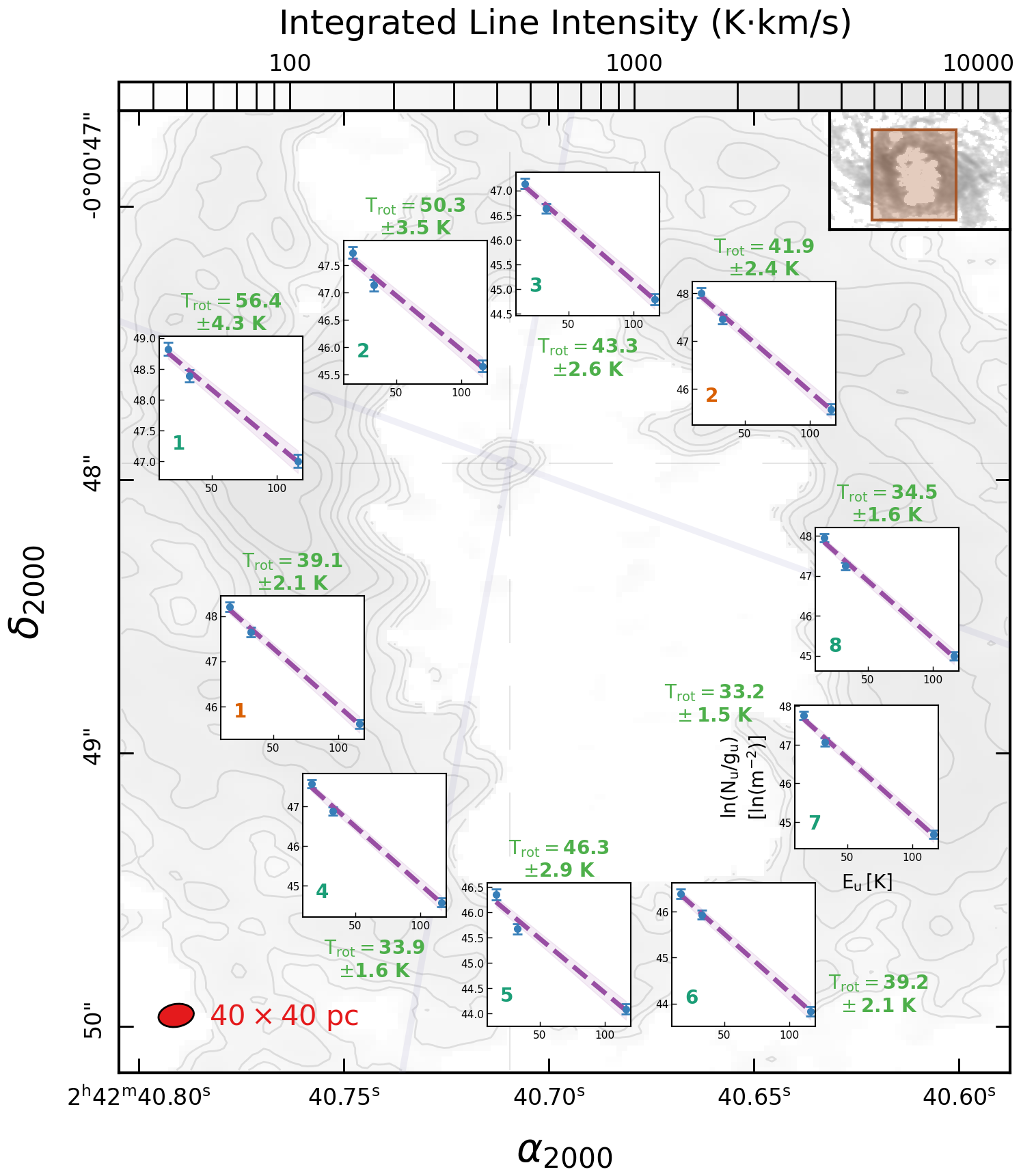}}
\caption{Rotational diagrams of all regions defined based on the outflow occupying the same locations on the background plot as assigned in the region definition in Sect. \ref{rsoutflow}. Region numbers are listed in Table \ref{tab2}. The y-axis of the rotational diagrams is in the natural logarithmic scale (contrary to the ``$\rm log_{10}$'' scale used in some other literature), while the x-axis is in the linear scale, and their labels are showcased for region G6\_40 (S). The blue data points with error bars within each plot are the weighted upper-level column densities from three CO transitions at their respective upper energy levels. The purple dashed line and the corresponding light purple shaded area are the fit based on the LTE condition and its uncertainty. The calculated rotational temperatures and their uncertainties are in green. The background plot is a cutout from Fig. \ref{fig:1}, and the cutout region is indicated in brown in the inset plot at the upper-right corner. The AGN wind bicone was added as shaded solid purple lines in the background plot, and the sampled region size is indicated at the lower-left corner. All the other symbols and markers of the background plot are the same as Fig. \ref{fig:1}.}

\label{fig:8}       % Give a unique label to the figure.
\end{figure*}

Following the above process, we first computed rotational temperatures for all the sampled regions by fitting the rotational diagrams. For calculating the column densities, the partition function was computed by fitting the available data from CDMS (The Cologne Database for Molecular Spectroscopy) before assigning the corresponding rotational temperature to the fitted function (Goorvitch \citeyear{goorvitch_etal_1994}; Gendriesch et al. \citeyear{gendriesch_etal_2009}). The total column densities were then calculated using the upper-level column density, the partition function, the rotational temperature, and the upper-level energy (also taken from CDMS) of each transition for each region. The error bars for column densities and rotational temperatures were propagated after assuming a 10\% calibration error combined with the RMS uncertainty on integrated line intensities.

\begin{table*}[!htbp]
\centering
\begin{tabular}{c c c c c}
\hline
\hline
Region & Rotational Temperature ($T_{\rm{rot}}$) & CO(2-1) & CO(3-2) & CO(6-5) \\ % Repeat this pattern 31 times
\hline
G1\_40 (N) & $\rm{56.4\pm4.3}$ & $44.1\pm5.8$ & $38.3\pm5.2$ & $42.0\pm8.5$ \\
G2\_40 (N) & $\rm{50.3\pm3.5}$ & $13.7\pm1.8$ & $10.5\pm1.4$ & $12.5\pm2.6$ \\
G3\_40 (N) & $\rm{43.3\pm2.6}$ & $6.86\pm0.85$ & $6.10\pm0.79$ & $6.6\pm1.4$ \\
G4\_40 (S) & $\rm{33.9\pm1.6}$ & $9.3\pm1.1$ & $7.65\pm0.96$ & $8.7\pm1.8$ \\
G5\_40 (S) & $\rm{46.3\pm2.9}$ & $3.27\pm0.41$ & $2.37\pm0.31$ & $2.91\pm0.58$ \\
G6\_40 (S) & $\rm{39.2\pm2.1}$ & $3.06\pm0.37$ & $2.97\pm0.38$ & $3.02\pm0.60$ \\
G7\_40 (S) & $\rm{33.2\pm1.5}$ & $11.0\pm1.3$ & $9.1\pm1.1$ & $10.2\pm2.0$ \\
G8\_40 (S) & $\rm{34.5\pm1.6}$ & $13.6\pm1.6$ & $10.9\pm1.4$ & $12.6\pm2.5$ \\
O1\_40 (E) & $\rm{39.1\pm2.1}$ & $18.8\pm2.3$ & $16.5\pm2.1$ & $17.9\pm3.6$ \\
O2\_40 (W) & $\rm{41.9\pm2.4}$ & $16.1\pm2.0$ & $13.8\pm1.8$ & $15.2\pm3.0$ \\
\hline
\end{tabular}
\caption{LTE rotational temperatures ($T_{\rm{rot}}$) and total CO column densities ($N$) from the original $40\times40\,\rm{pc}$ regions defined based on outflow in units of K and $10^{17}\,\rm{cm^{-2}}$ respectively. The total CO column density calculated from each CO transition is indicated by the name of each transition in the header. The location of each region within the CND is also provided in brackets (e.g., ``N'' stands for the northern CND).}
\label{tabb1}
\end{table*}

We produced rotational diagrams to visualize rotational temperatures and LTE conditions directly across the CND (see Fig. \ref{fig:8}). All regions defined based on outflow display well-fitted LTE models to the weighted upper-level column densities. Hence, gas within regions defined based on the outflow might be optically thin and in the LTE condition. 

\begin{figure}
\centering
\resizebox{0.475\textwidth}{!}{\includegraphics{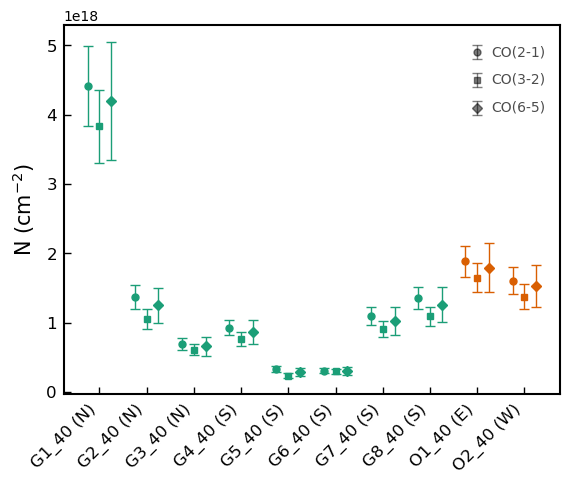}}
\caption{Total CO column densities ($N$) independently calculated from three CO transitions for all regions selected based on the outflow. The x-axis marks the region labels along with their locations around the CND (in brackets), while the y-axis indicates the total CO column densities in the unit of $\rm{cm^{-2}}$. The region code (along the x-axis) and color (the color of the data points) of each region match those in Table \ref{tab2} and are defined in Sect. \ref{rsoutflow}. The legend at the upper-right corner of the plot designates the symbols corresponding to the column densities calculated from three different CO transitions.}

\label{fig:7}       % Give a unique label to the figure.
\end{figure}

A similar conclusion can be drawn from the average total column density values (see Table \ref{tabb1} and Fig. \ref{fig:7}). We observe that for all regions defined based on the outflow, all three column densities calculated independently from three CO transitions roughly match the same level within each region. Column densities also vary across different regions. For regions sampled based on outflow, G1\_40 (N) has the highest column density and is higher than column densities from other regions by a factor of $\sim4$. This region resides at the location with the highest integrated line intensity near the eastern knot (E-knot; see Garc\'{i}a-Burillo et al. \citeyear{gb19} and references therein), as seen from velocity-integrated maps. 

\begin{figure}[!htbp]
\resizebox{0.475\textwidth}{!}{\includegraphics{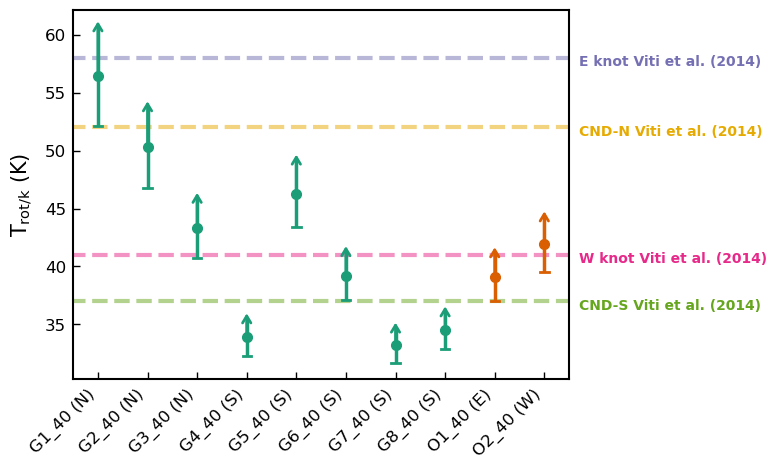}}
\caption{Summary of rotational temperatures for all regions sampled around the CND defined based on the outflow in Sect. \ref{rsoutflow}. The x-axis marks the region labels along with their locations around the CND (in brackets), while the y-axis indicates the rotational temperatures in the unit of K. The region code (along the x-axis) and color (the color of the data points) of each region match those in Table \ref{tab2} and are defined in Sect. \ref{rsoutflow}. The purple, yellow, pink, and lime dashed lines indicate the rotational temperatures at four different locations of the CND measured by Viti et al. (\citeyear{viti_etal_2014}).}

\label{fig:a23}       % Give a unique label to the figure.
\end{figure}

We also summarise rotational temperatures in Table \ref{tabb1} and Fig. \ref{fig:a23}. Overall, our rotation temperature results at different positions of the CND roughly match with lower resolution results from Viti et al. (\citeyear{viti_etal_2014}). The highest rotational temperature is at G1\_40 (N), corresponding to its high CO column density and high integrated line intensity, and, in general, rotational temperatures in the northern part of the CND are higher than the southern part of the CND, possibly indicating higher excitation due to the proximity of the molecular gas to the central engine\footnote{We note that the two regions O1\_40 (E) and O2\_40 (W) also have slightly higher temperatures relative to regions in the south despite being outside the bicone. The small temperature difference might be caused by the irradiation of molecular gas via high-energy photons/cosmic rays or ionized bubbles induced by the expanding radio jet.}. However, the difference between rotational temperature values in the northern and southern regions of the CND is not significant. In contrast, from Fig.\ref{fig:2}, the difference in velocity widths between the northeastern and the southern parts of the CND are considerable. This difference likely reflects the dynamic asymmetry of the molecular outflow. Specifically, the molecular gas in the northeastern CND appears to offer greater resistance to the ionized AGN wind or the radio jet, whereas the molecular gas in the southern CND has largely been dissociated. In fact, the morphology of the radio jet nebula imaged by the Very Large Array (VLA) at cm wavelengths (Gallimore et al. \citeyear{gallimore_etal_96}) and by ALMA at mm wavelengths (Garc\'{i}a-Burillo et al. \citeyear{gb17}) exhibits a pronounced northeast-to-south asymmetry. The northeastern radio lobe displays a distinct bow-shock arc structure and spans a smaller radial distance from the AGN compared to the southern radio lobe, which has a broader angle shape. These observations suggest that the radio jet in the southern CND is more evolved, indicating extensive past destruction of molecular gas in the southern CND.

\subsection{Non-LTE analysis}
\label{xLTE}

\begin{table*}[!htbp]
\centering
\begin{tabular}{c c c c c c}
\hline
\hline
Region & Line Width & $\rm H_2$ Number Density & CO(2-1) & CO(3-2) & CO(6-5) \\
 & ($\Delta{v_{\rm FWHM}}$; $\rm km\,s^{-1}$) & ($n_{\rm H_2}$; $\rm cm^{-3}$) & & & \\
\hline
% Add 20 more rows following this format
G1\_40 (N) & 160 & $10^5$ & 0.66 & 1.2\dag & 1.3\dag \\
 & & $10^6$ & 0.58 & 1.0\dag & 1.1\dag \\
G2\_40 (N) & 113 & $10^5$ & 0.33 & 0.60 & 0.50 \\
 & & $10^6$ & 0.29 & 0.51 & 0.46 \\
G3\_40 (N) & 80 & $10^5$ & 0.33 & 0.57 & 0.35 \\
 & & $10^6$ & 0.29 & 0.48 & 0.35 \\
G4\_40 (S) & 86 & $10^5$ & 0.59 & 0.91 & 0.33 \\
 & & $10^6$ & 0.54 & 0.81 & 0.36 \\
G5\_40 (S) & 73 & $10^5$ & 0.14 & 0.26 & 0.17 \\
 & & $10^6$ & 0.13 & 0.21 & 0.17 \\
G6\_40 (S) & 83 & $10^5$ & 0.17 & 0.29 & 0.14 \\
 & & $10^6$ & 0.15 & 0.25 & 0.15 \\
G7\_40 (S) & 85 & $10^5$ & 0.72 & 1.1\dag & 0.39 \\
 & & $10^6$ & 0.66 & 0.98 & 0.42 \\
G8\_40 (S) & 112 & $10^5$ & 0.64 & 0.98 & 0.38 \\
 & & $10^6$ & 0.58 & 0.88 & 0.41 \\
O1\_40 (E) & 99 & $10^5$ & 0.83 & 1.3\dag & 0.74 \\
 & & $10^6$ & 0.76 & 1.2\dag & 0.72 \\
O2\_40 (W) & 90 & $10^5$ & 0.70 & 1.2\dag & 0.72 \\
 & & $10^6$ & 0.63 & 1.0\dag & 0.70 \\
\hline
\end{tabular}
\caption{Optical depth values calculated from RADEX using the LVG method for all three CO transitions across all regions defined based on the outflow. The second column indicates the different FWHM line widths ($\Delta{v_{\rm FWHM}}$) applied to different regions. The third column lists the $\rm H_2$ number density values used for each region, with two different densities applied per region. For each region, the second row (corresponding to $n_{\rm H_2}=10^6\,\rm{cm^{-3}}$) follows the same region code and line width as the row above. The last three columns display the optical depth values for the CO transitions specified in the column headers. Optical depth values $\geq1$ are marked by the ``\dag'' symbol.}
\label{tabradex}
\end{table*}

To validate the well-fitted rotational diagrams and the matching column densities indicating the LTE condition of regions defined based on the outflow, we ran simple RADEX analyses (Van der Tak et al. \citeyear{vandertak_etal_07}) with the Large Velocity Gradient (LVG) method for each region using calculated rotational temperatures in the unit of K and total CO column densities in the unit of $\rm{cm^{-2}}$ to estimate corresponding optical depths, $\tau$, for all CO transitions for all regions (see Table \ref{tabradex}). We also calculated the line width spatially averaged within each region in FWHM and in the unit of $\rm{km\,s^{-1}}$ as an input parameter. For the total $\rm{H_2}$ density (the density of the major collision partner; $n_{\rm{H_2}}$), we set up a grid of density values, $n_{\rm{H_2}}=10^4\,\rm{cm^{-3}}$, $n_{\rm{H_2}}=10^5\,\rm{cm^{-3}}$, and $n_{\rm{H_2}}=10^6\,\rm{cm^{-3}}$, followed by restricting the value based on results from Table 6 in Viti et al. (\citeyear{viti_etal_2014}). While most of the transitions for most regions result in $\tau<1$, some of the regions near the E-knot, the W-knot, and CND-S display $\tau\gtrsim1$, especially for a $n_{\rm{H_2}}=10^4\,\rm{cm^{-3}}$. However, as calculated with a RADEX grid search in Viti et al. (\citeyear{viti_etal_2014}) using different line ratios of transitions from several available molecules, a $n_{\rm{H_2}}=10^4\,\rm{cm^{-3}}$ is unlikely within the CND, which leaves four regions with $\tau\gtrsim1$, regions G1\_40 (N), G7\_40 (S), O1\_40 (E), and O2\_40 (W). For region G1\_40 (N), optical depths calculated from the CO(3-2) and the CO(6-5) transitions with both $n_{\rm{H_2}}=10^5\,\rm{cm^{-3}}$ and $n_{\rm{H_2}}=10^6\,\rm{cm^{-3}}$ are slightly higher than 1 but below 1.5. For regions O1\_40 (E) and O2\_40 (W), only optical depths calculated from the CO(3-2) transition with both $n_{\rm{H_2}}=10^5\,\rm{cm^{-3}}$ and $n_{\rm{H_2}}=10^6\,\rm{cm^{-3}}$ are slightly higher than 1 but below 1.5. For region G7\_40 (S), only the optical depth calculated from the CO(3-2) transition with $n_{\rm{H_2}}=10^5\,\rm{cm^{-3}}$ is at $\sim1.1$. Regardless, the opacity of the majority of gas within the CND is low and roughly matches the LTE condition.

\subsection{Line ratios}
\label{lr}

Previous studies using lower spatial resolution data ($0.5''$) have shown that individual molecular line ratios are correlated with the UV and X-ray illumination of the molecular gas in the CND from the central engine which can be used to infer certain physical conditions of the molecular outflow, especially the excitation conditions of the gas (Garc\'{i}a-Burillo et al. \citeyear{gb14}; Viti et al. \citeyear{viti_etal_2014}). We, therefore, computed the CO(3-2)/(2-1), CO(6-5)/(2-1), and CO(6-5)/(3-2) ratio maps (Fig. \ref{fig:5}), using velocity-integrated intensity maps, to investigate the physical conditions of the CND. In general, due to a lack of emission of the CO(6-5) transition in the outer part of the CND, the inner CND has higher excitation compared to the outer CND. Within detected regions of the CND, the line ratios are higher in the northeast compared to the south, especially for the positions close to the northern and eastern inner edges of the structure. For the CO(3-2)/(2-1) ratio map, high excitation can also be spotted across the northeastern outskirt of the CND, further away from the central engine, as well as at a small section, located around $\rm{\alpha_{2000}=2^{h}42^{m}40.63^{s}}$ and $\rm{\delta_{2000}=-0^\circ00'50''}$, along the southern outer edge of the CND. For CO(6-5)/(2-1) and CO(6-5)/(3-2) ratio maps, high ratios also appear along the southern outer edge of the CND and around the northern outskirt of the structure. Again, higher ratios around the northeastern part of the CND when compared to the southern part of the CND might be a result of the proximity of the northeastern CND to the central engine.

\subsubsection{Region selections based on excitation}
\label{rsexcitation}

Besides selecting regions based on the outflow (see Sect. \ref{rsoutflow}), we also sampled regions based on excitation using the ratio maps (see Fig. \ref{fig:6} and Table \ref{tab3}). We separated the regions based on excitation levels relative to the root-mean-square (RMS) values from each ratio map. The dashed regions contain line ratios higher than 3 times the RMS value, while the solid regions contain ratios exceeding 2 times the RMS value. In the case of the CO(6-5)/(3-2) ratio, no region with ratios higher than 3 times the RMS value was found. Again, the green and orange colors indicate the location of each region with respect to the AGN wind bicone model, as discussed in Sect. \ref{rsoutflow}. We adopted a slightly different scheme for region definitions based on excitation in the form of ``high ratio''+``inside/outside bicone (in color)''+``region number''+``\_size''+``\_ratio'' (e.g., HRG1\_40\_3221 for region 1 inside or along the edge of the AGN wind bicone with high CO(3-2)/(2-1) ratio above 3 times the RMS value and HRO2\_40\_6532 for region 2 outside the bicone with high CO(6-5)/(3-2) ratio above 2 times the RMS value)\footnote{``region number''s here are used to distinguish high-excitation regions sampled under the same line ratio and are indicating different regions than in Sect. \ref{rsoutflow}.}.

\subsubsection{LTE analysis for high-ratio regions}
\label{lte_hr}

We likewise conducted the LTE analysis for gas within these high-excitation regions. Rotational diagrams indicate that LTE fits are worse for several regions sampled via high ratios than regions based on the outflow (see Fig. \ref{fig:a7}), which means some gas within high-ratio regions does not satisfy LTE conditions. We also found departures from LTE conditions based on the total CO column density values calculated independently using the 3 CO transitions (see Table \ref{tab3} and Fig. \ref{fig:a25}), especially for region HRG3\_40\_3221, where the column density can only be interpreted as its lower limit. For ratios associated with the CO(6-5) transition, larger discrepancies among 3 column densities were found in the northern regions compared to the southern regions. The departure from LTE conditions might imply that (i) the gas is optically thick and/or (ii) multiple gas components are enclosed within each region. Hence, rotational temperature values of these regions can only be interpreted as lower limits of their gas kinetic temperatures.

Variations of column densities and rotational temperatures (see Table \ref{tab3} and Fig. \ref{fig:a25}, \ref{fig:a24}) among different regions are also evident for regions selected based on excitation. Region HRG3\_40\_3221 has column density roughly 6 times higher than other regions with the same definition. The same region also has about 2 times higher rotational temperature when compared to regions further away from the central engine (HRG1\_40\_3221 and HRG5\_40\_3221), which might not be as dynamically or radiatively affected. For regions with high excitation in CO(6-5)/(2-1) and CO(6-5)/(3-2) ratios, their column densities are approximately at the same level, while, overall, the northern regions outside the AGN wind bicone have higher rotational temperatures than in the southern regions inside the bicone.

%--------------------------------------------------------------------
\section{Kinematics of the CND}
\label{kin}

Despite the CO gas being nearly in LTE across the CND, the average residual line profiles of several regions, as will be shown in this section, imply multiple gas components. We shall now investigate the kinematics of the CO gas in order to determine the influence of the ionized outflow on the molecular CND and the extent of the outflowing molecular gas. Kinematics analyses around the central engine of NGC 1068 have been carried out previously by multiple studies (Garc\'{i}a-Burillo et al. \citeyear{gb16}, \citeyear{gb19}; Imanishi et al. \citeyear{imanishi_etal_2020}; Impellizzeri et al. \citeyear{impellizzeri_etal_2019}). While Imanishi et al. (\citeyear{imanishi_etal_2020}) and Impellizzeri et al. (\citeyear{impellizzeri_etal_2019}) focus on kinematics of the torus, Garc\'{i}a-Burillo et al. (\citeyear{gb16}, \citeyear{gb19}) studied kinematics of the CND using position-velocity diagrams and kinematics models. In this section, we introduce the line profile analysis to systematically probe the kinematics of molecular gas in detail using the spectral information of the CO observations for all regions defined in Sect. \ref{rsoutflow} based on the outflow. 

\subsection{Single vs. Multi-component Gaussian Models}
\label{method}

We performed rotation curve subtraction, using the rotation curve from Garc\'{i}a-Burillo et al. (\citeyear{gb19}), on the original data from three CO line observations without clipping. The rotation curve was fitted based on the mean velocity distribution covering the central region of the galaxy in the CND coordinates of NGC 1068. We first projected the rotation curve to sky coordinates. After conversion to brightness temperature in units of $\rm{K\,km\,s^{-1}}$, emissions across all channels in three sets of CO observations were shifted to various velocities using the projected rotation map. We then spatially averaged emissions within each region defined based on the outflow while retaining the spectral dimension.

To systematize the line profile analysis, we adopted a similar technique as Imanishi et al. (\citeyear{imanishi_etal_2020}) and fitted all the averaged line profiles using three different models, a single Gaussian, a weighted double Gaussian, and a weighted triple Gaussian model. 

\begin{align}\label{eqn4}
    S_{\nu}(v) = 
    \left\{ 
    \begin{aligned}
    & A\mathcal{G}(\mu, \sigma)=\frac{A}{\sqrt{2\pi}\sigma}e^{-\frac{(v-\mu)^2}{2\sigma^2}} \hspace{85pt} (\text{single}) \\
    & A\left[w_1\mathcal{G}(\mu_1,\sigma_1)+w_2\mathcal{G}(\mu_2, \sigma_2)\right] \hspace{64pt} (\text{double}) \\
    & A\left[w_1\mathcal{G}(\mu_1,\sigma_1)+w_2\mathcal{G}(\mu_2, \sigma_2)+w_3\mathcal{G}(\mu_3,\sigma_3)\right] \hspace{5pt} (\text{triple})
    \end{aligned}
    \right.
\end{align}

We found that in some spectra, the secondary component of the weighted double and triple Gaussian models has a limited contribution to the overall fit. Hence, we first eliminated all emissions below 3 times the background noise level from all line profiles so that the background noise does not influence the fitting results. Afterward, we employed a hierarchical scheme to choose the optimal model based on the fit. The fitted double Gaussian model was first compared with the fitted single Gaussian model using two criteria. For the first criterion, we applied the extra-sum-of-squares F-test (Lupton \citeyear{lupton_1993}) similar to Sexton et al. (\citeyear{sexton_etal_2021}) and Kova\v{c}evi\'{c}-Doj\v{c}inovi\'{c}
 et al. (\citeyear{dojvcinovic_etal_2022}). This test assumes a null hypothesis that a simpler model with fewer parameters (e.g., the single Gaussian model) is more effective at describing the averaged line profile than a more complex model with more parameters (e.g., the weighted double Gaussian model). The test was done for all the regions that passed the first criterion, and we retained all the weighted double Gaussian fit when the P-value < 0.05. To prevent overfitting the line profiles, we also applied the Akaike Information Criterion (Akaike \citeyear{akaike_73}), $AIC=2\emph{k}\rm{-2ln}\hat{\emph{L}}$, where $k$ is the number of free parameters and $\hat{L}$ is the likelihood of the line profile data given the fit, to restrict the more complex model based on its complexity (i.e., the number of parameters). The difference $\Delta AIC=AIC_{\rm single}-AIC_{\rm double}$ between the fitting results of the two models measures the degree to which the weighted double Gaussian is favored over the single Gaussian fit. Following Wei et al. (\citeyear{WWM16}), the weighted double Gaussian fit is considered ``positive'' when $\Delta AIC>2$. When a weighted double Gaussian fit was rejected, the corresponding weighted triple Gaussian fit was also discarded. The same process was used to compare the remaining weighted double and triple Gaussian models.

\subsection{Line profiles behavior}
\label{lpb}

\begin{figure*}
\centering
\resizebox{0.75\textwidth}{!}{\includegraphics{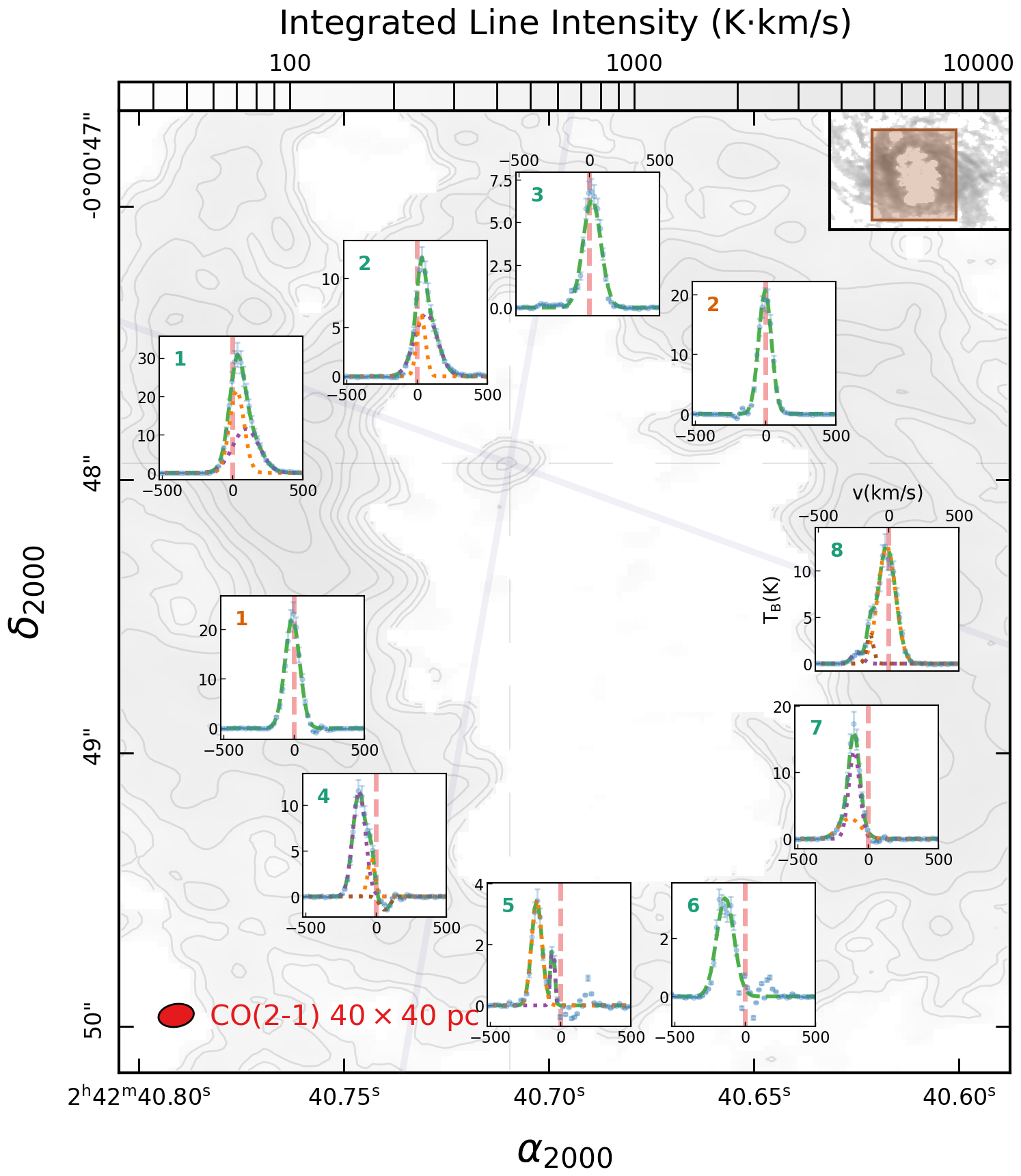}}
\caption{CO(2-1) transition line profiles of all regions defined based on the outflow occupying the exact locations on the background plot as assigned in the region definition in Sect. \ref{rsoutflow}. Region numbers, including their colors, are indicated at the upper left corner of each plot, corresponding to those appearing in the region codes in Table \ref{tab2} and are defined in Sect. \ref{rsoutflow}. The x-axis of all line profile plots indicates the radial velocity of the gas, $v$, in the unit of $\rm{km\,s^{-1}}$. The y-axes are for the amount of emission within each channel, $T_{\rm{B}}$, in the unit of K. The blue data points and error bars within each region are the observed averaged line profile overplotted by the optimal fit as a green dashed line. The mean velocity of the galaxy, $v_{\rm{sys}}$, is indicated by the vertical red dash line for each line profile plot. Suppose the averaged line profile within one region was fitted with a weighted multi-component Gaussian. All weighted Gaussian components are also provided in the plot as dotted lines using different colors (purple and orange for the first two components and brown for the third component). The transition of the line, CO(2-1), is indicated at the lower left corner of the background plot, and all other symbols are the same in the background plot as those in Fig. \ref{fig:8}.}

\label{fig:9}       % Give a unique label to the figure.
\end{figure*}

\begin{figure*}
\centering
\resizebox{0.75\textwidth}{!}{\includegraphics{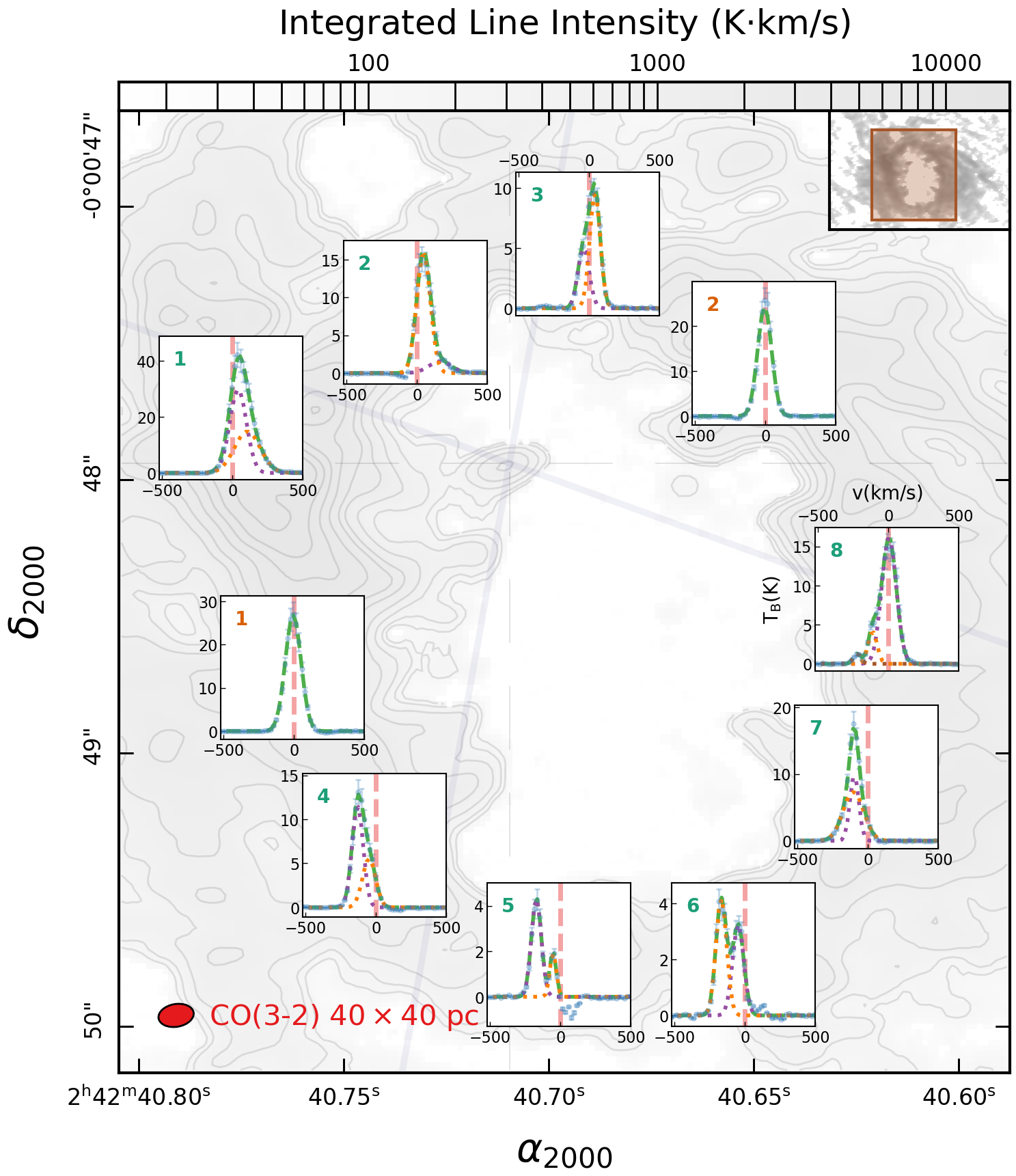}}
\caption{CO(3-2) transition line profiles of all regions defined based on the outflow occupying the exact locations on the background plot as assigned in the region definition in Sect. \ref{rsoutflow}. The background plot is the CO(3-2) velocity-integrated map from Fig. \ref{fig:a1}. The transition of the line CO(3-2) is indicated at the lower left corner of the background plot. All other symbols and markers are the same as Fig. \ref{fig:9}.}

\label{fig:10}       % Give a unique label to the figure.
\end{figure*}

\begin{figure*}
\centering
\resizebox{0.75\textwidth}{!}{\includegraphics{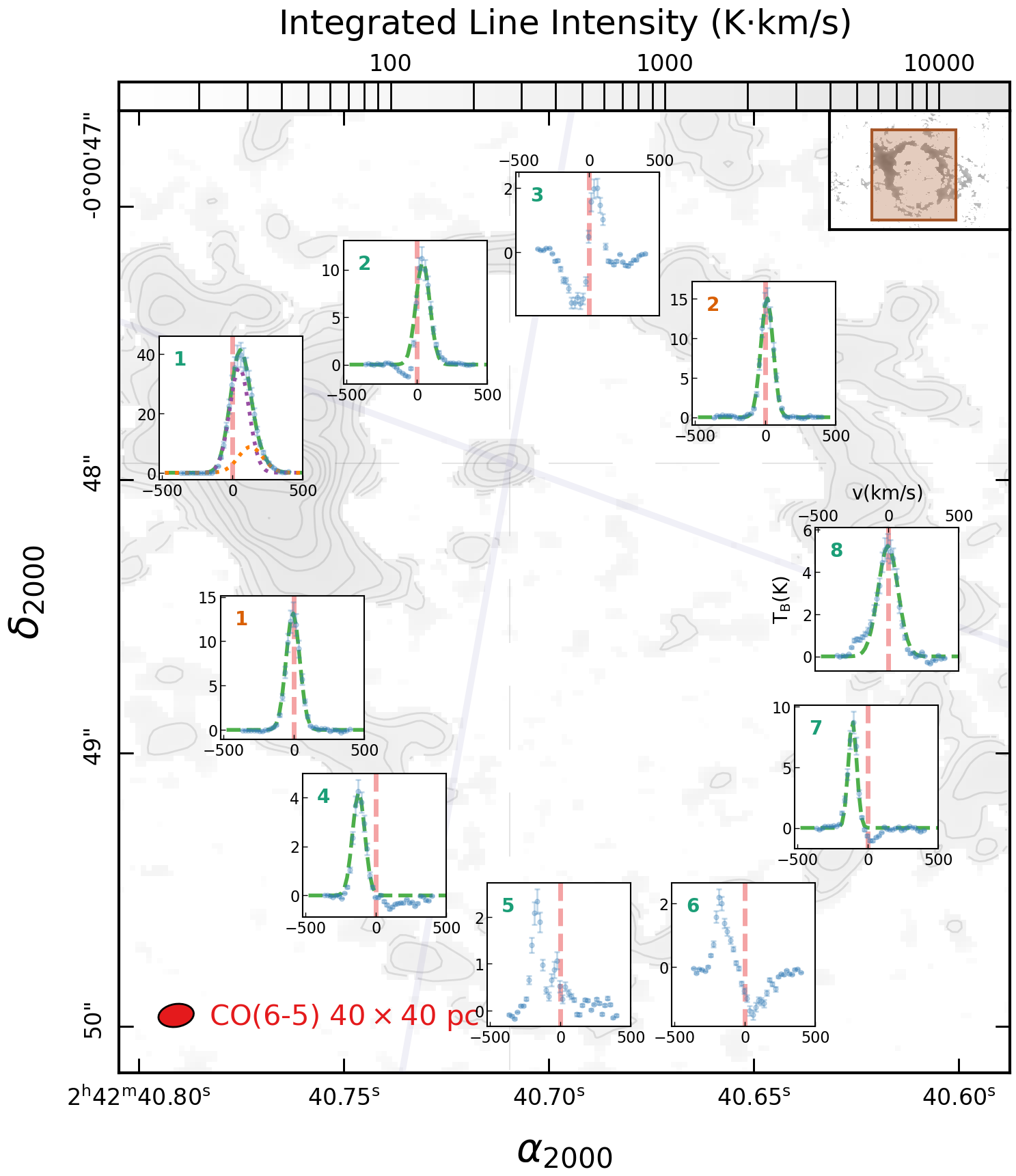}}
\caption{CO(6-5) transition line profiles of all regions defined based on the outflow occupying the exact locations on the background plot as assigned in the region definition in Sect. \ref{rsoutflow}. The background plot is the CO(6-5) velocity-integrated map from Fig. \ref{fig:a2}. The transition of the line CO(6-5) is indicated at the lower left corner of the background plot. All other symbols and markers are the same as Fig. \ref{fig:9}.}

\label{fig:13}       % Give a unique label to the figure.
\end{figure*}

The line profiles and their optimal fits for each region sampled based on the outflow are shown in Fig. \ref{fig:9}, \ref{fig:10}, \& \ref{fig:13}, and their fit results, including parameter values, are displayed in Table \ref{tabb3}. We will describe and discuss these line profile results in this section by separating them into different groups based on their behaviors and positions relative to the AGN wind bicone. When discussing a multi-component Gaussian fit, we define the first (major), secondary (minor), and third Gaussian components based on their weight parameters. Without the weight parameter, all Gaussian components are normalized based on the line profile. As a result, each Gaussian component captures the fraction of gas enclosed in each region proportional to the weight parameter.

\noindent\textbf{\underline{Regions G1\_40 (N) and G2\_40 (N)}:} 

\noindent Regions G1\_40 (N) and G2\_40 (N) (inset plots in the upper-left corners of Fig. \ref{fig:9}, \ref{fig:10}, and \ref{fig:13}) are located in the northeastern part of the CND, along and within the AGN wind bicone. In region G1\_40 (N), the emission line profiles for all three CO transitions—CO(2-1), CO(3-2), and CO(6-5)—roughly overlap in velocity space, with their bulk emission consistently appearing above the galaxy's systematic velocity at $\rm{v_{sys}}=1120\,\rm{km\,s^{-1}}$, as indicated by the red dashed lines in the figures. For all transitions, the two Gaussian components have peaks above $\rm{v_{sys}}$. 
The primary components, although closer to the mean velocity, remain redshifted by $\gtrsim30\,\rm{km\,s^{-1}}$ from $\rm{v_{sys}}$, and a spectral resolution of $20\,\rm{km\,s^{-1}}$ (channel width) is sufficient to confirm this departure. The CO(6-5) transition line profile in this region exhibits large error bars on fit parameters, potentially diverging from the results of other transitions. However, the CO(6-5) line profile shows an enhanced redshifted wing structure consistent with the CO(2-1) and CO(3-2) profiles. Similar patterns are evident in the fitting results for region G2\_40 (N), where the CO(2-1) and CO(3-2) transitions exhibit comparable redshifted features. In contrast, the CO(6-5) transition in this region is fitted with only a single redshifted Gaussian. Additionally, for region G2\_40 (N), the primary component of the CO(2-1) line profile is more significantly redshifted from the systematic velocity than its secondary component.

Several trends can be inferred from the line profiles in regions G1\_40 (N) and G2\_40 (N). Most of the gas in these regions is redshifted, with an additional redshifted wing structure characteristic of outflow dynamics. Notably, in both regions, the wing components (characterized by the widths of the fitted Gaussian profiles) exhibit broader widths than their corresponding components with smaller velocity departures. These broad-wing components likely trace the molecular outflow at the interface where the ionized AGN wind interacts with the molecular gas at the surface of the CND facing the observer. In contrast, the less redshifted components likely correspond to gas embedded deeper within the CND, dynamically influenced by the outflow but less directly exposed. Across transitions, as indicated by the weight parameters, the wing components of the CO(2-1) transition are the most prominent in both regions, while the less redshifted Gaussian components are generally stronger in higher-$\rm J$ transitions compared to lower-$\rm J$ transitions. These observations suggest that the gas associated with the wing component may be less dense than the gas corresponding to the lower-velocity component.

\noindent\textbf{\underline{Region G3\_40 (N):}}

\noindent Region G3\_40 (N), located along the northwestern edge of the AGN wind bicone, exhibits simpler line profile behavior compared to the more complex profiles seen in the northeastern regions G1\_40 (N) and G2\_40 (N). The CO(2-1) transition is well-described by a single Gaussian, with its mean velocity redshifted by less than one channel width from $\rm{v_{sys}}$. The width of this single Gaussian profile is also narrower than the wing components observed in regions G1\_40 (N) and G2\_40 (N). In contrast, the CO(3-2) line profile is best fitted with a double Gaussian, consisting of a redshifted primary component and a slightly blueshifted secondary component. The primary component likely traces gas at the edge of the molecular outflow, while the secondary component may represent gas outside the AGN wind bicone, dynamically independent from the outflow. For the CO(6-5) transition, no significant detection was observed, as the maximum or minimum value of the line profile remains below the $3\sigma$ noise threshold.

A spatial variation of line profiles is evident among the three northern regions, particularly highlighted by the differences between region G3\_40 (N) and the other two regions. The gas in region G3\_40 (N) aligns more closely with the galaxy's systematic velocity and appears less influenced by the outflow, as indicated by the CO(2-1) and CO(3-2) line profiles. This is consistent with a significant portion of the region lying outside the AGN wind bicone and its location at the edge of the line broadening and redshifted rotation-subtracted radial velocity features previously shown in Fig. \ref{fig:2} and \ref{fig:3}. In contrast, region G2\_40 (N), positioned near the center of the ionized AGN wind bicone, captures a more substantial amount of molecular outflow than region G3\_40 (N). Meanwhile, despite being located at the edge of the bicone, region G1\_40 (N), which coincides with line broadening and redshifted rotation-subtracted radial velocity features, also contains a considerable amount of outflowing molecular gas.

\noindent\textbf{\underline{Regions G4\_40 (S), G5\_40 (S), and G6\_40 (S):}}

\noindent The line profile behaviors in the southeastern part of the CND differ significantly from those in the northeastern regions. In all three southeastern regions, the CO(2-1) and CO(3-2) transitions exhibit similar line profile characteristics. For the CO(3-2) transition, the profiles are modeled using double Gaussians, with the primary components blueshifted by more than $120\,\rm{km\,s^{-1}}$ below $\rm{v_{sys}}$. The secondary components are less blueshifted but remain $\gtrsim35\,\rm{km\,s^{-1}}$ away from $\rm{v_{sys}}$, exceeding one channel width. A similar trend is observed in the CO(2-1) line profiles of regions G4\_40 (S) and G5\_40 (S). In contrast, the CO(2-1) line profile of region G6\_40 (S) is fitted with a single Gaussian, but its peak is blueshifted by more than $140\,\rm{km\,s^{-1}}$. For the CO(6-5) transition, only the line profile from region G4\_40 (S) is fitted with a significantly blueshifted single Gaussian, while the emission from the other two regions is weak relative to the noise level. Additionally, in region G4\_40 (S), a redshifted absorption feature is detected in the CO(2-1) line profile.

Contrary to the line profiles in the northeastern part of the CND, the secondary components of line profiles in regions G4\_40 (S), G5\_40 (S), and G6\_40 (S) are either more closely aligned with the galaxy's mean motion relative to the primary components or are entirely absent. This distinction between the northeastern and southeastern regions suggests that the ionized outflow in the southeastern part of the CND may be significantly stronger than in the northeast. However, high-velocity wing components are absent in the southeastern line profiles, replaced by two well-separated components in regions G5\_40 (S) and G6\_40 (S), indicative of a multi-component molecular outflow\footnote{A small redshifted peak in some line profiles may represent the redshifted constituent of the molecular outflow, consistent with the 3D geometry described by Garc\'{i}a-Burillo et al. (\citeyear{gb19}). Less redshifted absorption features could arise from the foreground blueshifted outflow obscuring the background redshifted outflow. These features are not further analyzed due to low detectability.}. Additionally, the widths of the more blueshifted components are comparable to those of the less blueshifted components in regions G4\_40 (S) and G5\_40 (S). Comparing the CO(3-2) line profiles from these regions, the line profile and its individual components exhibit the narrowest width in region G5\_40 (S). The narrower width may reflect a weaker emission, potentially due to significant molecular gas dissociation caused by intense interactions with ionized outflows. Region G5\_40 (S) appears to host the strongest ionized outflow among these regions, characterized by its high velocity and a relatively prominent, further blueshifted component. In fact, the farther blueshifted primary component of the CO(3-2) line profile is most prominent in region G5\_40 (S), as indicated by the weight parameters. Notably, G5\_40 (S) also roughly aligns with the southern molecular outflow direction identified by Garc\'{i}a-Burillo et al. (\citeyear{gb19}), which extends from the torus out into the CND.

\noindent\textbf{\underline{Regions G7\_40 (S) and G8\_40 (S):}}

\noindent Similar to region G4\_40 (S), the CO(6-5) transition line profile of region G7\_40 (S) is best fitted with a significantly blueshifted single Gaussian. In contrast, the CO(2-1) and CO(3-2) transition line profiles for region G7\_40 (S) are modeled with double Gaussians. Unlike the southeastern regions, the two Gaussian components in region G7\_40 (S) are closer in velocity. Specifically, the primary component of the CO(3-2) line profile may reflect the overall broadening of the emission line, while the corresponding minor component is offset by approximately $100\,\rm{km\,s^{-1}}$ below $\rm{v_{sys}}$. For the CO(2-1) line profile, the minor component corresponds to a blueshifted wing structure reminiscent of the redshifted wing components observed in the northeastern regions. In contrast, the CO(6-5) transition line profile in region G8\_40 (S) is centered around $\rm{v_{sys}}$. For this region, while the bulk of the CO(2-1) and CO(3-2) line profiles is also centered around the galaxy's mean motion and appears independent of the outflow, both transitions feature two minor components blueshifted by more than $100\,\rm{km\,s^{-1}}$ from $\rm{v_{sys}}$, suggesting the presence of gas entrained in the outflow.

When comparing these two regions to the southeastern regions, a lateral trend becomes apparent. The blueshifted velocity departure of the bulk CO(2-1) transition line profile reaches its maximum in region G5\_40 (S) and decreases, moving eastward toward region G4\_40 (S) and westward toward region G8\_40 (S). A similar trend is observed for the major component of the CO(3-2) transition line profiles, with its velocity departure and width both decreasing from region G5\_40 (S) compared to regions G4\_40 (S) and G7\_40 (S). Overall, the outflow strength appears to peak near region G5\_40 (S), but the amount of outflowing molecular gas decreases in this region, indicating a dynamic but less molecular gas-rich environment.

\noindent\textbf{\underline{Regions O1\_40 (E) and O2\_40 (W):}}

\noindent All line profiles in the two regions outside the AGN wind bicone, O1\_40 (E) and O2\_40 (W), are well-fitted with a single Gaussian centered around $\rm{v_{sys}}$, within one channel width. The broad widths of these profiles ($\rm \gtrsim200\,km\,s^{-1}$) may suggest a lateral outflow driven by a hot cocoon of ionized gas caused by the expanding radio jet (Couto et al. \citeyear{Couto17}; Balmaverde et al. \citeyear{Balmaverde19}; Venturi et al. \citeyear{venturi_etal_2021}; Balmaverde et al. \citeyear{Balmaverde22}).

\begin{figure*}
\centering
\resizebox{1\textwidth}{!}{\includegraphics{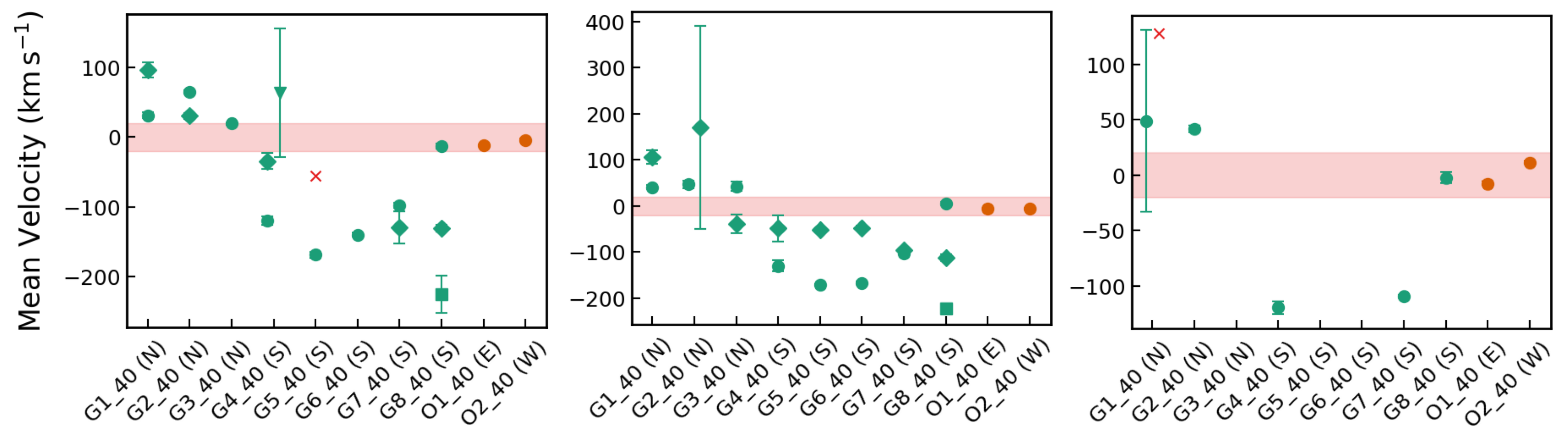}}
\caption{Mean velocity values of all fitted Gaussian components to the line profiles for the CO(2-1) (\textit{left panel}), the CO(3-2) (\textit{mid panel}), and the CO(6-5) (\textit{right panel}) transitions on the scale of $40\times40$ pc. For each panel, the red shaded region marks $\pm$ one channel width away from the systematic velocity of the galaxy, $\rm{v_{sys}}$. Based on the weight parameter, round symbols correspond to the largest component of each fit, the diamond symbols correspond to the secondary component of each fit, and the square symbols correspond to the smallest component of each fit. The downwards triangle symbol indicates the Gaussian component for absorption, and the ``$\times$'' symbol implies large error bars. All other schematics are the same as Fig. \ref{fig:7}.}

\label{fig:11}       % Give a unique label to the figure.
\end{figure*}

\begin{figure*}
\centering
\resizebox{1\textwidth}{!}{\includegraphics{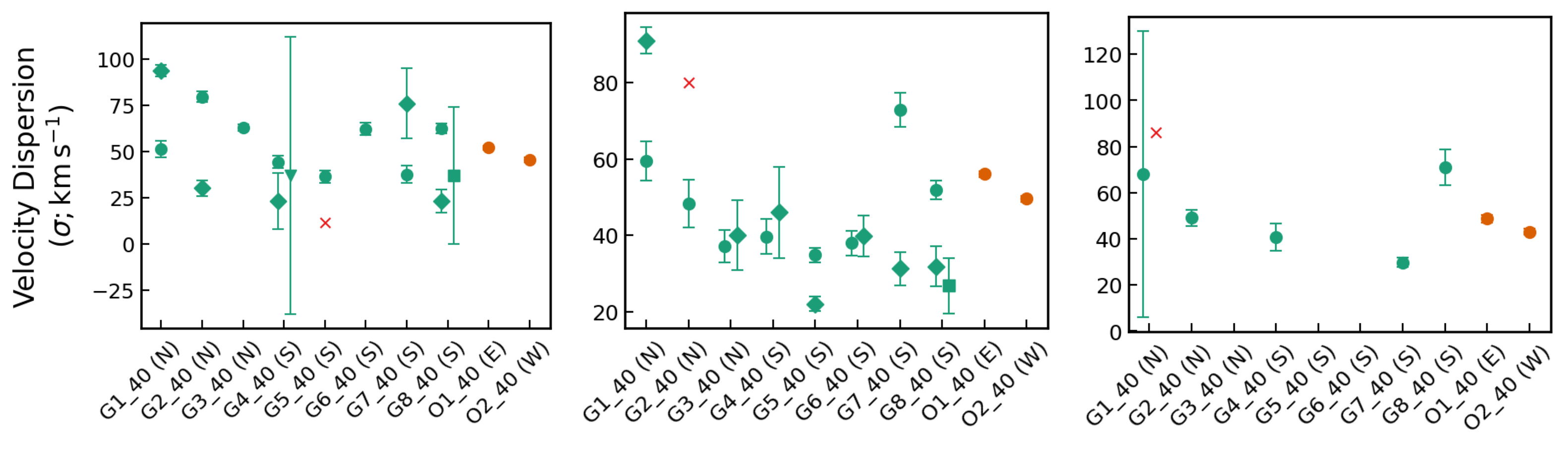}}
\caption{Velocity dispersion values of all fitted Gaussian components to the line profiles for the CO(2-1) (\textit{left panel}), the CO(3-2) (\textit{mid panel}), and the CO(6-5) (\textit{right panel}) transitions on the scale of $40\times40$ pc. All schematics follow that of Fig. \ref{fig:11}.}

\label{fig:12}       % Give a unique label to the figure.
\end{figure*}

\hfill

\noindent In summary, the CO(2-1) and CO(3-2) transition line profiles provide a comprehensive view of the gas kinematics at various positions within the CND, whereas the CO(6-5) transition often lacks sufficient emission to provide kinematic information for several regions. Overall, the majority of line profiles in the northeastern regions exhibit predominantly redshifted features with stronger emissions, while those in the southern regions are primarily blueshifted with comparatively weaker emissions, aligning with the gas dynamics illustrated in Fig. \ref{fig:1} \& \ref{fig:3}. As shown in Fig. \ref{fig:11}, the line-of-sight velocity departure of molecular gas within the outflow is larger in the southern part of the CND compared to the northern part. This, combined with the relative depletion of CO, suggests a stronger ionized outflow in the southern CND, potentially due to an outflow geometry where the southern AGN wind lies closer to the CND plane than the northern AGN wind. In the southern regions, molecular gas embedded deeper within the CND may be more strongly influenced by the ionized outflow. In particular, under the stronger outflow between regions G4\_40 (S) and G6\_40 (S), the high-velocity components from double Gaussian fits are more prominent, as indicated by larger weight parameters for these components compared to their corresponding low-velocity components in the southern CND. Spatial variations are also evident across regions within or along the AGN wind bicone. In the southern half of the bicone, the more blueshifted component is most prominent in region G5\_40 (S), with its prominence decreasing toward region G8\_40 (S) to the west and region G4\_40 (S) to the east. In the northern half, the wing components of line profiles are broader and more redshifted in region G1\_40 (N) than in G2\_40 (N) (also see Fig. \ref{fig:12}), while the wing component disappears entirely in region G3\_40 (N). These broad-wing structures indicate that molecular gas at the surface of the CND is being swept by the ionized AGN wind. Furthermore, in most regions with detectable outflows, low-velocity components (see the left and mid panels of Fig. \ref{fig:11}) are also significantly shifted from the galaxy’s systematic velocity, revealing multi-component outflows within the CND. In contrast, regions outside the AGN wind bicone show broad single-component emissions centered at the mean velocity of the galaxy, potentially implying influence from a jet-induced lateral outflow. 

We also examined the line profiles of regions defined based on excitation, which exhibit complex behaviors indicative of potential multiple gas components within each region. This complexity is consistent with the LTE analysis presented in Sect. \ref{lr}. However, the intricate nature of these line profiles poses challenges for a detailed kinematic analysis, which is beyond the scope of this study.

\subsection{Kinematics behavior at smaller scales}
\label{zsr}

To (1) investigate the scale at which the multiplicity of gas components breaks down (i.e., the scale at which a region encompasses only one gas component) and (2) attempt to isolate the outflow material entirely, we first sampled smaller regions on a scale of $20\times20$ pc within each of the previously defined regions, centered at their midpoints (see descriptions in Appendix \ref{lpz}). We then shifted these $20\times20$ pc regions to occupy different ``corners'' of their original $40\times40$ pc regions. The same line profile analysis procedure as described in Sect. \ref{method} was adopted, and the definition of Gaussian components for multi-component fits follows that of Sect. \ref{lpb}. The results are presented in Fig. \ref{fig:a11} to \ref{fig:a16} and in Table \ref{tabb6} to \ref{tabb9} in Appendix \ref{appe:lpzzs}.

\subsubsection{Definition of zoomed-in and shifted regions}
\label{rszs}

The discrepancies between smaller and larger regions mentioned in Appendix \ref{lpz} indicate the potential presence of multiple components of gas within some of the original $40\times40$ pc regions. To investigate these discrepancies further, we defined new regions by taking advantage of the high spatial resolution of the data, aiming to disentangle the different components of gas occupying distinct parts of the original $40\times40$ pc regions. For these new region definitions, we shifted the existing $20\times20$ pc regions, with the adjustments varying across different regions.
For positions G1 and G2, the $20\times20$ pc regions were shifted to the lower right (designated as DR) corners of the corresponding $40\times40$ pc regions to potentially capture stronger outflow signatures closer to the central engine. For position G7, a new $20\times20$ pc region was created to occupy the upper left (designated as UL) corner of G7\_40 (S), aiming to include stronger outflow along the inner edge of the CND. At position G8, located along the edge of the bicone, the new $20\times20$ pc region was selected to occupy the lower left (designated as DL) corner of G8\_40 (S), contrasting with position G4, where the $20\times20$ pc region was shifted to cover the entire lower right (DR) corner of G4\_40 (S). For the southern positions G5 and G6, the new $20\times20$ pc regions were chosen further from the AGN position to potentially capture less gas entrained in strong outflows. For G5, the $20\times20$ pc region was shifted to occupy the lower (designated as D) part of G5\_40 (S), centered at the right ascension coordinate of the original region, while for G6, the lower right (DR) corner was selected. Returning to the northern CND, for region G3\_40 (N), most of which lies outside the AGN wind bicone, the $20\times20$ pc region was shifted to the eastern part of G3\_40 (N), centered on its declination coordinate, to search for outflow signatures while avoiding the upper left corner, which lacks CO(6-5) emission. Additionally, since a significant fraction of G4\_40 (S) with a prominent highly blueshifted Gaussian component lies outside the AGN wind bicone, we also sampled the lower right corner (DR) of region O1\_40 (E) as a zoomed-in and shifted region. A shifted region was not defined for position O2 outside the bicone. The naming scheme for the new regions includes the position code, the size of the region, and the direction of the shift (e.g.,  for a $20\times20$ pc region occupying the lower right corner of G1\_40 (N), the region code is G1\_20\_DR (N)).

\subsubsection{Kinematics of zoomed-in and shifted regions}
\label{lpbzsr}

Line profiles and their optimal fits of the shifted regions are shown in Fig. \ref{fig:a14} for the CO(2-1) transition, Fig. \ref{fig:a15} for the CO(3-2) transition, and Fig. \ref{fig:a16} for the CO(6-5) transition. A summary of the fitting results is displayed in Table \ref{tabb9}. Detailed descriptions of the behavior of these line profiles can be found in Appendix \ref{lpzs}.

\begin{figure}
\centering
\resizebox{0.5\textwidth}{!}{\includegraphics{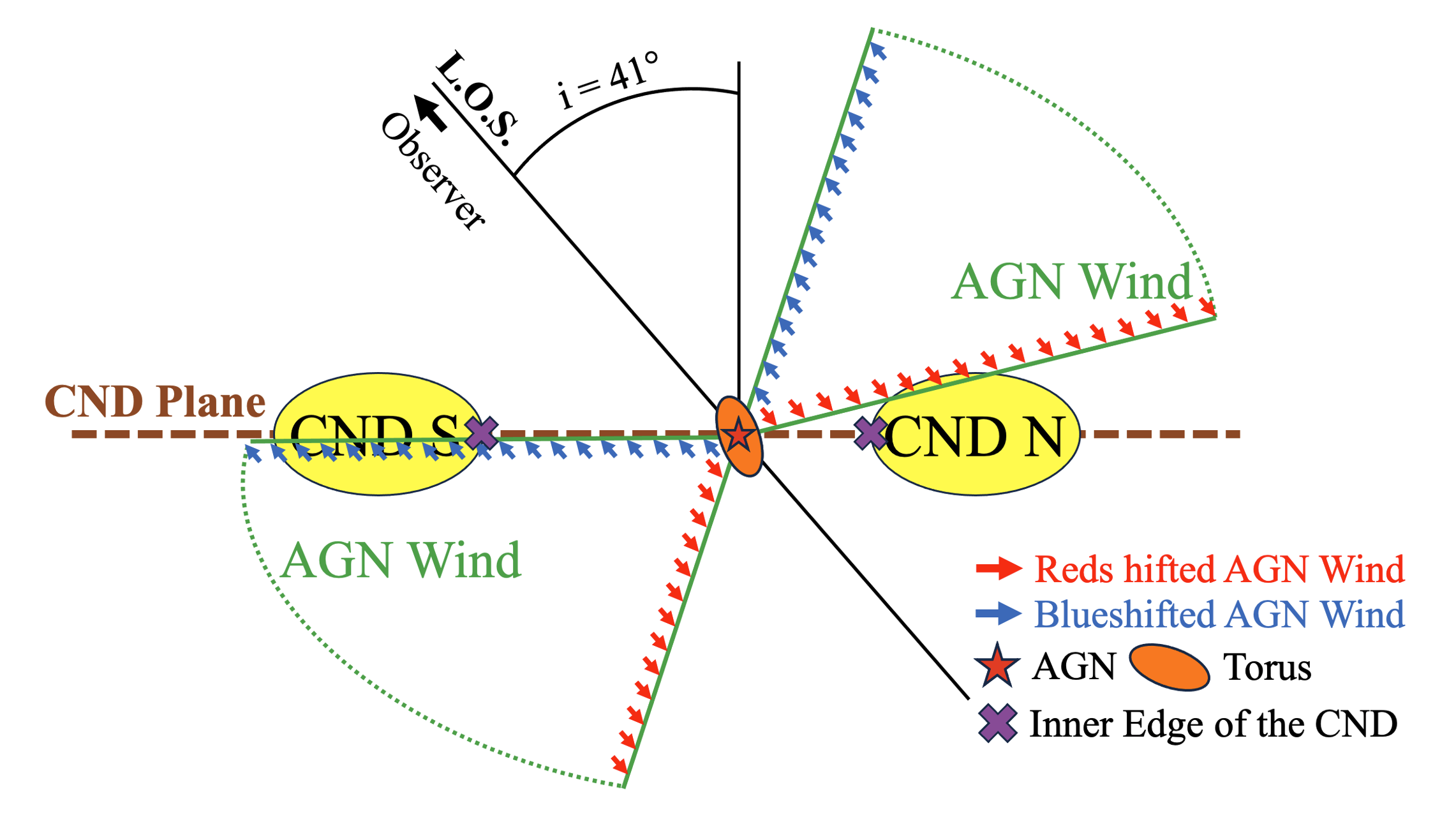}}
\caption{Scheme of the narrow-line region (NLR) in NGC 1068 (not to scale). The ionized AGN wind (green) is blueshifted (blue arrows) on the front face and redshifted (red arrows) on the back, as viewed along the line of sight. Interaction between the AGN wind and the CND (yellow ellipses) drives molecular outflows, blueshifted in the south and redshifted in the north, as observed by ALMA. The tilt of the AGN wind relative to the CND affects outflow locations: in the north, it interacts farther from the AGN than the inner edge (marked by purple crosses) of the CND, while in the south, interaction occurs at the inner edge. As a result, the strongest molecular outflows emerge at the inner edge of the southern CND and farther from the inner edge in the northern CND.}

\label{fig:diagram}       % Give a unique label to the figure.
\end{figure}

Based on line profiles of the zoomed-in and shifted regions (see descriptions in Appendix \ref{lpzs}), more molecular gas is entrained in the outflow closer to or inside the AGN wind bicone. In the southern CND, the ionized outflow is stronger along the inner edge of the CND, which causes more molecular gas to be entrained in the high-velocity component of the outflow. In the northeastern CND, the ionized outflow is stronger further away from the inner edge of the CND, characterized by broad wing components within the line profiles. One possible explanation for the different line profile behaviors in the northern and southern CND is the 3D outflow geometry (Garc\'{i}a-Burillo et al. \citeyear{gb19}; see Fig. \ref{fig:diagram}). The ionized outflow is closer to the plane of the CND in the south, where the ionized outflow contacts the inner edge of the CND first, than in the north, where the ionized outflow first contacts the top of the CND facing the observer away from the inner edge and sweeps the molecular gas at the surface of the CND. A substantial amount of molecular gas was also found extending outside the AGN wind bicone into the eastern part of the CND. Considering positions G1 and G5 might coincide with the strongest interactions between the molecular gas and the ionized outflow, a possible ``misalignment'' might exist between the ionized AGN wind bicone and the molecular outflow. This asymmetry of the molecular outflow also matches the structure of the radio jet previously observed in ionization bubbles by May \& Steiner (\citeyear{ms17}).

\subsection{Temperature and CO column density behaviors at smaller scales}
\label{tdss}

In addition to the line profile investigation on the scale of $20\times20$ pc and for the shifted regions, as discussed in Sect. \ref{zsr} and \ref{lpbzsr}, we created rotational diagrams to check the LTE condition and temperature structure for these smaller regions. The results are shown in Fig. \ref{fig:a20} \& \ref{fig:a21} and are displayed in Table \ref{tabb12} \& \ref{tabb13}. Summaries of rotational temperatures and total CO column densities from all these regions are also provided in Fig. \ref{fig:a30} \& \ref{fig:a31}. Same as on the large $40\times40$ pc scale, gas within all smaller regions satisfies the LTE condition. Rotational temperatures from most of these regions also match with each other, and the values derived from gas within larger regions (within the error bar range), barring regions G3\_20 (N), G5\_40 (S), and all regions of position G6 in the southern CND. The inconsistency among the temperature values at the southern positions might be due to a stronger and multi-component outflow centered around position G5, while region G3\_20 (N) might not contain as much outflowing gas with higher temperature as regions G3\_40 (N) and G3\_20\_L (N). For the total CO column densities, their values from three transitions for each $20\times20$ pc or shifted region are consistent. The average CO column densities are slightly larger than those from the corresponding $40\times40$ pc regions but remain within the same order of magnitude. This increase in average column density may suggest a higher concentration of CO in the smaller regions. Overall, the density structure observed on the larger scale (as shown in Fig. \ref{fig:7}) is preserved. Similar temperature and density behaviors observed across small and large scales may suggest that the clumpy CO gas within the CND is dynamically influenced by the ionized outflow. This interaction could lead to the stirring of molecular gas, promoting mixing and resulting in CO gas with low opacity. Optically thin CO emissions have been previously reported in similar systems, such as IC 5063 and NGC 3256-S, by Dasyra et al. (\citeyear{dasyra_etal_2016}) and Pereira-Santaella et al. (\citeyear{PS23}).

%--------------------------------------------------------------------

\section{Impact of the outflow}
\label{mdot}

To understand the impact of the ionized outflow on the molecular gas inside the CND and the power of the molecular outflow, we calculated the mass outflow rate, $\rm{d}\mathit{M}/\rm{d}\mathit{t}$, for each $40\times40$ pc region defined based on the outflow (see Sect. \ref{rsoutflow}) and compared the values between the northeastern and the southern CND. In this section, we will first outline the method of estimating $\rm{d}\mathit{M}/\rm{d}\mathit{t}$ values, followed by their implications on the impact of the outflow across the CND.

\subsection{Estimation of mass outflow rates}
\label{calc_mdot}

We first evaluate the total molecular mass impacted by the ionized outflow using the CO emission line brightness to molecular hydrogen column density conversion, $X_{\rm{CO}} = N(\rm{H_2})/\mathit{I}_{\rm{CO(1-0)}}$. Considering the low optical depth values calculated using RADEX in Sect. \ref{LTE}, in combination with relatively high excitation in the inner part of the CND (see Fig. 5), $X_{\rm{CO}}$ inside the CND might be much lower than in the Milky Way or possibly lower than the value of $1/(4^{+6}_{-1})\times \rm{X^{MW}_{CO}}$ obtained in Garc\'{i}a-Burillo et al. (\citeyear{gb14}), where $\rm{X^{MW}_{CO}}$ is the Milky Way value. A smaller $X_{\rm{CO}}$ factor could imply a lower molecular mass within the outflow component of the CND. To calculate the average $X_{\rm{CO}}$ values, we assume a range of CO abundances\footnote{We also considered using $n_{\rm{H_2}}$ from RADEX. However, constraining a $n_{\rm{H_2}}$ value for each region is challenging due to the degeneracy between $n_{\rm{H_2}}$ and $T_{\rm{k}}$}, $\rm{[CO]/[H_2]\sim5\times10^{-5}}$ to $\sim10^{-4}$ (Usero et al. \citeyear{Usero04}). The values are computed based on $X_{\rm{CO}}=N_{\rm{CO}}/(\rm{[CO]/[H_2]}\,\mathit{I}_{\rm{CO(1-0)}})$, and the $I_{\rm{CO(1-0)}}$ values are extrapolated through rotational diagrams for each region. The results (see Table \ref{tab:4}) are on average $4.8\pm0.4$ -- $9.6\pm0.9$ times smaller than the canonical Milky Way value, $\rm{X^{MW}_{CO}}=2\times10^{20}\,cm^{-2}\,(K\,km\,s^{-1})^{-1}$, and are slightly lower than that of Usero et al. (\citeyear{Usero04}), $\sim4$ -- $8$ times smaller than the canonical value. Regardless, the $X_{\rm CO}$ values from this work are roughly consistent with values from Usero et al. (\citeyear{Usero04}), Garc\'{i}a-Burillo et al. (\citeyear{gb14}), and Israel (\citeyear{Israel09A}, \citeyear{Israel09B}).

\begin{table}[h!]
\centering
\begin{tabular}{lcc}
\hline
\hline
Region & Max $X_{\rm CO}$ & Min $X_{\rm CO}$ \\
\hline
G1\_40 (N) & $5.4\pm1.6 \times 10^{19}$ & $2.71\pm0.80 \times 10^{19}$ \\
G2\_40 (N) & $4.9\pm1.5 \times 10^{19}$ & $2.46\pm0.73 \times 10^{19}$ \\
G3\_40 (N) & $4.3\pm1.2 \times 10^{19}$ & $2.15\pm0.62 \times 10^{19}$ \\
G4\_40 (S) & $3.5\pm1.0 \times 10^{19}$ & $1.75\pm0.50 \times 10^{19}$ \\
G5\_40 (S) & $4.3\pm1.3 \times 10^{19}$ & $2.29\pm0.66 \times 10^{19}$ \\
G6\_40 (S) & $3.9\pm1.1 \times 10^{19}$ & $1.97\pm0.56 \times 10^{19}$ \\
G7\_40 (S) & $3.4\pm1.0 \times 10^{19}$ & $1.72\pm0.48 \times 10^{19}$ \\
G8\_40 (S) & $3.6\pm1.0 \times 10^{19}$ & $1.78\pm0.50 \times 10^{19}$ \\
O1\_40 (E) & $3.9\pm1.1 \times 10^{19}$ & $1.97\pm0.55 \times 10^{19}$ \\
O2\_40 (W) & $4.2\pm1.2 \times 10^{19}$ & $2.09\pm0.59 \times 10^{19}$ \\
\hline
\end{tabular}
\caption{Estimated lower and upper $X_{\rm CO}$ factors for all $40\times40$ pc regions defined based on the outflow (see Sect. \ref{rsoutflow}) in the unit of $\rm cm^{-2}\,(K\,km\,s^{-1})^{-1}$. The region codes are displayed in the first column along with corresponding region locations (e.g., ``N'' stands for the northern CND).}
\label{tab:4}
\end{table}

To derive the mass outflow rate, we need to assume a specific outflow geometry. Maiolino et al. (\citeyear{Maiolino12}) and subsequently Cicone et al. (\citeyear{cicone_etal_2014}) provided two simple geometries of an AGN outflow, a conical/multi-conical outflow or a fragmented shell geometry, uniformly filled with outflowing clouds. Given that the $40\times40$ pc regions subtend a substantial radial span away from the AGN position (i.e., the ``tip'' of the outflowing cone), we adopted the fragmented shell geometry without approximation, and the mass outflow rate is given by

\begin{equation}\label{eqnof}
\frac{\rm{d}\mathit{M}}{\rm{d}\mathit{t}} = \frac{3 \times v_{\text{out}} \times r_{\text{out}}^2 \times M_{\text{out}}}{r_{\text{out}}^3 - r_{\text{in}}^3}\,,
\end{equation}

\noindent where $v_{\rm{out}}$ is the average outflow speed of the region (velocity directs radially away from the AGN in the NGC 1068 galactic frame), $M_{\rm{out}}$ is the outflowing molecular mass, and $r_{\rm{out}}$ and $r_{\rm{in}}$ are the outer and inner radii of the shell. 

The outflow mass, $M_{\rm{out}}$, can be calculated from the residual CO image cubes (without rotation of the CND and mean motion of the galaxy). Following Garc\'{i}a-Burillo et al. (\citeyear{gb14}), all emission within $\langle v_{\text{res}} \rangle = [-50, +50] \, \text{km s}^{-1}$ was subtracted from residual CO image cubes to eliminate the contribution from non-outflowing gas with virial motions around the rotation ($v_{\text{res}}$ is the residual velocity of the gas excluding rotation and the mean motion of the galaxy). The resulting cubes containing mostly outflowing gas were applied $3\sigma$ clipping before being used to produce an outflow velocity-integrated map (see Fig. \ref{fig:21outflow}, \ref{fig:32outflow}, \& \ref{fig:65outflow}), which contains the total emission from the outflow across the CND. Ratios between the emission from the outflow and the total emission were then calculated. We ignored the ratios from the CO(6-5) transition due to a lack of detection in some regions. The average ratios were used to infer the amount of $\mathit{I}_{\rm{CO(1-0)}}$ corresponding to the outflowing gas, $\mathit{I}_{\rm{CO(1-0)}, out}$. In turn, $\mathit{I}_{\rm{CO(1-0)}, out}$ was used to evaluate $N(\rm{H_2})_{out}$, the average column density of outflowing gas within each region and, subsequently, $M_{\rm{out}}$. 

\begin{figure}
\centering
\resizebox{0.5\textwidth}{!}{\includegraphics{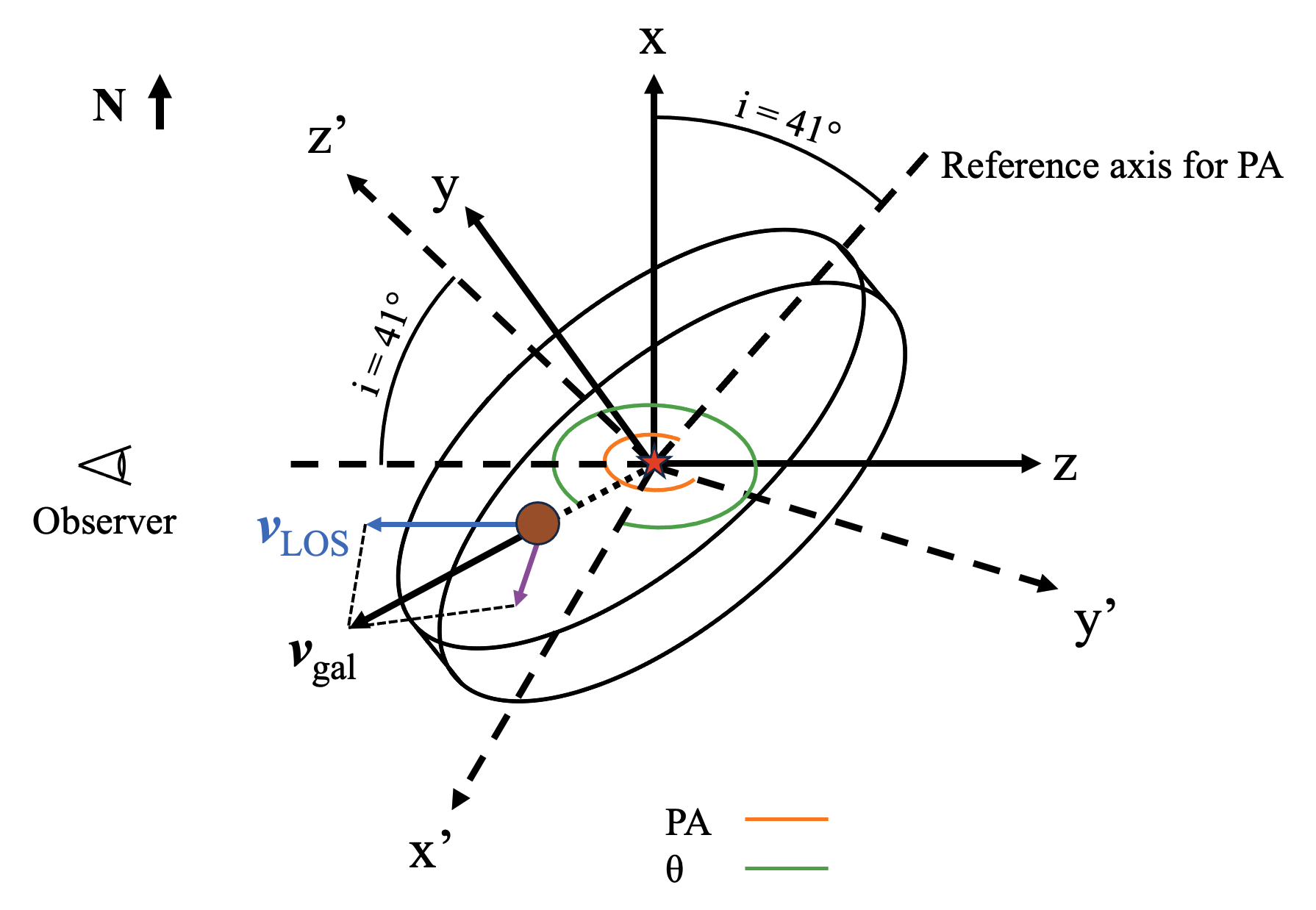}}
\caption{Projection of velocity ($\boldsymbol{v}_{\rm{gal}}$) from NGC 1068 coordinates (primed) to sky coordinates (unprimed). The primed frame results from a $i=41^{\circ}$ inclination rotation along the line of sight (from the negative $\rm z$-axis to the $\rm z'$-axis) and a $PA=289^{\circ}$ rotation around $\rm z'$-axis. The $\rm x'$-axis aligns with the minor axis of the CND, matching the reference azimuthal direction ($\rm \theta'_0$), with the $90^\circ$ difference between $\rm x'$ and $\rm y'$-axes corresponding to the difference between PA and the $\rm \theta'_0$ direction. The primed system follows a rightward primary axis convention, while the sky frame follows a northward reference. The PA reference axis is the $\rm x$-axis after inclination rotation. The brown ball marks the molecular gas region, with outflow velocity $\boldsymbol{v}_{\rm{gal}}$ assumed to lie in the CND plane, directed radially from the AGN (red star). It decomposes as $v_{\rm{out}}\cos(\theta)$ along $\rm x'$-axis and $v_{\rm{out}}\sin(\theta)$ along $\rm y'$-axis, where $\theta$ is measured from $\rm x'$-axis. Using Eqn.~\ref{eqn6}, $\boldsymbol{v}_{\rm{gal}}$ is projected to sky velocity $\boldsymbol{v}_{\rm{sky}}$, decomposed into the observed radial velocity $v_{\rm{LOS}}$ and a transverse component (purple arrow), which further splits into $v_{\rm{x}}$ and $v_{\rm{y}}$.}

\label{fig:projection}       % Give a unique label to the figure.
\end{figure}

Similar to $M_{\rm{out}}$, the outflow speed values, $v_{\rm{out}}$, were also estimated from the residual CO image cubes. After $3\sigma$ clipping, the resulting cubes were used for producing outflow radial-velocity fields for all transitions, which contain the mean line of sight velocity of the outflow, $v_{\rm{LOS}}$. To calculate $v_{\rm{out}}$, we need to de-project $v_{\rm{LOS}}$ which can be considered as the line of sight component of the outflow velocity in the sky coordinates, $\boldsymbol{v}_{\rm{sky}}$ (see Fig. \ref{fig:projection} for visualization). Likewise, $v_{\rm{out}}$ can be considered as the radial component outwards from the AGN position in the plane of the CND of the outflow velocity in the CND coordinates, $\boldsymbol{v}_{\rm{gal}}$. When projecting $\boldsymbol{v}_{\rm{gal}}$ to $\boldsymbol{v}_{\rm{sky}}$ (i.e., from the CND coordinates to sky), two rotation matrices were applied to correct for the position angle and inclination, 

\begin{align}\label{eqn6}
\boldsymbol{v}_{\rm{sky}} &= \boldsymbol{\rm{R}}_{\mathit{i}}\,\boldsymbol{\rm{R}}_{\mathit{PA}}\,\boldsymbol{v}_{\rm{gal}} \notag \\
&= \begin{bmatrix}
\cos(\mathit{PA}) & -\sin(\mathit{PA}) & 0 \\
\cos(i)\sin(\mathit{PA}) & \cos(i)\cos(\mathit{PA}) & -\sin(i) \\
\sin(i)\sin(\mathit{PA}) & \sin(i)\cos(\mathit{PA}) & \cos(i) \\
\end{bmatrix}
\,\boldsymbol{v}_{\rm{gal}}\,.
\end{align}

\noindent $\boldsymbol{v}_{\rm{gal}}$ is assumed to be completely contributed from the outflow (and inflow\footnote{This is technically all motions excluding the virial motion, CND rotation, and mean motion of the galaxy.}), and 

\begin{equation}\label{eqn7}
    \boldsymbol{v}_{\rm{gal}} = 
    \begin{bmatrix}
    v_{\rm{out}}\cos(\theta) \\
    v_{\rm{out}}\sin(\theta) \\
    v_{\perp}\sim0 \\
    \end{bmatrix}\,,
\end{equation}

\noindent where $\theta$ is the angle of the position of the gas with respect to the reference azimuthal direction in the CND coordinates, $\theta_0'=\rm{PA}-90^{\circ}=199^{\circ}$ in the sky coordinates. We effectively assume that the outflow radially outwards from the AGN position (or the inflow radially towards the AGN position) is the only velocity contributing to the residual velocity in the sky coordinates, $\boldsymbol{v}_{\rm{sky}}$. We also assume that the main outflow velocity is within the plane of the CND, and $v_{\rm{out}} \gg v_{\perp}$, where $v_{\perp}$ is the potential outflow velocity component perpendicular to the CND plane. $\boldsymbol{v}_{\rm{sky}}$ takes the form of 

\begin{equation}\label{eqn8}
    \boldsymbol{v}_{\rm{sky}} = 
    \begin{bmatrix}
    v_{\rm{x}} \\
    v_{\rm{y}} \\
    v_{\rm{LOS}} \\
    \end{bmatrix}\,,
\end{equation}

\noindent where $v_{\rm{x}}$ and $v_{\rm{y}}$ are contributions from the outflow to velocity perpendicular to the line of sight. By combining Eqn. \ref{eqn6} to \ref{eqn8}, the relation between the outflow speed and the mean line of sight radial velocity could be expressed as 

\begin{equation}\label{eqnv}
    v_{\rm{out}} = \frac{v_{\rm{LOS}}}{\sin(i)\,\sin(PA+\theta)}\,.
\end{equation}

\noindent We first de-projected the entire line of sight velocity field to the outflow speed map for each transition. For this process, we set $PA=289^{\circ}$ and $i=41^{\circ}$, following that of Garc\'{i}a-Burillo et al. (\citeyear{gb19}). To avoid extreme outflow rates caused by velocity outliers within each region, we retained only the inner 68\% of outflow speed values for each region and CO transition before computing the average outflow speed. $1\sigma$ uncertainties of outflow speed values were obtained through bootstrapping the average outflow speed within 68\% of all values for each region.

To determine $r_{\rm{in}}$ and $r_{\rm{out}}$, we first projected galactic radii from the AGN position to the sky coordinates. The projected radii map was also ``clipped'' for each transition to account only for the radii of emission above $3\sigma$ level, followed by calculating the 68\% percentiles of radii within each region as $r_{\rm{in}}$ and $r_{\rm{out}}$. Error bars of $r_{\rm{in}}$ and $r_{\rm{out}}$ were also computed via bootstrapping.

\subsection{Mass outflow rates and outflow velocities across the CND}
\label{res_mdot}

\begin{table*}[h!]
    \centering
    \begin{tabular}{ c c c c }
        \hline
        \hline
        Region & CO(2-1) & CO(3-2) & CO(6-5) \\ 
        \hline
        G1\_40 (N) & $17.8\pm5.8$--$36\pm12$ & $18.6\pm6.1$--$37\pm12$ & $17.7\pm5.8$--$35\pm12$ \\
        G2\_40 (N) & $1.54\pm0.51$--$3.1\pm1.0$ & $1.48\pm0.49$--$2.96\pm0.97$ & $1.74\pm0.57$--$3.5\pm1.2$ \\
        G3\_40 (N) & $0.176\pm0.060$--$0.35\pm0.12$ & $0.071\pm0.028$--$0.142\pm0.056$ & $0.80\pm0.26$--$1.59\pm0.53$\dag \\
        G4\_40 (S) & $1.27\pm0.40$--$2.54\pm0.81$ & $1.27\pm0.40$--$2.55\pm0.81$ & $1.99\pm0.64$--$4.0\pm1.3$ \\
        G5\_40 (S) & $0.57\pm0.18$--$1.14\pm0.37$ & $0.56\pm0.18$--$1.13\pm0.36$ & $1.27\pm0.43$--$2.54\pm0.86$\dag \\
        G6\_40 (S) & $0.54\pm0.17$--$1.08\pm0.34$ & $0.54\pm0.17$--$1.08\pm0.34$ & $2.02\pm0.66$--$4.1\pm1.3$\dag \\
        G7\_40 (S) & $2.08\pm0.65$--$4.2\pm1.3$ & $2.15\pm0.67$--$4.3\pm1.3$ & $2.73\pm0.86$--$5.5\pm1.7$ \\
        G8\_40 (S) & $1.03\pm0.33$--$2.06\pm0.66$ & $0.47\pm0.16$--$0.93\pm0.33$ & $0.090\pm0.120$--$0.18\pm0.24$ \\
        O1\_40 (E) & $0.32\pm0.11$--$0.65\pm0.21$ & $0.159\pm0.058$--$0.32\pm0.12$ & $0.125\pm0.066$--$0.25\pm0.13$ \\
        O2\_40 (W) & $-0.76\pm0.25$--$-0.38\pm0.13$ & $-0.90\pm0.30$--$-0.45\pm0.15$ & $1.34\pm0.44$--$2.68\pm0.87$ \\
        \hline
    \end{tabular}
    \caption{Estimated mass outflow rates and corresponding uncertainties for all regions defined based on the outflow (see Sect. \ref{rsoutflow}) and for all transitions in the unit of $\rm{M_{\odot}\,yr^{-1}}$. The region codes are indicated in the first column along with corresponding region locations (e.g., ``N'' stands for the northern CND), and the transitions are shown in the top row. For each region and transition, the estimated minimum outflow rate is displayed first, followed by the maximum outflow rate. The values denoted with ``\dag'' correspond to regions with limited detection.}
    \label{tab:5}
\end{table*}

Following the recipe outlined above, we calculated the upper and lower limits of the mass outflow rate for all regions and transitions based on the range of $X_{\rm{CO}}$ for each region. Uncertainties of all outflow rate values were propagated from associated variables. The results are recorded in Table \ref{tab:5}. We also display average outflow speeds for each region and transition in Table \ref{tab:6}.

\begin{table}[h!]
    \centering
    \begin{tabular}{ c c c c }
        \hline
        \hline
        Region & CO(2-1) & CO(3-2) & CO(6-5) \\ 
        \hline
        G1\_40 (N) & $569\pm17$ & $593\pm17$ & $580\pm19$ \\
        G2\_40 (N) & $176.2\pm2.8$ & $176.1\pm2.3$ & $135.1\pm1.1$ \\
        G3\_40 (N) & $53.8\pm5.3$ & $21.9\pm4.8$ & $110.0\pm1.5$\dag \\
        G4\_40 (S) & $188.46\pm0.96$ & $188.5\pm1.1$ & $181.4\pm1.8$ \\
        G5\_40 (S) & $220.4\pm3.8$ & $229.5\pm3.2$ & $273.7\pm3.1$\dag \\
        G6\_40 (S) & $227.1\pm2.3$ & $232.7\pm2.6$ & $310.3\pm3.0$\dag \\
        G7\_40 (S) & $245.8\pm1.3$ & $249.3\pm1.7$ & $226.3\pm1.1$ \\
        G8\_40 (S) & $156.8\pm8.3$ & $72\pm11$ & $15\pm19$ \\
        O1\_40 (E) & $41.6\pm3.4$ & $20.5\pm3.8$ & $15.1\pm6.3$ \\
        O2\_40 (W) & $-66.8\pm7.0$ & $-79.3\pm7.5$ & $193\pm13$ \\
        \hline
    \end{tabular}
    \caption{Outflow speed away from the AGN position in the plane of the CND and their corresponding uncertainties for all regions defined based on the outflow (see Sect. \ref{rsoutflow}) and for all transitions in the unit of $\rm km\,s^{-1}$. The region codes are indicated in the first column along with corresponding region locations (e.g., ``N'' stands for the northern CND), and the transitions are shown in the top row. The values denoted with ``\dag'' correspond to regions with limited detection.}
    \label{tab:6}
\end{table}

Region G1\_40 (N) (see Fig. \ref{fig:4}) has a significantly larger molecular mass outflow rate compared to other regions, which reveals that the mass outflow rate mainly depends on the total enclosed mass of molecular gas within each region. Considering the numerical difference between $r_{\rm in}$ and $r_{\rm out}$ is much larger than $v_{\rm out}$, the remainder of Eqn. \ref{eqnof} is substantially smaller than the values of $M_{\rm out}$, which are in the order of $\sim10^5$(mainly in the southern CND) to $\sim10^6\,\rm M_{\odot}$ (mainly in the northeastern CND)\footnote{$r_{\rm in}$ and $r_{\rm out}$ are $\gtrsim10\,\rm pc$, while $10\lesssim v_{\rm out}\lesssim10^3\, \rm km\,s^{-1}$.}, and $M_{\rm out}$ for each region is about $\sim10$--$20$ times smaller than the total outflowing molecular mass inside the CND, $1.8^{+0.6}_{-1.1}\times10^7\,\rm M_{\odot}$ (Garc\'{i}a-Burillo et al. \citeyear{gb14}). A lower molecular mass budget in the southern CND than in the northeast indicates the depletion of molecular gas in the south, which contributes to the relatively low molecular mass outflow rates. When comparing outflow speed values between northern and southern regions, their values from regions G5\_40 (S), G6\_40 (S), and G7\_40 (S) are larger than the two northern regions, G2\_40 (N) and G3\_40 (N). However, region G1\_40 (N) in the northeast has the largest outflow speed, more than 2 times higher than that of the southern regions. Outside the AGN wind bicone, region O1\_40 (E) exhibits a lower mass outflow rate compared to the southern regions, along with a low outflow speed below the virial speed of $50\,\rm km\,s^{-1}$, suggesting a limited impact from the outflow in this area. On the other side of the CND, region O2\_40 (W) shows a negative mass outflow rate and a negative outflow speed of $\rm \lesssim50\,km\,s^{-1}$, which could indicate an inflow of molecular gas. However, the negative outflow speed in this region may be an artifact introduced by the asymptotic behavior of the inverse $\sin$ function in Eqn. \ref{eqnv}.

We averaged molecular mass outflow rates among the northern and southern regions separately, excluding the CO(6-5) transitions of regions G3\_40 (N), G5\_40 (S), and G6\_40 (S) due to a lack of detection. In the northeastern CND, the average outflow rate within a $40\times40$ pc region ranges from $8.5\pm1.5$ to $17.0\pm3.0\,\rm{M_{\odot}yr^{-1}}$. These numbers decrease to $0.72\pm0.13$--$1.44\pm0.26\,\rm{M_{\odot}yr^{-1}}$ without region G1\_40 (N). The average outflow rate in the southern CND is $1.35\pm0.14$--$2.71\pm0.29\,\rm{M_{\odot}yr^{-1}}$. The average outflow rate of the northeastern CND is higher than that of the southern CND, implying a stronger molecular outflow in the northeast, especially around region G1\_40 (N). We likewise calculated the average outflow speed among the northern and southern regions. The average outflow speed in the northeast is $288.2\pm2.8\,\rm km\,s^{-1}$, and the average outflow speed in the south is $187.1\pm1.9\,\rm km\,s^{-1}$. If we exclude region G1\_40 (N), the average outflow speed in the north drops to $70.4\pm1.0\,\rm km\,s^{-1}$. The substantial molecular gas reservoir near region G1\_40 (N) appears to be more strongly influenced by the ionized outflow compared to the rest of the CND. This significant molecular mass outflow suggests a greater resistance to the ionized AGN wind in this area. Additionally, the outflow impacts a broader region in the southern CND than in the northeastern CND, where the molecular outflow is primarily concentrated around region G1\_40 (N).

We further calculated the total molecular mass outflow rate of all regions, and the total amounts to $38.4\pm3.9\,\rm{M_{\odot}yr^{-1}}$, which is around 60\% the value for the entire CND stated in Garc\'{i}a-Burillo et al. (\citeyear{gb14}), $63^{+21}_{-37}\,\rm{M_{\odot}yr^{-1}}$. This fraction exceeds the ratio of the total sampled area to the geometrical extent of the CND. The discrepancy might be owing to different %$X_{\rm{CO}}$ values and 
geometries used for the mass outflow rate calculation. While Garc\'{i}a-Burillo et al. (\citeyear{gb14}) adopted the conical model for a more conservative value, we decided to calculate more precise outflow rates for the comparisons between regions. Regardless, when following the approach using the conical model, the total outflow rate from all regions is $27.9\pm3.0\,\rm{M_{\odot}yr^{-1}}$, which counts about 45\% of the mass outflow rate of the entire CND. Considering the physical scale of the CND is more than double that of the sampled regions combined, the majority of the molecular outflow power comes from the inner part of the CND, especially around region G1\_40 (N).

\section{Conclusions}
\label{dis}

We have selected a number of regions within the CND of NGC 1068 and investigated the temperature and density distribution of molecular gas within the CND using high spatial resolution ($\sim0.1''$ or $\sim7\,\rm pc$) data from the CO(2-1), CO(3-2), and CO(6-5) transitions previously observed by ALMA. We characterized the kinematics of gas by analyzing residual line profile behaviors within the selected regions after subtracting rotation of the CND from the data, followed by calculating corresponding molecular mass outflow rates and outflow speed values in the plane of the CND. Our conclusions are summarised as follows:

\begin{itemize}
    \item[i] Most of the CO gas within the CND of NGC 1068 is optically thin and exhibits LTE conditions, as concluded from both the LTE and the RADEX analyses. Departures from LTE conditions are only observed for CO gas with high excitation, indicating either an optically thick condition or multiple gas components within these regions.
    \item[ii] The rotational (or kinetic) temperatures of CO across the CND match the lower resolution values reported by Viti et al. (\citeyear{viti_etal_2014}) and range between $\sim30\,\rm K$ and $\sim60\,\rm K$, with slightly higher temperatures ($\gtrsim40\,\rm K$) observed in the northeastern regions. In contrast, CO gas in the northeastern CND displays much broader line widths compared to the southern CND. This difference in line widths suggests a dynamic asymmetry between the northeastern and southern outflows, with the molecular gas in the northeast offering greater resistance to the ionized AGN wind.
    \item[iii] Line profiles from the 40$\times$40 pc regions within or along the edge of the bicone display an increasing fraction of high-velocity outflowing molecular gas towards a position offset east of the southern bicone center from the southeastern and southwestern boundaries of the bicone, as well as towards the east from the west in the northern part of the bicone. These regions on the eastern side of the AGN wind bicone show a higher prevalence of high-velocity molecular outflows, suggesting a potential misalignment between the molecular and ionized outflows.
    \item[iv] Line profiles from further zoomed-in and shifted $20\times20$ pc regions outside or along the edge of the AGN wind bicone show an increasing fraction of outflowing molecular gas closer to or further inside the AGN wind bicone. A significant amount of outflowing molecular gas was also detected outside the AGN wind bicone in the eastern part of the CND, implying a mismatch in the spread between the molecular and ionized outflows.
    \item[v] From the zoomed-in and shifted $20\times20$ pc regions, we also observe a larger fraction of outflowing molecular gas near the inner edge of the CND in the south, contrary to a smaller fraction of outflowing molecular gas along the northern inner edge of the CND. This suggests that the ionized AGN wind may be tilted farther away from the plane of the CND in the northeast compared to the south.
    \item[vi] For regions inside the AGN wind bicone or along the eastern edge of the AGN wind bicone with line profiles fitted with multi-Gaussian models, their low-velocity components are shifted more than one channel width ($20\,\rm km\,s^{-1}$) away from $\rm v_{sys}$, indicating a multi-component molecular outflow.
    \item[vii] We calculated $X_{\rm CO}$ values across the CND, and the values are $4.8\pm0.4$--$9.6\pm0.9$ times smaller than the canonical Milky Way value, $\rm{X^{MW}_{CO}}=2\times10^{20}\,cm^{-2}\,(K\,km\,s^{-1})^{-1}$.
    \item[viii] We calculated the molecular mass outflow rate and speed for regions within the CND by de-projecting line-of-sight velocities to outflow velocities in the plane of the CND and estimating the outflow molecular mass using $X_{\rm CO}$. The molecular mass outflow rate in region G1\_40 (N) is significantly higher than in the rest of the CND. Outside region G1\_40 (N), molecular mass outflow rates do not exceed $\sim5.5\,\rm M_{\odot}\,yr^{-1}$ within each $40\times40$ pc region. This suggests that the ionized outflow has a minimal kinematic impact on the molecular gas reservoir in most of the CND.
    \item[ix] The total molecular mass outflow rate of all sampled regions within the CND exceeds 60\% that of Garc\'{i}a-Burillo et al. (\citeyear{gb14}), $63^{+21}_{-37}\,\rm{M_{\odot}yr^{-1}}$, with region G1\_40 (N) contributing to most of the outflow rate among these regions.
\end{itemize}

\noindent Our study further confirms the 3D outflow geometry proposed by Garc\'{i}a-Burillo et al. (\citeyear{gb19}) as a viable explanation for the outflow within the CND. This is supported by the observed asymmetry of the outflow. The ionized AGN wind appears to be located in front of the plane of the CND in the northeast, where it intersects with the molecular gas farther from the AGN position, near the geometrically thickest part of the structure. This interpretation aligns with a recent study by Hagiwara et al. (\citeyear{hagiwara_24}) on dense molecular gas (HCN and $\rm HCO^+$) and $\rm H_2O$ maser spots within the streamer connecting the northern CND to the torus near component C of the 5 GHz continuum (see Gallimore et al. \citeyear{gallimore_etal_96}, \citeyear{gallimore_etal_04}). In the northeast, the ionized outflow may only sweep the surface of the molecular CND facing the observer, producing broad-wing structures in the line profiles observed in this region. Conversely, in the south, the ionized AGN wind seems to be angled closer to, but slightly behind, the plane of the CND, where it interacts with the molecular gas closer to the inner edge of the disc. The mass outflow rate is larger in the northeast, particularly near region G1\_40 (N), indicating greater resistance from the molecular gas to the ionized outflow, especially when considering its high molecular gas density (Viti et al. \citeyear{viti_etal_2014}). This suggests the outflow in the northeast is less dynamically evolved compared to the south, as further evidenced by the asymmetric morphology of the radio jet (Gallimore et al. \citeyear{gallimore_etal_96}). Beyond the E-knot, the impact of the ionized outflow on molecular gas is limited due to the predominantly optically thin nature of the CO gas within the CND. In these regions, especially in the southern CND, molecular gas dissociation reduces the kinematic influence of the ionized outflow.

Outside the AGN wind bicone, the origin of the broad line profiles is still unknown. Uncertainties remain regarding the molecular mass outflow rate and outflow speed in regions outside the bicone due to the simplified de-projection of line-of-sight velocity. The current velocity de-projection method does not account for motion outside the plane of the CND, potentially leading to inaccuracies. A more sophisticated kinematic model is required to address these limitations and could also provide a clearer explanation for the observed negative outflow rate in the western part of the CND.

Future ALMA observations might also benefit the understanding of molecular gas kinematics close to the AGN in NGC 1068. Despite the large variance in kinematic behaviors observed at small spatial scales (e.g.,  $20\times20$ pc), our study of the CND highlights the wealth of additional information that can be extracted using high spatial resolution data, particularly by comparing CO gas kinematics across different scales.

\begin{acknowledgements}
We thank the anonymous referee for the useful suggestions for improving this paper. This project has received funding from the European Research Council (ERC) under the European Union's Horizon 2020 research and innovation programme (MOPPEX; Grant agreement No. 833460). YZZ, SV, and KYH acknowledge assistance from Allegro, the European ALMA Regional Center node in the Netherlands. This paper makes use of the following ALMA data: ADS/JAO.ALMA\#2013.1.00055.S, ADS/JAO.ALMA\#2016.1.00232.S. ALMA is a partnership of ESO (representing its member states), NSF (USA) and NINS (Japan), together with NRC (Canada), NSTC and ASIAA (Taiwan), and KASI (Republic of Korea), in cooperation with the Republic of Chile. The Joint ALMA Observatory is operated by ESO, AUI/NRAO and NAOJ. SGB acknowledges support from the Spanish grant PID2022-138560NB-I00, funded by MCIN/AEI/10.13039/501100011033/FEDER, EU. We acknowledge Louise Lamblin for her previous work on the project.
\end{acknowledgements}

\bibliographystyle{aa} % style aa.bst
\bibliography{cite.bib} % your references Yourfile.bib

\begin{appendix}

\section{Additional CO Maps}
\label{appa:addcom}

\begin{figure}[!htbp]
\resizebox{0.475\textwidth}{!}{\includegraphics{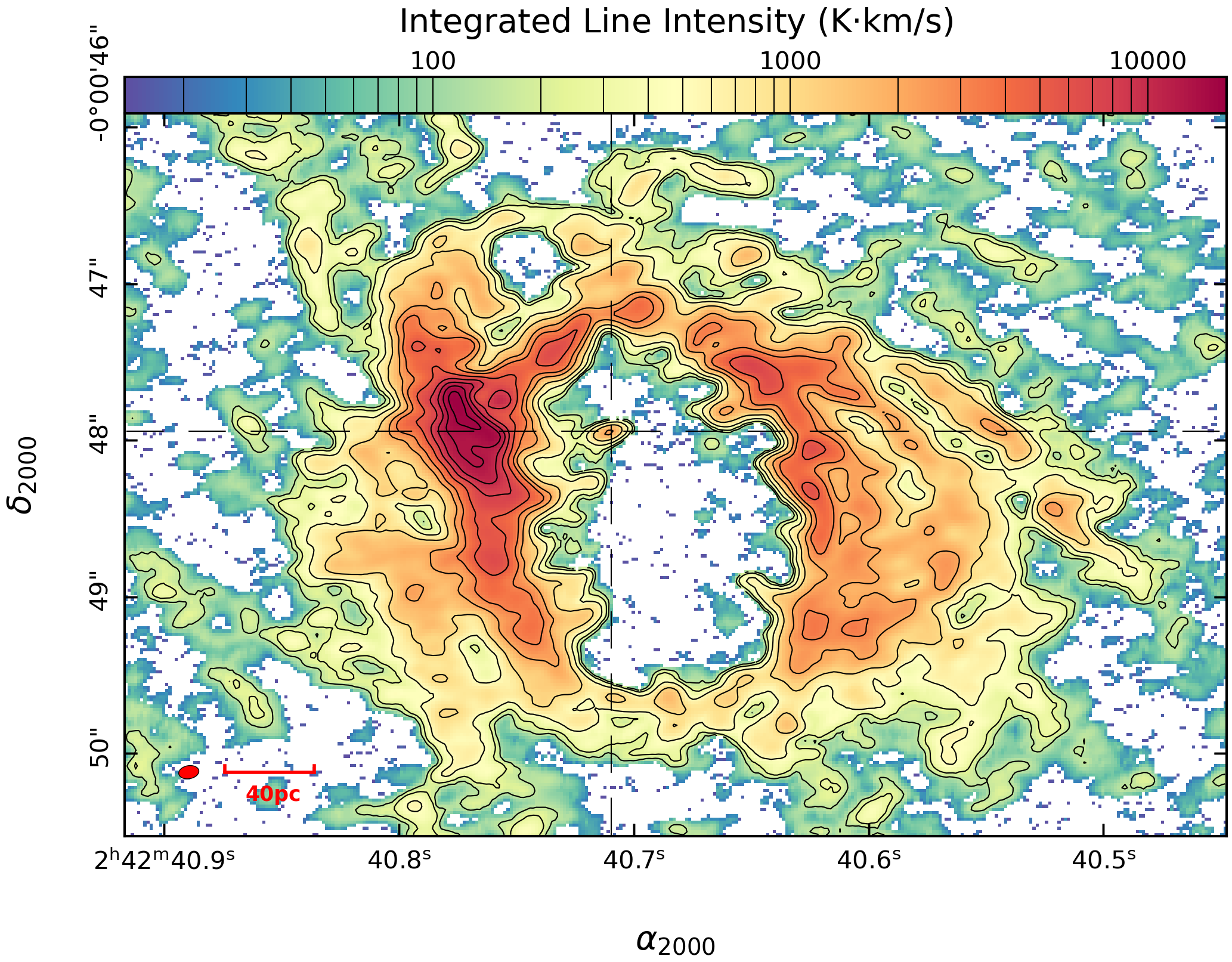}}
\caption{The same as Fig. \ref{fig:1} but for the CO(3-2) transition. The contours cover the same extent as the color bar with levels 3$\sigma$, 5$\sigma$, 10$\sigma$, 20$\sigma$, 40$\sigma$, 60$\sigma$, 100$\sigma$--400$\sigma$ in steps of $50\sigma$, where $1\sigma = 49.5\,\rm{K\,km\,s^{-1}}$.}

\label{fig:a1}       % Give a unique label to the figure.
\end{figure}

\begin{figure}[!htbp]
\resizebox{0.475\textwidth}{!}{\includegraphics{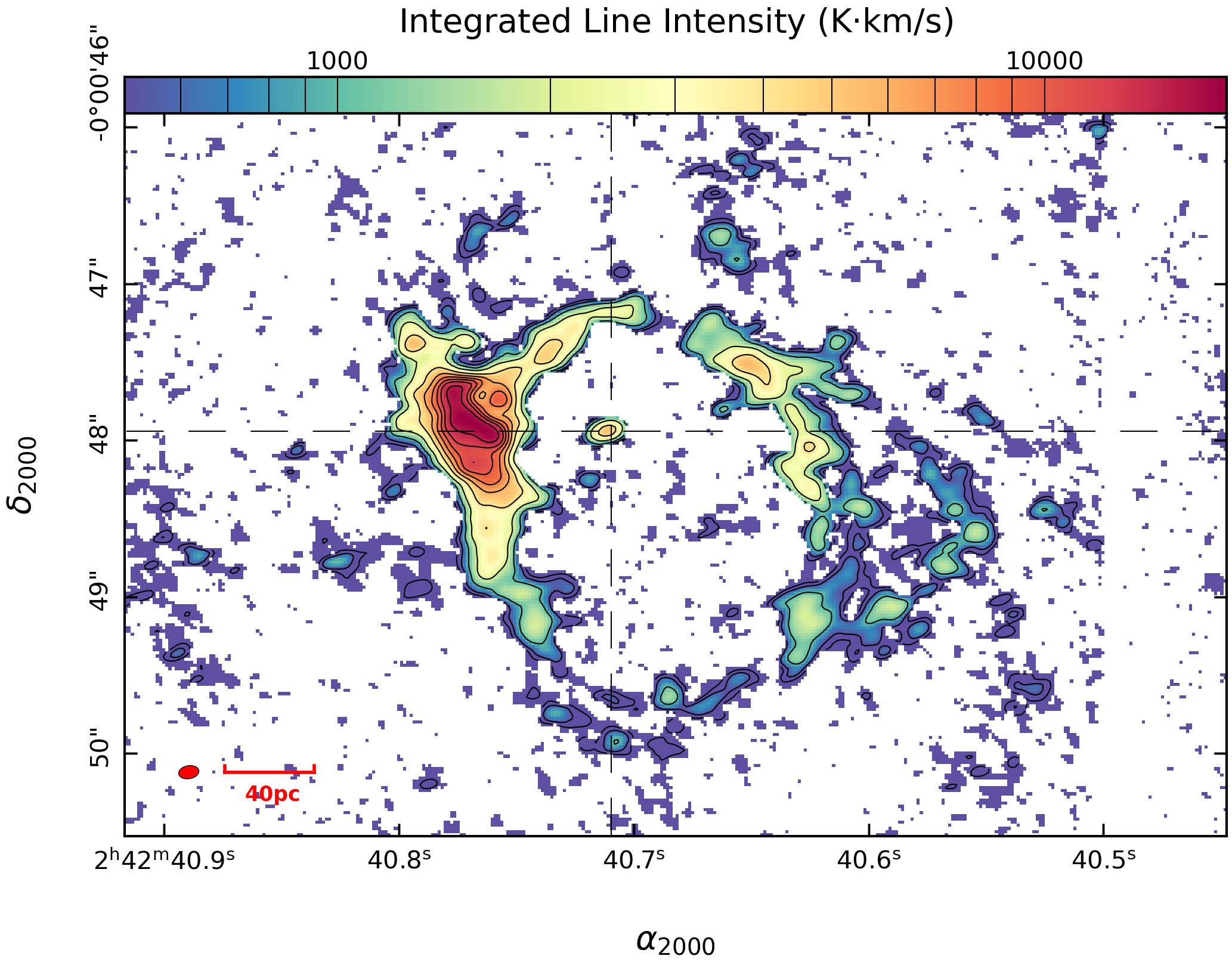}}
\caption{The same as Fig. \ref{fig:1} but for the CO(6-5) transition. The color ranges from $500\,\rm K\,km\,s^{-1}$ to the maximum value in logarithmic scale, and the contours have levels 3$\sigma$, 5$\sigma$, 10$\sigma$, 20$\sigma$, 40$\sigma$, 60$\sigma$, and 80$\sigma$, where $1\sigma=99.9\,\rm K\,km\,s^{-1}$.}

\label{fig:a2}       % Give a unique label to the figure.
\end{figure}

\begin{figure}[!htbp]
\resizebox{0.475\textwidth}{!}{\includegraphics{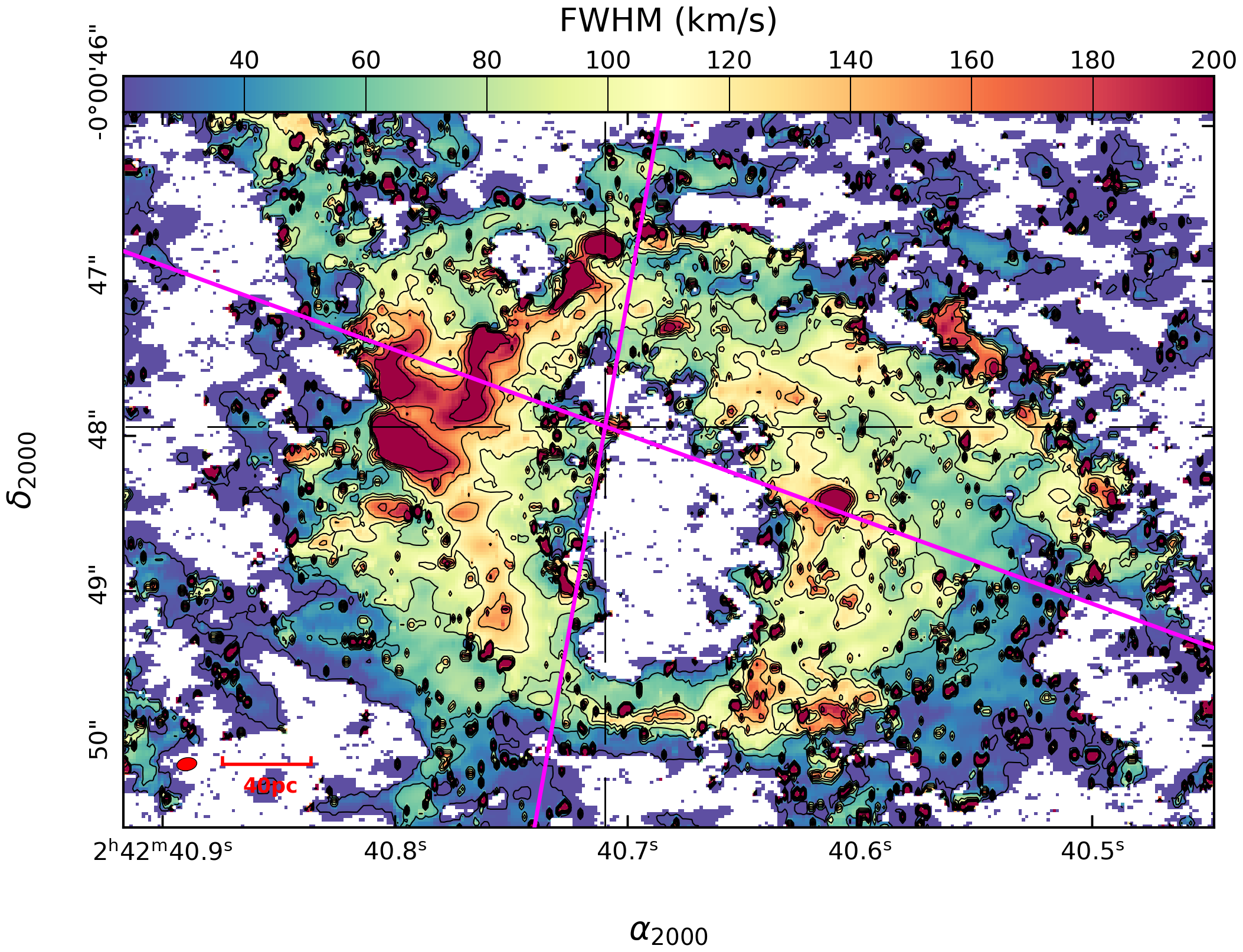}}
\caption{The same as Fig. \ref{fig:2} but for the CO(3-2) transition.}

\label{fig:a3}       % Give a unique label to the figure.
\end{figure}

\begin{figure}[!htbp]
\resizebox{0.475\textwidth}{!}{\includegraphics{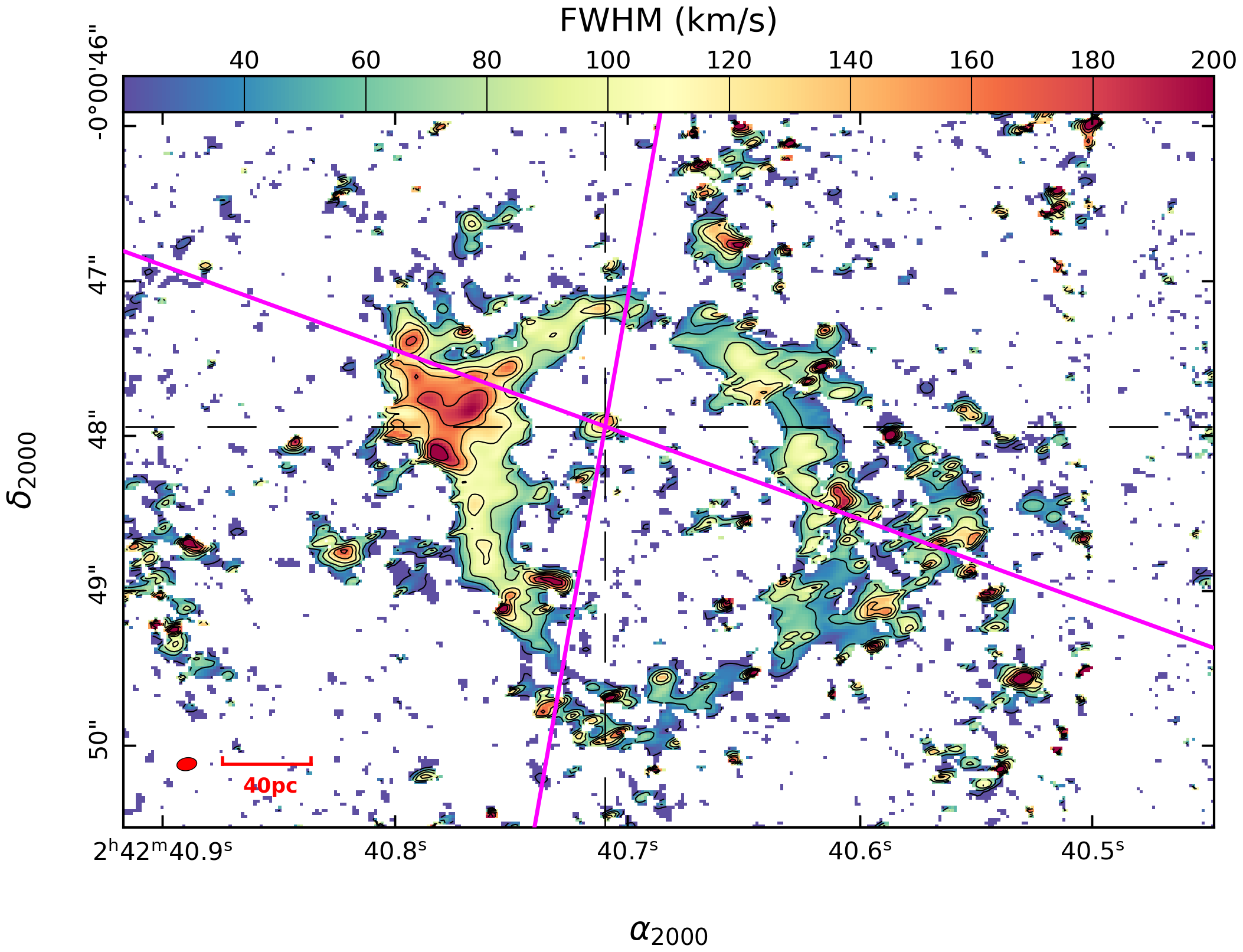}}
\caption{The same as Fig. \ref{fig:2} but for the CO(6-5) transition.}

\label{fig:a4}       % Give a unique label to the figure.
\end{figure}

\begin{figure}[!htbp]
\resizebox{0.475\textwidth}{!}{\includegraphics{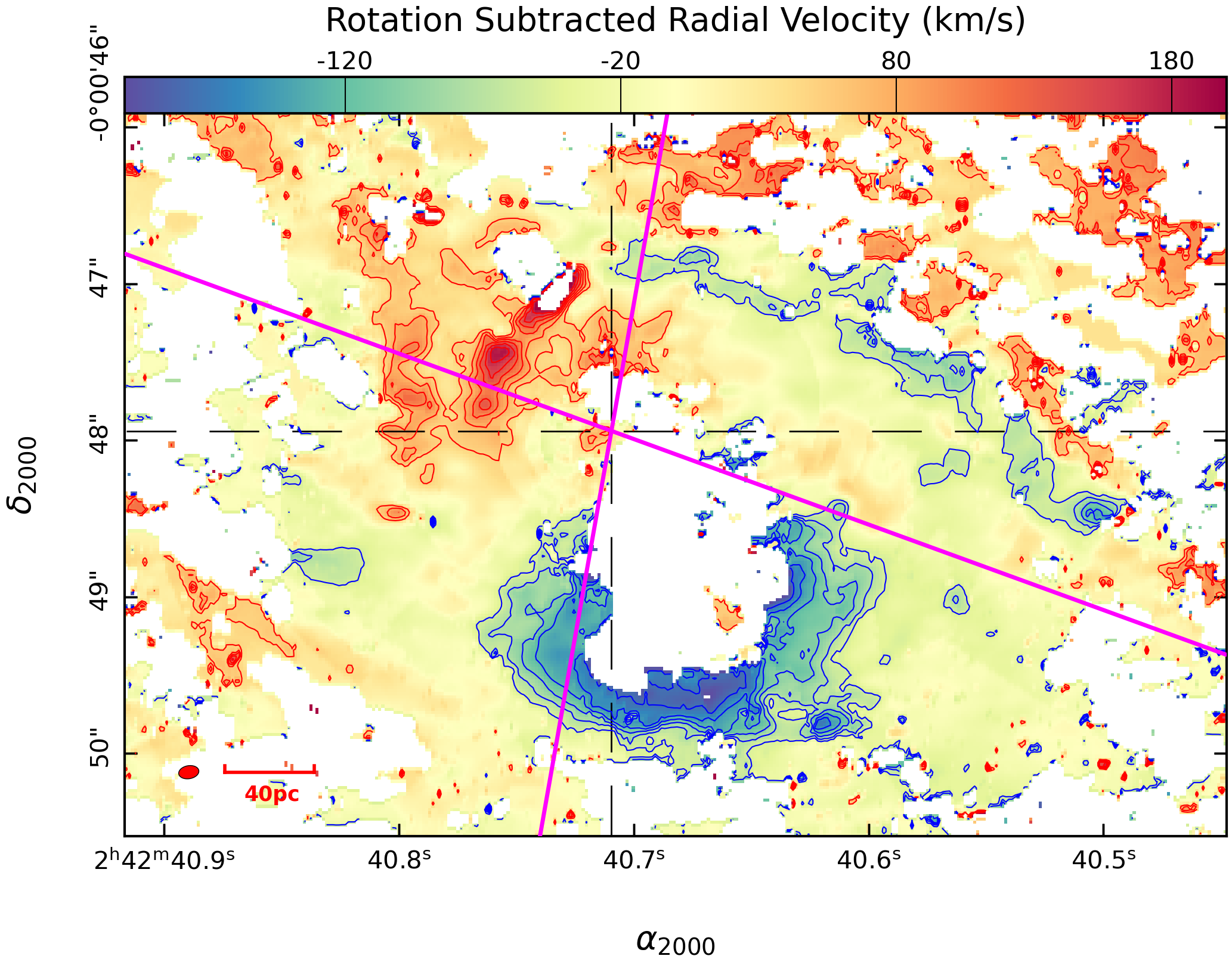}}
\caption{The same as Fig. \ref{fig:3} but for the CO(3-2) transition.}

\label{fig:a5}       % Give a unique label to the figure.
\end{figure}

\newpage

\begin{figure}[!htbp]
\resizebox{0.475\textwidth}{!}{\includegraphics{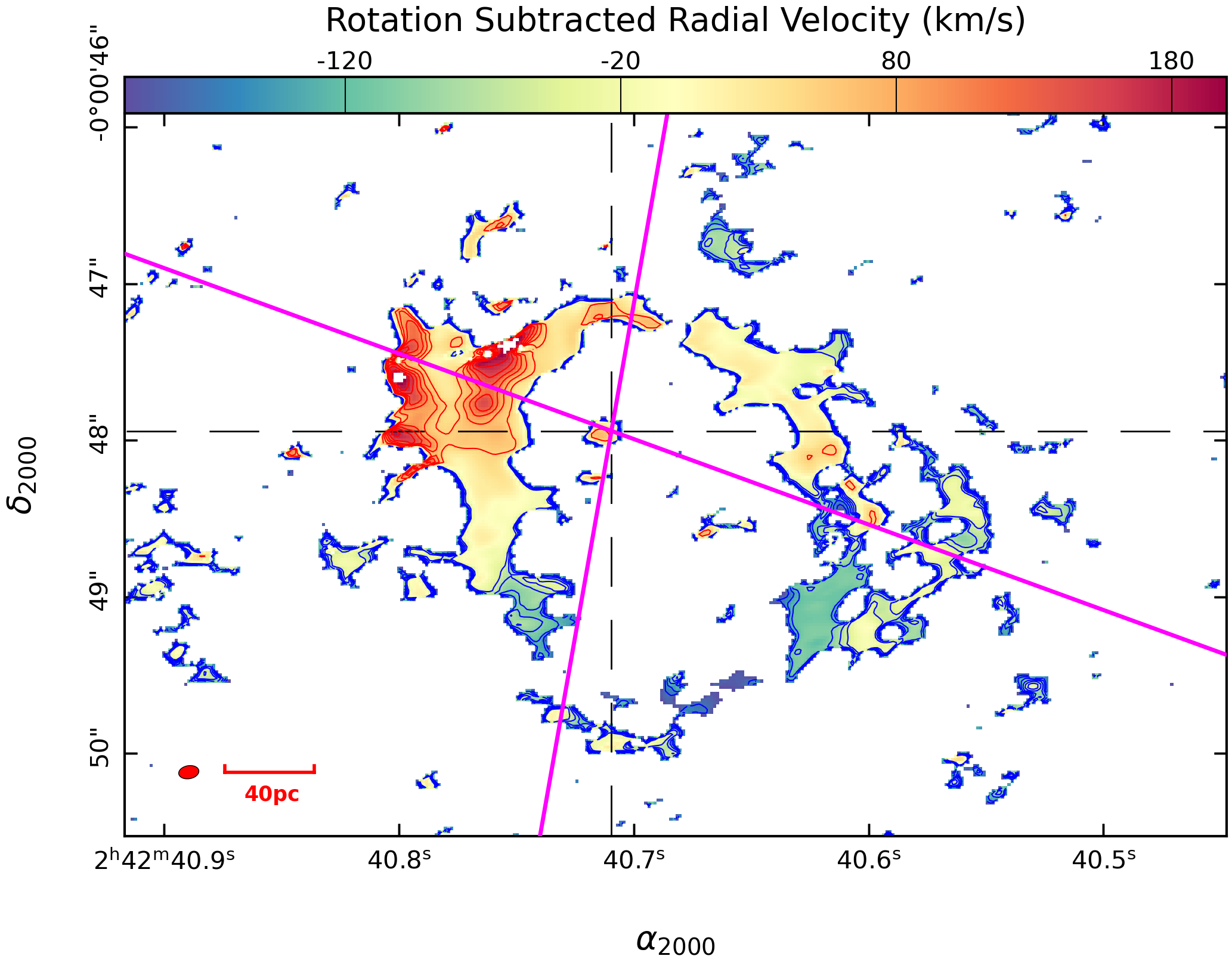}}
\caption{The same as Fig. \ref{fig:3} but for the CO(6-5) transition.}

\label{fig:a6}       % Give a unique label to the figure.
\end{figure}

\clearpage

\section{Line ratio maps and LTE analysis results of regions based on high excitation}
\label{appb:lrlteres}

\begin{figure*}
\centering
\resizebox{1\textwidth}{!}{\includegraphics{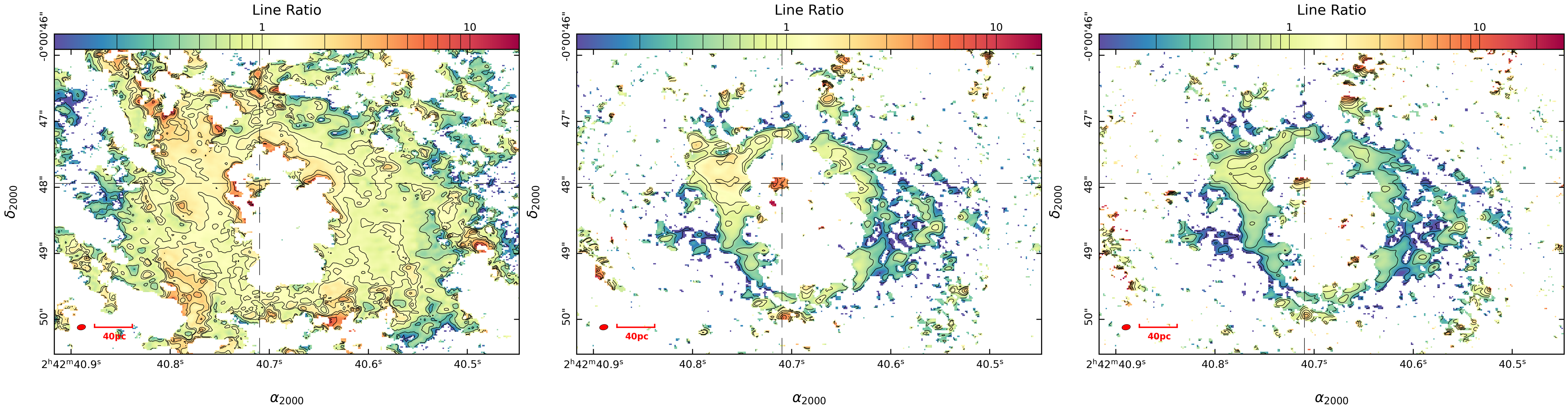}}
\caption{CO(3-2)/CO(2-1), CO(6-5)/CO(2-1), and CO(6-5)/CO(3-2) line ratio maps shown in the left, middle and right panels respectively. The color ranges from 0.1 to the maximum ratio for each plot. In the left panel, the contours span the same extent as the color bar with levels 0.1$\sigma$, 0.2$\sigma$, 0.4$\sigma$, 0.8$\sigma$, 1$\sigma$, 1.5$\sigma$, 2$\sigma$, 4$\sigma$, 6$\sigma$, 8$\sigma$, and 10$\sigma$ where $1\sigma=2.141273$, the root-mean-square value from each ratio map. For the middle panel, the contour levels are 0.1$\sigma$, 0.2$\sigma$, 0.4$\sigma$, 0.8$\sigma$, 1$\sigma$, 1.5$\sigma$, 2$\sigma$, 3$\sigma$, 4$\sigma$, and 5$\sigma$ with $1\sigma=4.712347$. The right panel contains contours with levels 0.05$\sigma$, 0.1$\sigma$, 0.2$\sigma$, 0.4$\sigma$, 0.6$\sigma$, 1$\sigma$, 1.5$\sigma$, 2$\sigma$, 3$\sigma$, 4$\sigma$, and 6$\sigma$, where $1\sigma=5.144535$. The symbols and markers for all panels are the same as in Fig. \ref{fig:1}.}

\label{fig:5}       % Give a unique label to the figure.
\end{figure*}

\begin{figure*}
\centering
\resizebox{1\textwidth}{!}{\includegraphics{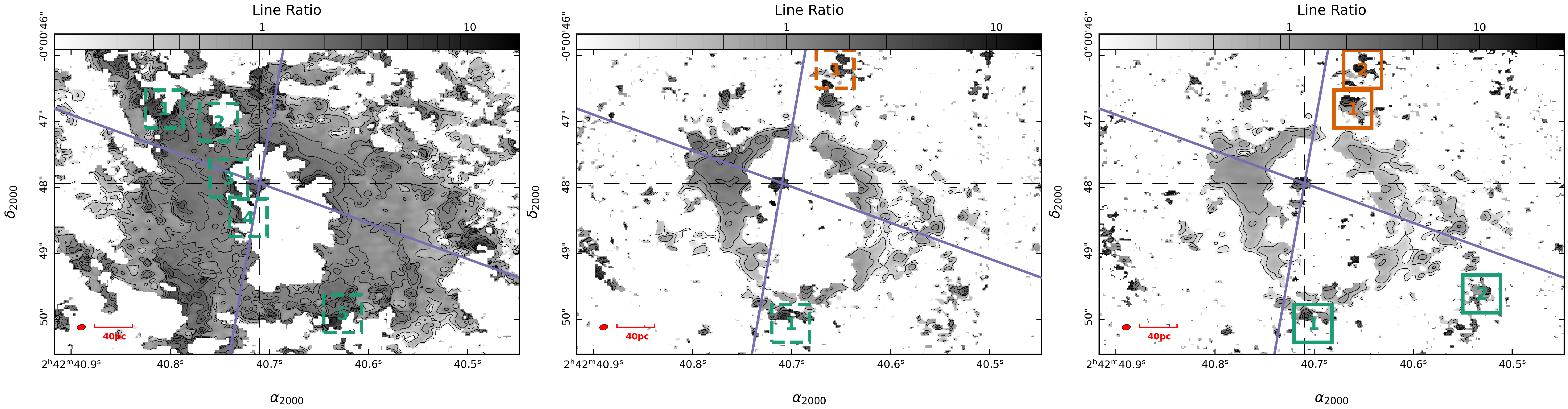}}
\caption{Region selection based on the AGN wind bicone, which is indicated by the purple lines, and excitation levels from the line ratio maps. The left, middle, and right panels are the region selections based on the CO(3-2)/CO(2-1), CO(6-5)/CO(2-1), and CO(6-5)/CO(3-2) ratio maps respectively. The number and color of each region correspond to those in the equivalent region code displayed in Table \ref{tab3}. The dashed regions contain line ratios $\geq3\sigma$, while the solid regions only include line ratios $\geq2\sigma$. Other schematics of the region definitions are the same as Fig. \ref{fig:4}. The underlying plots for all three panels are the corresponding line ratio maps from Fig. \ref{fig:5}.}

\label{fig:6}       % Give a unique label to the figure.
\end{figure*}

\clearpage

\begin{sidewaystable}[!htbp]
\centering
\begin{tabular}{c c c c c c c c}
\hline
\hline
Region & Excitation & Right Ascension & Declination & Rotational Temperature ($T_{\rm{rot}}$) & CO(2-1) & CO(3-2) & CO(6-5) \\ % Repeat this pattern 31 times
\hline
HRG1\_40\_3221 & $\geq3\sigma$ & 02h42m40.806s & $\rm{-00^{\circ}00'46.814''}$ & $30.9\pm1.4$ & $1.23\pm0.15$ & $1.81\pm0.24$ & $1.42\pm0.30$ \\
HRG2\_40\_3221 & $\geq3\sigma$ & 02h42m40.751s & $\rm{-00^{\circ}00'47.014''}$ & $44.2\pm3.1$ & $3.58\pm0.47$ & $3.11\pm0.43$ & $3.42\pm0.81$ \\
HRG3\_40\_3221 & $\geq3\sigma$ & 02h42m40.741s & $\rm{-00^{\circ}00'47.864''}$ & $73.0\pm7.6$ & $26.3\pm4.0$ & $16.5\pm2.6$ & $22.3\pm5.0$ \\
HRG4\_40\_3221 & $\geq3\sigma$ & 02h42m40.721s & $\rm{-00^{\circ}00'48.464''}$ & $48.5\pm3.7$ & $2.96\pm0.45$ & $2.06\pm0.30$ & $2.53\pm0.59$ \\
HRG5\_40\_3221 & $\geq3\sigma$ & 02h42m40.626s & $\rm{-00^{\circ}00'49.914''}$ & $29.3\pm1.3$ & $1.45\pm0.18$ & $1.80\pm0.23$ & $1.55\pm0.35$ \\
HRG1\_40\_6521 & $\geq3\sigma$ & 02h42m40.701s & $\rm{-00^{\circ}00'50.064''}$ & $49.8\pm3.5$ & $2.12\pm0.28$ & $1.42\pm0.19$ & $1.83\pm0.38$ \\
HRO1\_40\_6521 & $\geq3\sigma$ & 02h42m40.656s & $\rm{-00^{\circ}00'46.214''}$ & $56.1\pm4.4$ & $2.08\pm0.28$ & $1.03\pm0.15$ & $1.61\pm0.34$ \\
HRG1\_40\_6532 & $\geq2\sigma$ & 02h42m40.701s  & $\rm{-00^{\circ}00'50.064''}$ & $49.8\pm3.5$ & $2.12\pm0.28$ & $1.42\pm0.19$ & $1.83\pm0.38$ \\
HRG2\_40\_6532 & $\geq2\sigma$ & 02h42m40.531s & $\rm{-00^{\circ}00'49.614''}$ & $43.9\pm2.7$ & $2.30\pm0.28$ & $1.20\pm0.16$ & $1.84\pm0.38$ \\
HRO1\_40\_6532 & $\geq2\sigma$ & 02h42m40.661s & $\rm{-00^{\circ}00'46.814''}$ & $52.3\pm3.8$ & $3.72\pm0.48$ & $1.84\pm0.25$ & $2.91\pm0.60$ \\
HRO2\_40\_6532 & $\geq2\sigma$ & 02h42m40.651s & $\rm{-00^{\circ}00'46.214''}$ & $60.3\pm5.1$ & $2.02\pm0.28$ & $0.93\pm0.13$ & $1.53\pm0.32$ \\
\hline
\end{tabular}
\caption{LTE rotational temperatures ($T_{\rm{rot}}$) and total CO column densities ($N$) from the original $40\times40\,\rm{pc}$ regions defined based on excitation in units of K and $10^{17}\,\rm{cm^{-2}}$ respectively. We also provide coordinates in right ascension and declination at the centers of these regions, and definitions of the region numbers (in the first column) are stated in Sect. \ref{rsexcitation}. The total CO column density calculated from each CO transition is indicated by the name of each transition in the header. The excitation level (above 3 or 2 times the RMS value) is also provided under the ``Excitation'' column.}
\label{tab3}
\end{sidewaystable}

\clearpage

\begin{figure*}[!htbp]
\resizebox{\textwidth}{!}{\includegraphics{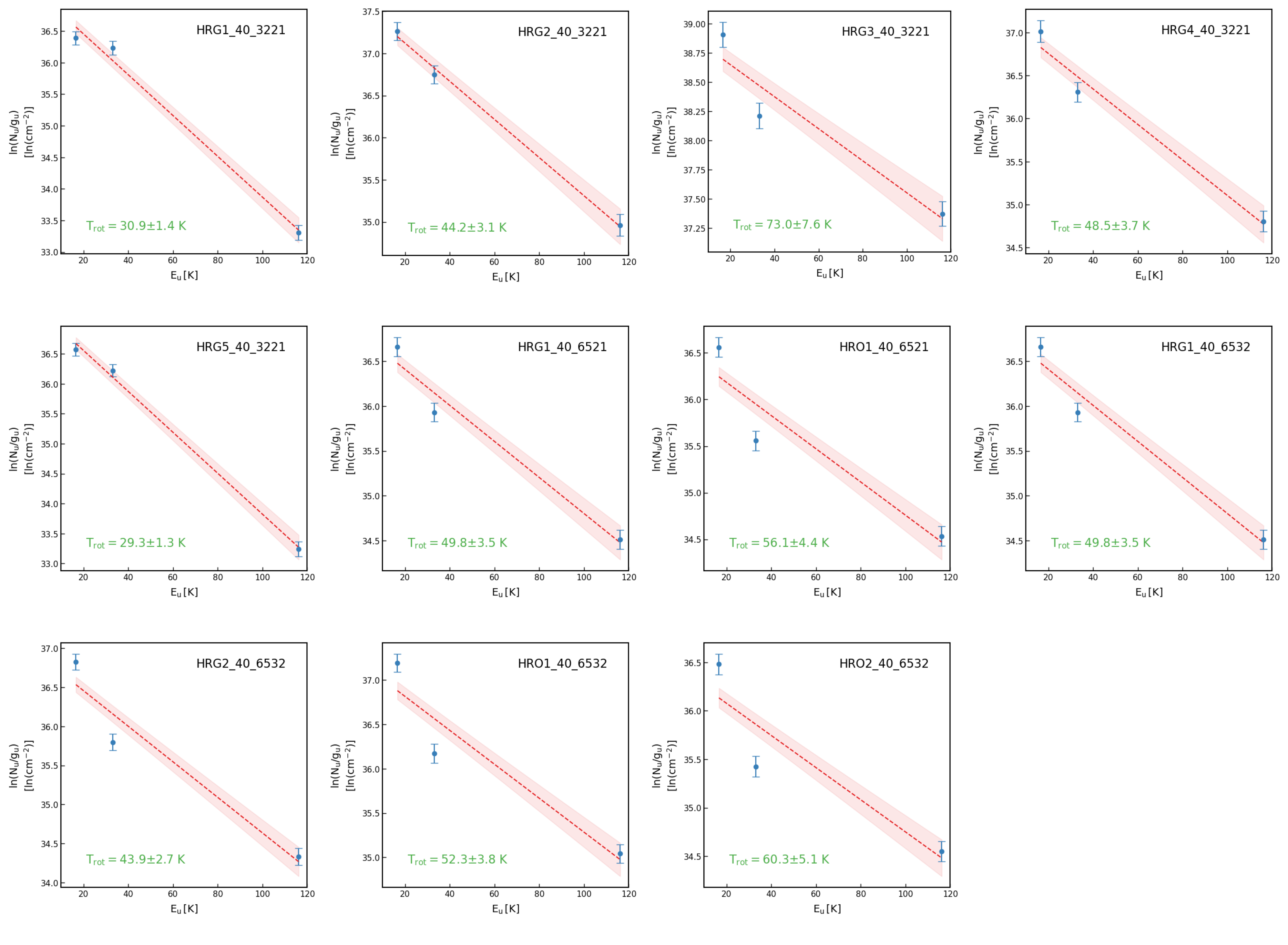}}
\caption{Rotational diagrams of regions defined based on line ratios. Region numbers are indicated at the upper right corner of each diagram. The rotational temperature values and their uncertainties are displayed in the lower-left corner of each diagram. All other symbols and markers of each rotational diagram are the same as inset plots in Fig. \ref{fig:8}, except that the linear fit and its uncertainty are in red instead of purple.}

\label{fig:a7}       % Give a unique label to the figure.
\end{figure*}

\begin{figure*}[!htbp]
\resizebox{1\textwidth}{!}{\includegraphics{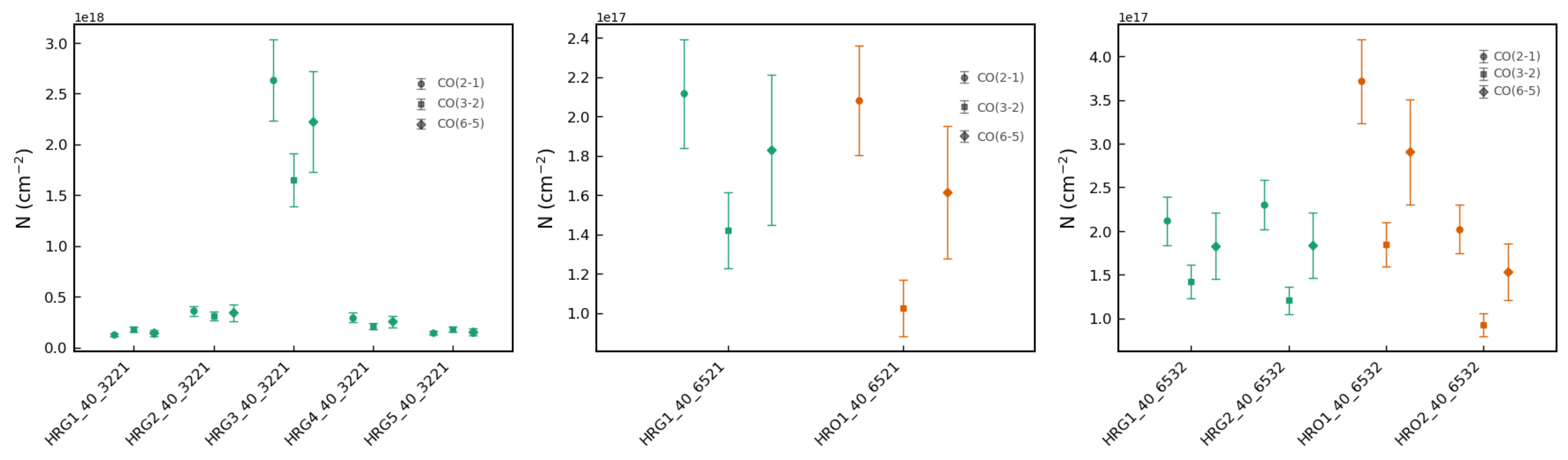}}
\caption{Total CO column densities ($N$) independently calculated from three CO transitions for all regions selected based on the CO(3-2)/CO(2-1) (\textit{left panel}), CO(6-5)/CO(2-1) (\textit{middle panel}), and CO(6-5)/CO(3-2) (\textit{right panel}) line ratio. For each panel, the x-axis marks the region labels, while the y-axis indicates the total CO column densities in the unit of $\rm{cm^{-2}}$. The region code (along the x-axis) and color (the color of the data points) of each region match those in Table \ref{tab3}. All other symbols and markers are the same as in Fig. \ref{fig:7}.}

\label{fig:a25}       % Give a unique label to the figure.
\end{figure*}

\begin{figure*}[!htbp]
\resizebox{1\textwidth}{!}{\includegraphics{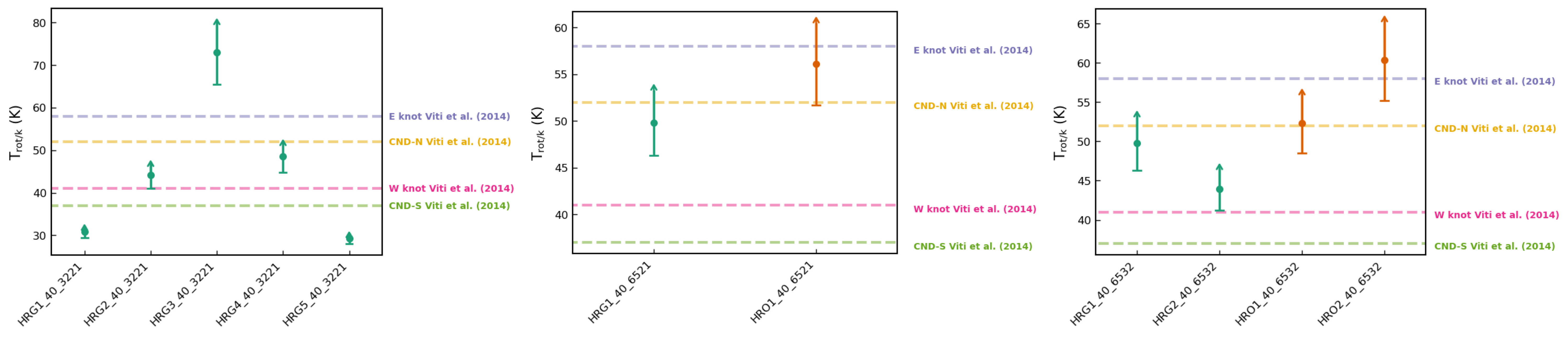}}
\caption{Summary of rotational temperatures for all regions sampled around the CND defined based on the CO(3-2)/CO(2-1) (\textit{left panel}), CO(6-5)/CO(2-1) (\textit{middle panel}), and CO(6-5)/CO(3-2) (\textit{right panel}) line ratio. For each panel, the x-axis marks the region labels, while the y-axis indicates the rotational temperatures in the unit of K. The region code (along the x-axis) and color (the color of the data points) of each region match those in Table \ref{tab3}. All other symbols and markers are the same as in Fig. \ref{fig:a23}.}

\label{fig:a24}       % Give a unique label to the figure.
\end{figure*}

\clearpage

\section{Fitting results of line profiles for $40\times40$ pc regions defined based on the outflow}
\label{appc:lpfit}

\newpage

\begin{sidewaystable*}[!htbp]
%\begin{table*}[!htbp]
\centering
\begin{tabular}{c c c c c c c c c c c c}
\hline
\hline
Transition & Region & \# & $w_1$ & $w_2$ & $w_3$ & $\mu_1$ & $\sigma_1$ & $\mu_2$ & $\sigma_2$ & $\mu_3$ & $\sigma_3$ \\ % Repeat this pattern 31 times
\hline
CO(2-1) & G1\_40 (N) & 2 & $0.507\pm0.079$ & $0.493\pm0.079$ &  & $1150.7\pm3.9$ & $51.4\pm4.6$ & $1216\pm11$ & $93.6\pm3.2$ &  &  \\
 & G2\_40 (N) & 2 & $0.730\pm0.049$ & $0.270\pm0.053$ &  & $1184.3\pm2.5$ & $79.6\pm2.9$ & $1149.8\pm4.2$ & $30.2\pm4.2$ &  &  \\
 & G3\_40 (N) & 1 &  &  &  & $1139.1\pm1.9$ & $63.1\pm1.8$ &  &  &  &  \\
 & G4\_40 (S) & 3 & $0.84\pm0.08$ & $0.16\pm0.16$ & ``$-$'' & $999.9\pm5.9$ & $44.4\pm3.5$ & $1085.1\pm11.5$ & $23.3\pm15.1$ & $1183\pm92$ & $37\pm75$ \\
 & G5\_40 (S) & 2 & $\sim0.83$ & $\sim0.17$ &  & $951.1\pm3.8$ & $36.4\pm3.5$ & $\sim1064$ & $\sim12$ &  &  \\
 & G6\_40 (S) & 1 &  &  &  & $980.0\pm2.8$ & $62.1\pm3.4$ &  &  &  &  \\
 & G7\_40 (S) & 2 & $0.68\pm0.17$ & $0.32\pm0.16$ &  & $1021.8\pm3.2$ & $37.7\pm4.5$ & $990\pm23$ & $76\pm19$ &  &  \\
 & G8\_40 (S) & 3 & $0.868\pm0.038$ & $0.082\pm0.031$ & $0.050\pm0.052$ & $1107.1\pm3.8$ & $62.4\pm2.6$ & $989.1\pm5.2$ & $23.2\pm6.1$ & $895\pm27$ & $37\pm37$ \\
 & O1\_40 (E) & 1 &  &  &  & $1107.8\pm1.2$ & $52.1\pm1.0$ &  &  &  &  \\
 & O2\_40 (W) & 1 &  &  &  & $1115.4\pm1.7$ & $45.6\pm1.4$ &  &  &  &  \\
CO(3-2) & G1\_40 (N) & 2 & $0.57\pm0.11$ & $0.43\pm0.11$ &  & $1160.3\pm4.7$ & $59.4\pm5.1$ & $1226\pm15$ & $91.0\pm3.5$ &  &  \\
 & G2\_40 (N) & 2 & $0.86\pm0.40$ & $0.14\pm0.48$ &  & $1166.8\pm7.9$ & $48.3\pm6.3$ & $1290\pm220$ & $80\pm210$ &  &  \\
 & G3\_40 (N) & 2 & $0.64\pm0.17$ & $0.36\pm0.17$ &  & $1162\pm10$ & $37.1\pm4.2$ & $1081\pm20$ & $40.0\pm9.1$ &  &  \\
 & G4\_40 (S) & 2 & $0.64\pm0.23$ & $0.36\pm0.23$ &  & $990\pm12$ & $39.7\pm4.6$ & $1071\pm29$ & $46\pm12$ &  &  \\
 & G5\_40 (S) & 2 & $0.776\pm0.035$ & $0.224\pm0.016$ &  & $949.1\pm1.6$ & $34.9\pm1.9$ & $1067.9\pm1.7$ & $22.0\pm1.9$ &  &  \\
 & G6\_40 (S) & 2 & $0.550\pm0.043$ & $0.450\pm0.049$ &  & $952.4\pm3.7$ & $38.0\pm3.2$ & $1072.5\pm4.0$ & $39.9\pm5.4$ &  &  \\
 & G7\_40 (S) & 2 & $0.633\pm0.080$ & $0.367\pm0.087$ &  & $1016.8\pm2.1$ & $72.9\pm4.5$ & $1023.9\pm3.1$ & $31.3\pm4.4$ &  &  \\
 & G8\_40 (S) & 3 & $0.835\pm0.042$ & $0.130\pm0.033$ & $0.0348\pm0.0072$ & $1124.4\pm3.8$ & $51.9\pm2.4$ & $1007.9\pm7.3$ & $31.9\pm5.2$ & $897.8\pm4.0$ & $26.8\pm7.3$ \\
 & O1\_40 (E) & 1 &  &  &  & $1114.4\pm1.0$ & $56.02\pm0.75$ &  &  &  &  \\
 & O2\_40 (W) & 1 &  &  &  & $1114.4\pm1.0$ & $49.64\pm0.78$ &  &  &  &  \\
CO(6-5) & G1\_40 (N) & 2 & $0.8\pm5.6$ & $0.2\pm5.6$ &  & $1169\pm82$ & $68\pm62$ & $1300\pm1600$ & $90\pm440$ &  &  \\
 & G2\_40 (N) & 1 &  &  &  & $1161.9\pm3.1$ & $49.2\pm3.5$ &  &  &  &  \\
 & G4\_40 (S) & 1 &  &  &  & $1000.4\pm5.9$ & $40.7\pm5.8$ &  &  &  &  \\
 & G7\_40 (S) & 1 &  &  &  & $1010.7\pm1.6$ & $29.8\pm1.9$ &  &  &  &  \\
 & G8\_40 (S) & 1 &  &  &  & $1117.7\pm5.0$ & $71.0\pm7.6$ &  &  &  &  \\
 & O1\_40 (E) & 1 &  &  &  & $1112.5\pm1.7$ & $48.7\pm1.8$ &  &  &  &  \\
 & O2\_40 (W) & 1 &  &  &  & $1131.5\pm1.3$ & $43.1\pm1.2$ &  &  &  &  \\
\hline
\end{tabular}
\caption{Fitting results of CO(2-1), CO(3-2), and CO(6-5) line profiles sampled from the original $40\times40$ regions defined based on outflow. The transition is indicated in the first column of the table and follows the transition above if not specified. The number of Gaussian components fitted is listed under the column indicated by the ``\#'' sign. $\rm{w_1}$ and $\rm{w_2}$ are the two weights associated with the Gaussian components, as shown in Equation \ref{eqn4}. These two parameters are not given when only a single Gaussian was fitted to the line profile, and only $\rm{w_1}$ is given as the sole weight when a weighted double Gaussian was fitted. The rest of the parameters are the mean and standard deviation of three different Gaussian components. Only the first one or two sets are indicated if a single or weighted double Gaussian was fitted to the line profile. The ``-'' sign indicates the absorption Gaussian component. Only the positive Gaussian components are considered for the weight calculation. The ``$\sim$'' sign is used when the fitting result contains a large error bar (not provided in the table). The location of each region within the CND is also provided in brackets after the corresponding region code in column 2 (e.g., ``N'' stands for the northern CND).}
\label{tabb3}
%\end{table*}
\end{sidewaystable*}

\clearpage

\section{Line profile behavior at smaller scales}
\label{appd:lps}

\subsection{Line profile behavior of zoomed-in regions}
\label{lpz}

We address the discrepancies between the fitting results for the $40\times40$ pc and $20\times20$ pc regions. \textbf{\underline{G8\_20 (S):}} Located along the southwestern edge of the AGN wind bicone, region G8\_20 (S) exhibits a distinct behavior for the CO(6-5) transition compared to region G8\_40 (S). While the latter profile is fitted with a single Gaussian, at the smaller scale, it is modeled with a double Gaussian. The minor component is significantly blueshifted from $\rm{v_{sys}}$, indicating outflowing gas not captured at the larger scale due to the inclusion of non-outflowing gas outside the AGN wind bicone. \textbf{\underline{G6\_20 (S) \& G7\_20 (S):}} For these regions, the number of Gaussian components remains unchanged between the two scales; however, the more blueshifted component becomes more prominent in the CO(3-2) line profile at the smaller scale. This likely reflects the exclusion of less blueshifted gas farther from the inner edge of the CND in the smaller regions. \textbf{\underline{G5\_20 (S):}} In the southeastern CND, region G5\_20 (S) isolates the most blueshifted outflowing gas within G5\_40 (S), resulting in a single Gaussian fit for the CO(2-1) and CO(3-2) line profiles. The stronger average emission within the $20\times20$ pc region also allows the CO(6-5) line profile to be fitted with a single Gaussian. \textbf{\underline{G1\_20 (N) \& G2\_20 (N):}} Unlike the southern regions, the $20\times20$ pc regions in the northern CND, compared to their $40\times40$ pc counterparts, show smaller or absent wing components. For instance, the CO(3-2) line profile in G2\_20 (N) and the CO(6-5) line profile in G1\_20 (N) lose their wing components entirely, likely due to the exclusion of gas from the E-knot (see Garc\'{i}a-Burillo et al. \citeyear{gb19}) near the edges of the collimated jet (Gallimore et al. \citeyear{gallimore_etal_96}, \citeyear{gallimore_etal_04}). \textbf{\underline{G3\_20 (N):}} Along the northwestern edge of the AGN wind bicone, the CO(3-2) line profile of region G3\_20 (N) is fitted with a redshifted single Gaussian, lacking the slightly blueshifted component observed at the larger scale. The smaller $20\times20$ pc region likely excludes more gas independent of the outflow that is located farther from the AGN wind bicone.

\subsection{Line profile behavior of zoomed-in and shifted regions}
\label{lpzs}

For the zoomed-in and shifted regions, results from the new line profiles and their fits further refine the identification of regions contributing to different Gaussian components within the line profiles of the original $40\times40$ pc regions. \textbf{\underline{G1\_20\_DR (N):}} The fitting results for region G1\_20\_DR (N) are consistent with those of G1\_40 (N) across all three CO transitions, featuring a double Gaussian with a broad redshifted wing component. The weight parameters for the wing components in G1\_20\_DR (N) are larger than those of G1\_20 (N) and comparable to G1\_40 (N), indicating that G1\_20\_DR (N), situated near the E-knot (see Sect. \ref{zsr}), captures most of the emission from the stronger ionized outflow that dominates G1\_40 (N). \textbf{\underline{G2\_20\_DR (N):}} For G2\_20\_DR (N), the CO(2-1) and CO(3-2) line profiles show less outflowing gas compared to G2\_40 (N) and G2\_20 (N). The CO(2-1) profile is fitted with a single redshifted Gaussian without a broad wing, while the CO(3-2) profile includes an additional Gaussian component centered at $\rm v_{sys}$. This reduced outflow may result from the geometry of the northern ionized outflow, which is tilted further from the CND plane, causing the wind to intersect molecular gas further from the AGN, near the geometrically thickest part of the CND. \textbf{\underline{G3\_20\_L (N):}} The CO(2-1) line profile of G3\_20\_L (N) exhibits a more redshifted component and a small blueshifted component, unlike the single Gaussian fits for the same transition in larger regions. With a larger fraction of G3\_20\_L (N) lying within the AGN wind bicone, more outflowing gas is detected. \textbf{\underline{G4\_20\_DR (S):}} Closer to the AGN wind bicone, G4\_20\_DR (S) contains more outflowing gas than G4\_40 (S) and G4\_20 (S). The CO(2-1) and CO(3-2) profiles show more prominent blueshifted components, with the CO(3-2) profile reducible to a single Gaussian centered at the velocity of the previous farther blueshifted component. \textbf{\underline{G5\_20\_D (S) \& G6\_20\_DR (S):}} For G5\_20\_D (S) and G6\_20\_DR (S), further from the CND's inner edge, the CO(3-2) profiles indicate less outflowing gas than previous fits, with smaller weight parameters for the farther blueshifted components. The CO(2-1) emission is too weak for optimal fits, while the CO(6-5) profile in G6\_20\_DR (S) shows almost no emission. In G5\_20\_D (S), the CO(6-5) profile is fitted with a less blueshifted single Gaussian, indicating weaker high-velocity outflows. \textbf{\underline{G7\_20\_UL (S):}} Near the CND's inner edge, G7\_20\_UL (S) shows more high-velocity outflowing gas than G7\_40 (S) and G7\_20 (S). The new CO(2-1) and CO(3-2) profiles feature a farther blueshifted Gaussian component with a larger weight parameter, lacking the less blueshifted wing components seen previously. Together with G5\_20\_D (S) and G6\_20\_DR (S), these regions within the southern AGN wind bicone reveal more molecular gas entrained in high-velocity outflows along the inner edge of the CND, contrasting with the line profile behavior in G2\_20\_DR (N). The southern AGN wind, angled closer to the CND plane, likely interacts with molecular gas at the inner edge and deeper within the CND, driving blueshifted multi-component outflows. In the north, the AGN wind primarily sweeps surface molecular gas, resulting in redshifted broad-wing structures. \textbf{\underline{G8\_20\_DL (S):}} Like G4\_20\_DR (S), G8\_20\_DL (S) contains more outflowing gas than G8\_40 (S) and G8\_20 (S), with CO(2-1) and CO(6-5) profiles fitted with a single blueshifted Gaussian rather than multi-component models. The CO(3-2) profile includes a major blueshifted component and a minor component near $\rm v_{sys}$. \textbf{\underline{O1\_20\_DR (E):}} Region O1\_20\_DR (E) captures outflowing gas extending beyond the AGN wind bicone into the eastern CND. Unlike O1\_40 (E) and O1\_20 (E), its line profile is fitted with a double Gaussian, including a minor blueshifted component.

\clearpage

\section{Line profiles and their fitting results of zoomed-in $20\times20$ pc or zoomed-in and shifted regions}
\label{appe:lpzzs}

\begin{figure*}[!htbp]
\resizebox{1\textwidth}{!}{\includegraphics{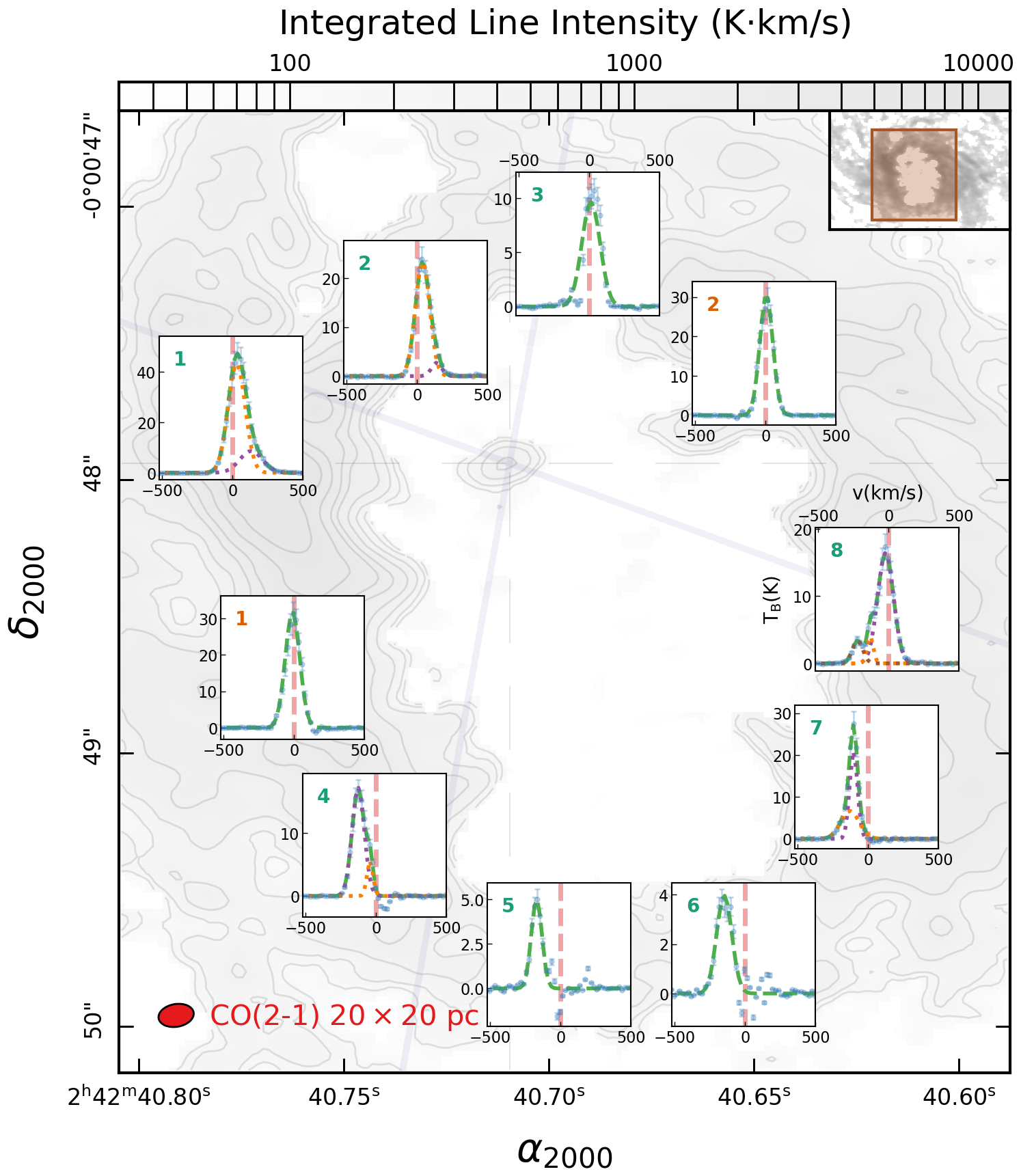}}
\caption{Zoomed in CO(2-1) transition line profiles of all $20\times20$ pc regions defined based on the outflow occupying the exact locations on the background plot as their corresponding original $40\times40$ pc regions assigned in the region definition in Sect. \ref{rsoutflow} (positions of the inset plots only represent the rough locations of the sampled $20\times20$ pc regions). Region numbers, including their colors, are indicated at the upper left corner of each plot, corresponding to those appearing in the region codes in Table \ref{tabb6} and are defined in Sect. \ref{zsr}. The size of the regions, $20\times20$ pc, is indicated at the lower left corner of the background plot, and all other symbols are the same in the background plot and line profile diagrams as those in Fig. \ref{fig:9}.}

\label{fig:a11}       % Give a unique label to the figure.
\end{figure*}

\begin{figure*}[!htbp]
\resizebox{1\textwidth}{!}{\includegraphics{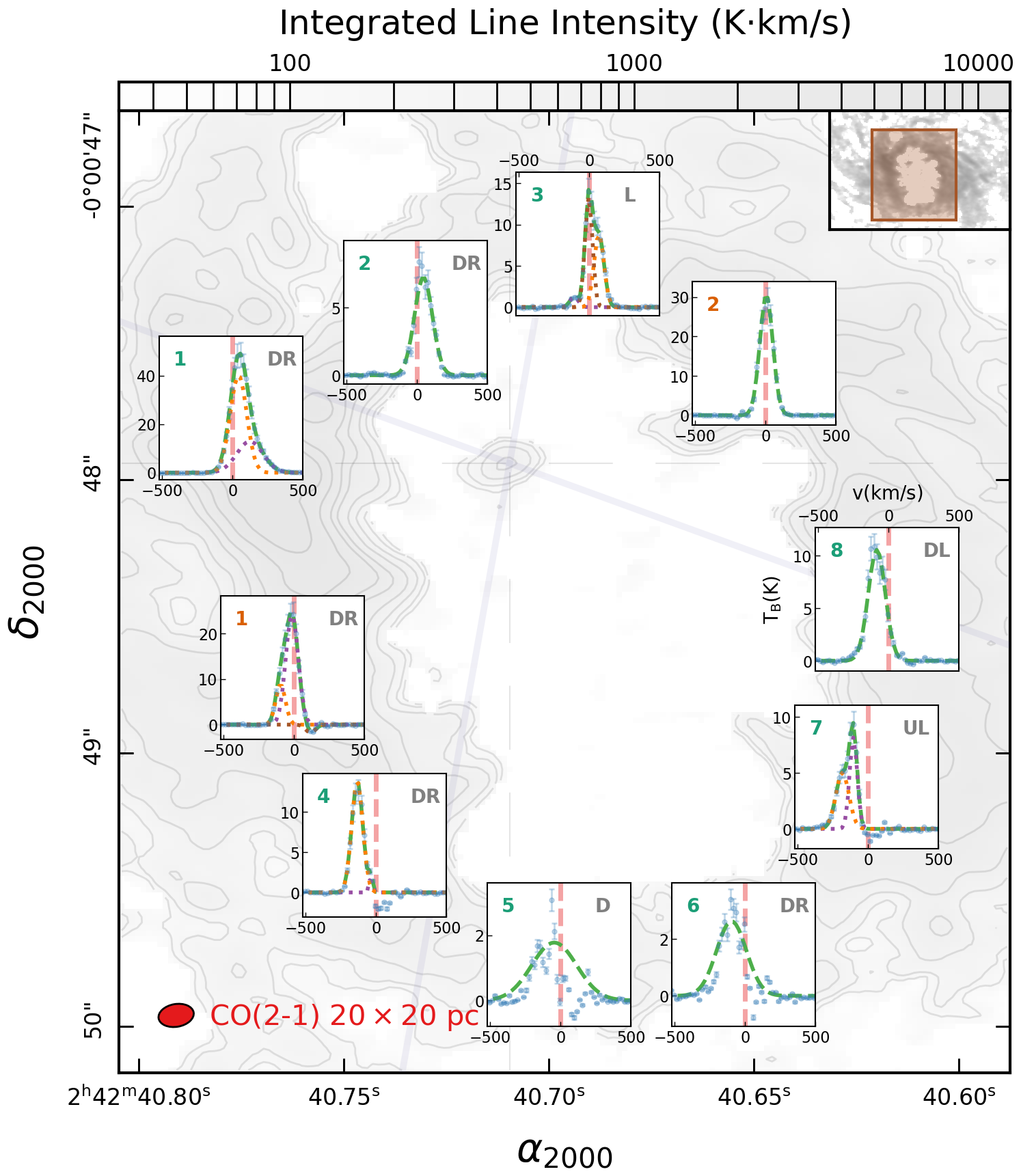}}
\caption{CO(2-1) transition line profiles of all further zoomed-in and shifted regions defined based on the outflow occupying the exact locations on the background plot as their corresponding original $40\times40$ pc regions assigned in the region definition in Sect. \ref{rsoutflow} (positions of the inset plots only represent the rough locations of the sampled zoomed-in and shifted regions). Region numbers, including their colors, are indicated at the upper left corner of each plot, corresponding to those appearing in the region codes in Table \ref{tabb9} and are defined in Sect. \ref{rszs}. Different definitions of the regions (sizes and their locations within corresponding $40\times40$ pc regions) are indicated at the upper right corner of each line profile diagram, and all other symbols are the same in the background plot and line profile diagrams as those in Fig. \ref{fig:9}.}

\label{fig:a14}       % Give a unique label to the figure.
\end{figure*}

\begin{figure*}[!htbp]
\resizebox{1\textwidth}{!}{\includegraphics{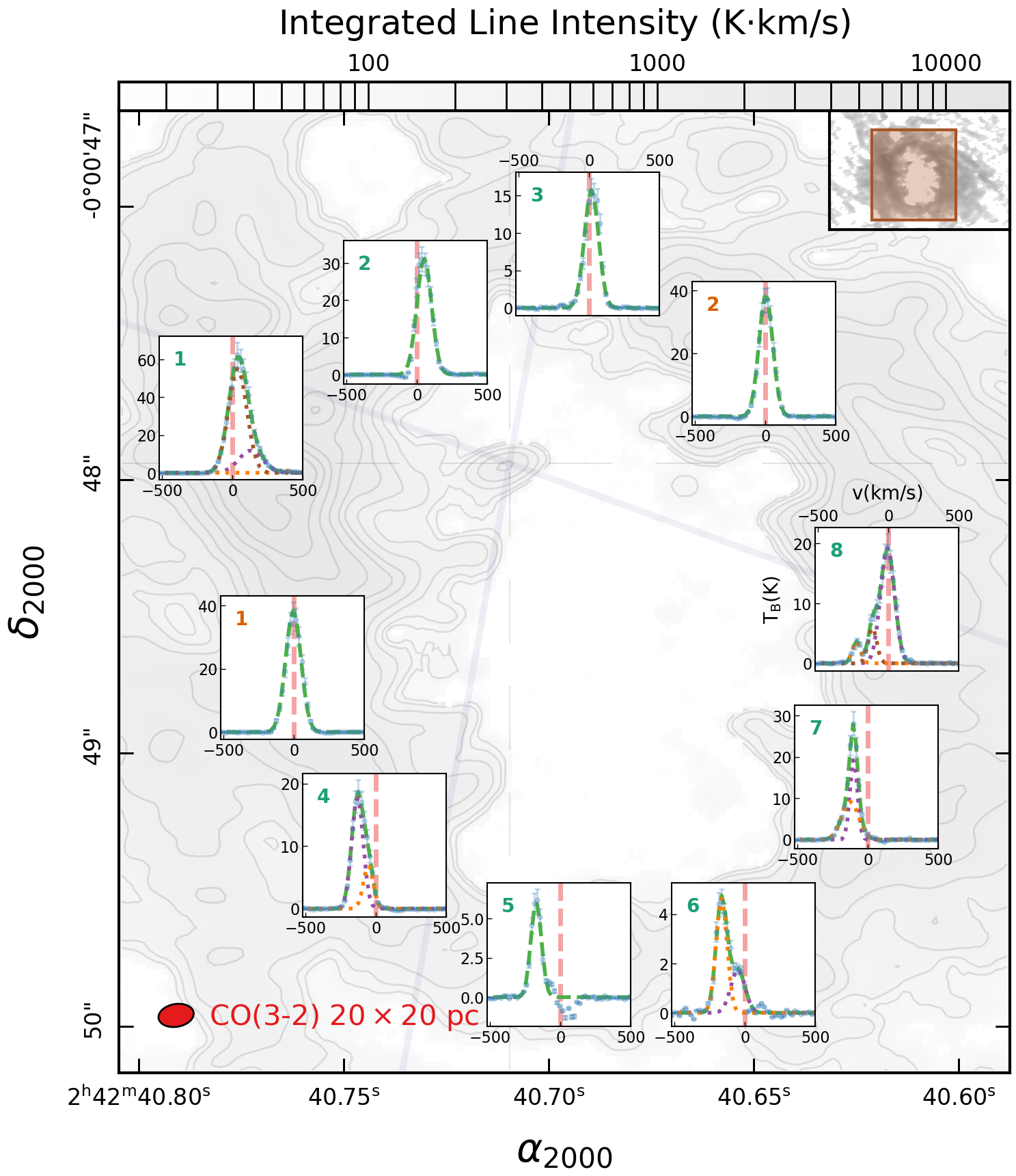}}
\caption{Zoomed in CO(3-2) transition line profiles of all $20\times20$ pc regions defined based on the outflow occupying the exact locations on the background plot as their corresponding original $40\times40$ pc regions assigned in the region definition in Sect. \ref{rsoutflow} (positions of the inset plots only represent the rough locations of the sampled $20\times20$ pc regions). Region numbers, including their colors, are indicated at the upper left corner of each plot, corresponding to those appearing in the region codes in Table \ref{tabb6} and are defined in Sect. \ref{zsr}. The size of the regions, $20\times20$ pc, is indicated at the lower left corner of the background plot, and all other symbols are the same in the background plot and line profile diagrams as those in Fig. \ref{fig:10}.}

\label{fig:a12}       % Give a unique label to the figure.
\end{figure*}

\begin{figure*}[!htbp]
\resizebox{1\textwidth}{!}{\includegraphics{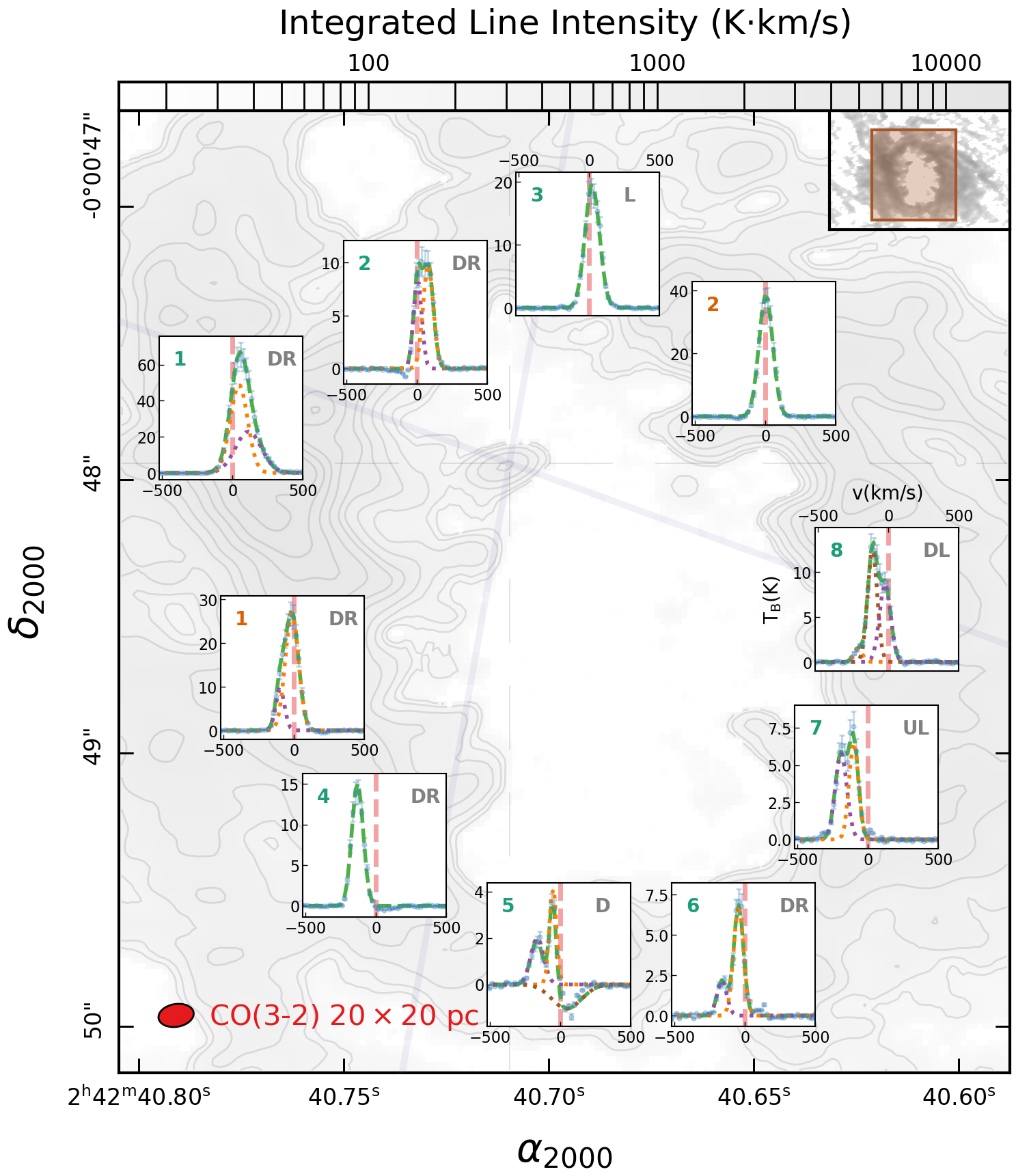}}
\caption{CO(3-2) transition line profiles of all further zoomed-in and shifted regions defined based on the outflow occupying the exact locations on the background plot as their corresponding original $40\times40$ pc regions assigned in the region definition in Sect. \ref{rsoutflow} (positions of the inset plots only represent the rough locations of the sampled zoomed-in and shifted regions). Region numbers, including their colors, are indicated at the upper left corner of each plot, corresponding to those appearing in the region codes in Table \ref{tabb9} and are defined in Sect. \ref{rszs}. Different definitions of the regions (sizes and their locations within corresponding $40\times40$ pc regions) are indicated at the upper right corner of each line profile diagram, and all other symbols are the same in the background plot and line profile diagrams as those in Fig. \ref{fig:10}.}

\label{fig:a15}       % Give a unique label to the figure.
\end{figure*}

\clearpage

\begin{figure*}[!htbp]
\resizebox{1\textwidth}{!}{\includegraphics{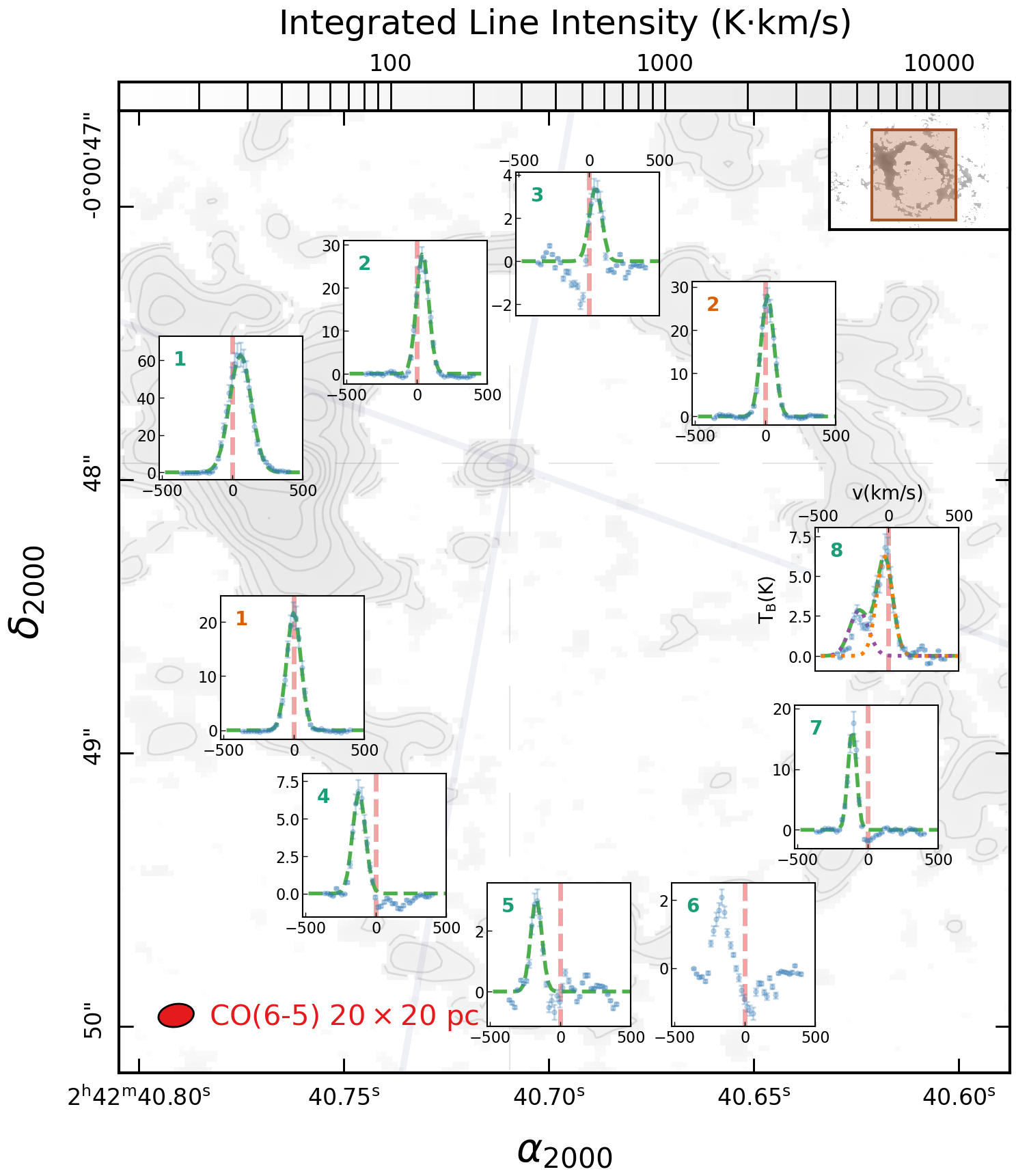}}
\caption{Zoomed in CO(6-5) transition line profiles of all $20\times20$ pc regions defined based on the outflow occupying the exact locations on the background plot as their corresponding original $40\times40$ pc regions assigned in the region definition in Sect. \ref{rsoutflow} (positions of the inset plots only represent the rough locations of the sampled $20\times20$ pc regions). Region numbers, including their colors, are indicated at the upper left corner of each plot, corresponding to those appearing in the region codes in Table \ref{tabb6} and are defined in Sect. \ref{zsr}. The size of the regions, $20\times20$ pc, is indicated at the lower left corner of the background plot, and all other symbols are the same in the background plot and line profile diagrams as those in Fig. \ref{fig:13}.}

\label{fig:a13}       % Give a unique label to the figure.
\end{figure*}

\begin{figure*}[!htbp]
\resizebox{1\textwidth}{!}{\includegraphics{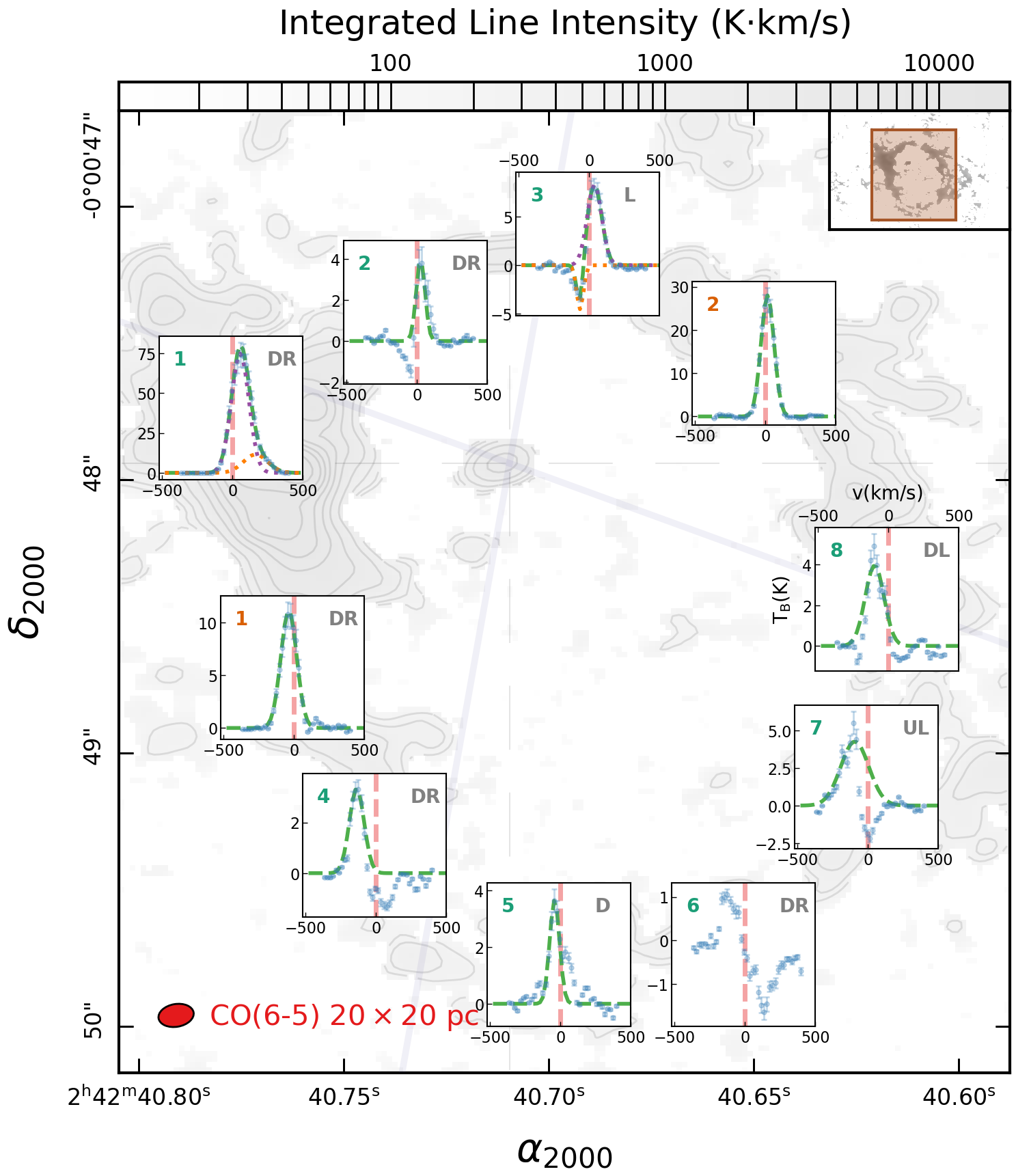}}
\caption{CO(6-5) transition line profiles of all further zoomed-in and shifted regions defined based on the outflow occupying the exact locations on the background plot as their corresponding original $40\times40$ pc regions assigned in the region definition in Sect. \ref{rsoutflow} (positions of the inset plots only represent the rough locations of the sampled zoomed-in and shifted regions). Region numbers, including their colors, are indicated at the upper left corner of each plot, corresponding to those appearing in the region codes in Table \ref{tabb9} and are defined in Sect. \ref{rszs}. Different definitions of the regions (sizes and their locations within corresponding $40\times40$ pc regions) are indicated at the upper right corner of each line profile diagram, and all other symbols are the same in the background plot and line profile diagrams as those in Fig. \ref{fig:13}.}

\label{fig:a16}       % Give a unique label to the figure.
\end{figure*}

\clearpage

\clearpage

\begin{sidewaystable*}[!htbp]
%\begin{table*}[!htbp]
\centering
\begin{tabular}{c c c c c c c c c c c c}
\hline
\hline
Transition & Region & \# & $w_1$ & $w_2$ & $w_3$ & $\mu_1$ & $\sigma_1$ & $\mu_2$ & $\sigma_2$ & $\mu_3$ & $\sigma_3$ \\ % Repeat this pattern 31 times
\hline
CO(2-1) & G1\_20 (N) & 2 & $0.79\pm0.25$ & $0.21\pm0.24$ &  & $1149.9\pm9.1$ & $60.6\pm3.8$ & $1252\pm86$ & $80\pm28$ &  &  \\
 & G2\_20 (N) & 2 & $0.923\pm0.070$ & $0.077\pm0.060$ &  & $1159.3\pm5.1$ & $50.4\pm2.3$ & $1259\pm50$ & $36\pm10$ &  &  \\
 & G3\_20 (N) & 1 &  &  &  & $1132.3\pm2.4$ & $65.1\pm2.1$ &  &  &  &  \\
 & G4\_20 (S) & 2 & $0.865\pm0.053$ & $0.135\pm0.041$ &  & $993.4\pm3.7$ & $43.5\pm2.1$ & $1077.1\pm4.8$ & $20.6\pm3.4$ &  &  \\
 & G5\_20 (S) & 1 &  &  &  & $950.0\pm1.5$ & $37.8\pm1.6$ &  &  &  &  \\
 & G6\_20 (S) & 1 &  &  &  & $972.0\pm2.3$ & $54.4\pm2.4$ &  &  &  &  \\
 & G7\_20 (S) & 2 & $0.571\pm0.062$ & $0.429\pm0.055$ &  & $1017.0\pm2.1$ & $27.5\pm2.2$ & $991.3\pm4.6$ & $65.2\pm3.8$ &  &  \\
 & G8\_20 (S) & 3 & $0.812\pm0.037$ & $0.109\pm0.011$ & $0.079\pm0.028$ & $1098.9\pm3.6$ & $56.5\pm2.7$ & $899.9\pm4.0$ & $36.8\pm5.1$ & $989.3\pm5.0$ & $21.9\pm5.1$ \\
 & O1\_20 (E) & 1 &  &  &  & $1109.1\pm1.1$ & $51.47\pm0.78$ &  &  &  &  \\
 & O2\_20 (W) & 1 &  &  &  & $1125.5\pm1.0$ & $46.25\pm0.76$ &  &  &  &  \\
CO(3-2) & G1\_20 (N) & 3 & $0.78\pm0.73$ & $0.22\pm0.73$ & $0.0065\pm0.0043$ & $1157\pm19$ & $65.5\pm9.5$ & $1250\pm240$ & $85\pm72$ & $1494\pm27$ & $23\pm46$ \\
 & G2\_20 (N) & 1 &  &  &  & $1172.5\pm1.3$ & $48.43\pm0.97$ &  &  &  &  \\
 & G3\_20 (N) & 1 &  &  &  & $1136.8\pm1.3$ & $50.26\pm0.97$ &  &  &  &  \\
 & G4\_20 (S) & 2 & $0.75\pm0.14$ & $0.25\pm0.14$ &  & $988.5\pm8.8$ & $40.8\pm3.5$ & $1065\pm17$ & $34.6\pm6.5$ &  &  \\
 & G5\_20 (S) & 1 &  &  &  & $947.6\pm1.5$ & $37.6\pm1.5$ &  &  &  &  \\
 & G6\_20 (S) & 2 & $0.67\pm0.10$ & $0.33\pm0.12$ &  & $954.3\pm5.9$ & $39.1\pm4.2$ & $1070\pm14$ & $50\pm17$ &  &  \\
 & G7\_20 (S) & 2 & $0.517\pm0.052$ & $0.483\pm0.061$ &  & $994.3\pm2.2$ & $59.5\pm2.3$ & $1019.0\pm2.4$ & $26.3\pm2.2$ &  &  \\
 & G8\_20 (S) & 3 & $0.768\pm0.046$ & $0.148\pm0.038$ & $0.0840\pm0.0061$ & $1113.7\pm4.2$ & $49.7\pm2.2$ & $1006.2\pm6.9$ & $30.6\pm5.0$ & $892.9\pm2.3$ & $28.7\pm2.7$ \\
 & O1\_20 (E) & 1 &  &  &  & $1115.3\pm1.1$ & $54.51\pm0.71$ &  &  &  &  \\
 & O2\_20 (W) & 1 &  &  &  & $1123.22\pm0.94$ & $48.75\pm0.61$ &  &  &  &  \\
CO(6-5) & G1\_20 (N) & 1 &  &  &  & $1172.5\pm1.6$ & $76.6\pm1.2$ &  &  &  &  \\
 & G2\_20 (N) & 1 &  &  &  & $1160.1\pm1.2$ & $43.49\pm0.94$ &  &  &  &  \\
 & G3\_20 (N) & 1 &  &  &  & $1165.4\pm7.6$ & $45\pm18$ &  &  &  &  \\
 & G4\_20 (S) & 1 &  &  &  & $999.5\pm3.0$ & $45.7\pm4.5$ &  &  &  &  \\
 & G5\_20 (S) & 1 &  &  &  & $949\pm11$ & $40\pm24$ &  &  &  &  \\
 & G7\_20 (S) & 1 &  &  &  & $1008.8\pm1.2$ & $30.3\pm1.2$ &  &  &  &  \\
 & O8\_20 (S) & 2 & $0.65\pm0.11$ & $0.35\pm0.16$ &  & $1093.2\pm8.3$ & $56.5\pm9.7$ & $911\pm23$ & $67\pm36$ &  &  \\
 & O1\_20 (E) & 1 &  &  &  & $1118.6\pm1.4$ & $49.9\pm1.3$ &  &  &  &  \\
 & O2\_20 (W) & 1 &  &  &  & $1135.1\pm1.2$ & $46.21\pm0.94$ &  &  &  &  \\
\hline
\end{tabular}
\caption{Fitting results of CO(2-1), CO(3-2), and CO(6-5) line profiles sampled from the zoomed-in $20\times20$ pc regions defined based on the outflow, centered at the coordinates provided in Table \ref{tab2}. The schematics of the table are the same as Table \ref{tabb3}.}
\label{tabb6}
%\end{table*}
\end{sidewaystable*}

\clearpage

\begin{sidewaystable*}[!htbp]
%\begin{table*}[!htbp]
\centering
\begin{tabular}{c c c c c c c c c c c c c}
\hline
\hline
Transition & Region & \# & $w_1$ & $w_2$ & $w_3$ & $\mu_1$ & $\sigma_1$ & $\mu_2$ & $\sigma_2$ & $\mu_3$ & $\sigma_3$ \\ % Repeat this pattern 31 times
\hline
CO(2-1) & G1\_20\_DR (N) & 2 & $0.663\pm0.099$ & $0.337\pm0.099$ &  & $1162.7\pm3.8$ & $58.7\pm4.0$ & $1244\pm21$ & $93.0\pm5.4$ &  &  \\
 & G2\_20\_DR (N) & 1 &  &  &  & $1165.4\pm1.9$ & $64.2\pm1.9$ &  &  &  &  \\
 & G3\_20\_L (N) & 3 & $0.505\pm0.094$ & $0.442\pm0.090$ & $0.053\pm0.016$ & $1116.4\pm4.7$ & $26.7\pm3.1$ & $1189.2\pm8.8$ & $34.9\pm4.3$ & $1012.3\pm5.3$ & $30\pm11$ \\
 & G4\_20\_DR (S) & 2 & $0.9\pm1.9$ & $\sim0.1$ &  & $986.7\pm1.7$ & $39.4\pm1.6$ & $\sim1085$ & $\sim5$ &  &  \\
 & G5\_20\_D (S) & 1 &  &  &  & $1075\pm81$ & $160\pm76$ &  &  &  &  \\
 & G6\_20\_DR (S) & 1 &  &  &  & $1023.7\pm6.9$ & $106\pm15$ &  &  &  &  \\
 & G7\_20\_UL (S) & 2 & $0.52\pm0.12$ & $0.48\pm0.12$ &  & $938\pm13$ & $46.8\pm7.1$ & $1015.8\pm4.2$ & $27.0\pm3.7$ &  &  \\
 & G8\_20\_DL (S) & 1 &  &  &  & $1036.4\pm1.7$ & $60.8\pm1.5$ &  &  &  &  \\
 & O1\_20\_DR (E) & 3 & $0.78\pm0.24$ & $0.22\pm0.21$ & ``$-$'' & $1105\pm13$ & $44\pm11$ & $1025\pm21$ & $34.0\pm6.9$ & $1238\pm22$ & $34\pm30$ \\
CO(3-2) & G1\_20\_DR (N) & 2 & $0.59\pm0.13$ & $0.41\pm0.13$ &  & $1166.6\pm4.8$ & $61.3\pm5.3$ & $1234\pm19$ & $91.7\pm4.0$ &  &  \\
 & G2\_20\_DR (N) & 2 & $0.59\pm0.12$ & $0.41\pm0.11$ &  & $1200.0\pm8.6$ & $35.8\pm3.9$ & $1125.2\pm8.2$ & $28.5\pm4.3$ &  &  \\
 & G3\_20\_L (N) & 1 &  &  &  & $1142.1\pm1.1$ & $51.07\pm0.83$ &  &  &  &  \\
 & G4\_20\_DR (S) & 1 &  &  &  & $989.1\pm1.1$ & $40.27\pm0.80$ &  &  &  &  \\
 & G5\_20\_D (S) & 3 & $0.516\pm0.058$ & $0.484\pm0.030$ & ``$-$'' & $1065.4\pm1.9$ & $22.3\pm1.9$ & $951.3\pm2.8$ & $43.4\pm3.8$ & $1182.5\pm7.9$ & $53\pm21$ \\
 & G6\_20\_DR (S) & 2 & $0.740\pm0.037$ & $0.260\pm0.024$ &  & $1075.8\pm1.7$ & $31.1\pm1.4$ & $960.3\pm3.6$ & $38.2\pm4.6$ &  &  \\
 & G7\_20\_UL (S) & 2 & $0.524\pm0.093$ & $0.476\pm0.094$ &  & $929.4\pm8.8$ & $41.0\pm4.9$ & $1019.7\pm6.8$ & $33.0\pm4.0$ &  &  \\
 & G8\_20\_DL (S) & 3 & $0.521\pm0.098$ & $0.423\pm0.087$ & $0.056\pm0.021$ & $1006.6\pm5.7$ & $33.2\pm5.0$ & $1096.4\pm9.8$ & $39.2\pm4.8$ & $904\pm10$ & $30\pm10$ \\
 & O1\_20\_DR (E) & 2 & $0.810\pm0.089$ & $0.190\pm0.084$ &  & $1105.3\pm6.8$ & $46.4\pm3.1$ & $1021.8\pm9.7$ & $30.8\pm4.0$ &  &  \\
CO(6-5) & G1\_20\_DR (N) & 2 & $0.83\pm0.38$ & $0.17\pm0.38$ &  & $1175\pm12$ & $62.4\pm5.6$ & $1280\pm180$ & $85\pm70$ &  &  \\
 & G2\_20\_DR (N) & 1 &  &  &  & $1146\pm13$ & $32\pm21$ &  &  &  &  \\
 & G3\_20\_L (N) & 2 & ``$+$'' & ``$-$'' &  & $1158\pm11$ & $51.8\pm8.5$ & $1055\pm21$ & $25\pm23$ &  &  \\
 & G4\_20\_DR (S) & 1 &  &  &  & $983.4\pm7.2$ & $52\pm21$ &  &  &  &  \\
 & G5\_20\_D (S) & 1 &  &  &  & $1078.6\pm5.8$ & $35\pm15$ &  &  &  &  \\
 & G7\_20\_UL (S) & 1 &  &  &  & $1027\pm49$ & $100\pm55$ &  &  &  &  \\
 & G8\_20\_DL (S) & 1 &  &  &  & $1022.1\pm6.4$ & $67\pm11$ &  &  &  &  \\
 & O1\_20\_DR (E) & 1 &  &  &  & $1083.4\pm2.3$ & $56.5\pm2.4$ &  &  &  &  \\
\hline
\end{tabular}
\caption{Fitting results of CO(2-1), CO(3-2), and CO(6-5) line profiles sampled from the zoomed-in and shifted $20\times20$ pc regions defined based on outflow. The definition of the zoom and shift for each region is stated in Sect. \ref{rszs}. The ``+'' and ``-'' signs indicate Gaussian components of the emission and absorption, respectively. Only the positive Gaussian components are considered for the weight calculation. The ``$\sim$'' sign is used when the fitting result contains a large error bar (not provided in the table). The rest of the schematics is the same as Table \ref{tabb3}.}
\label{tabb9}
%\end{table*}
\end{sidewaystable*}

\clearpage

\section{LTE analysis results of zoomed-in $20\times20$ pc or zoomed-in and shifted regions}
\label{appf:ltezzs}

\begin{figure*}[!htbp]
\resizebox{1\textwidth}{!}{\includegraphics{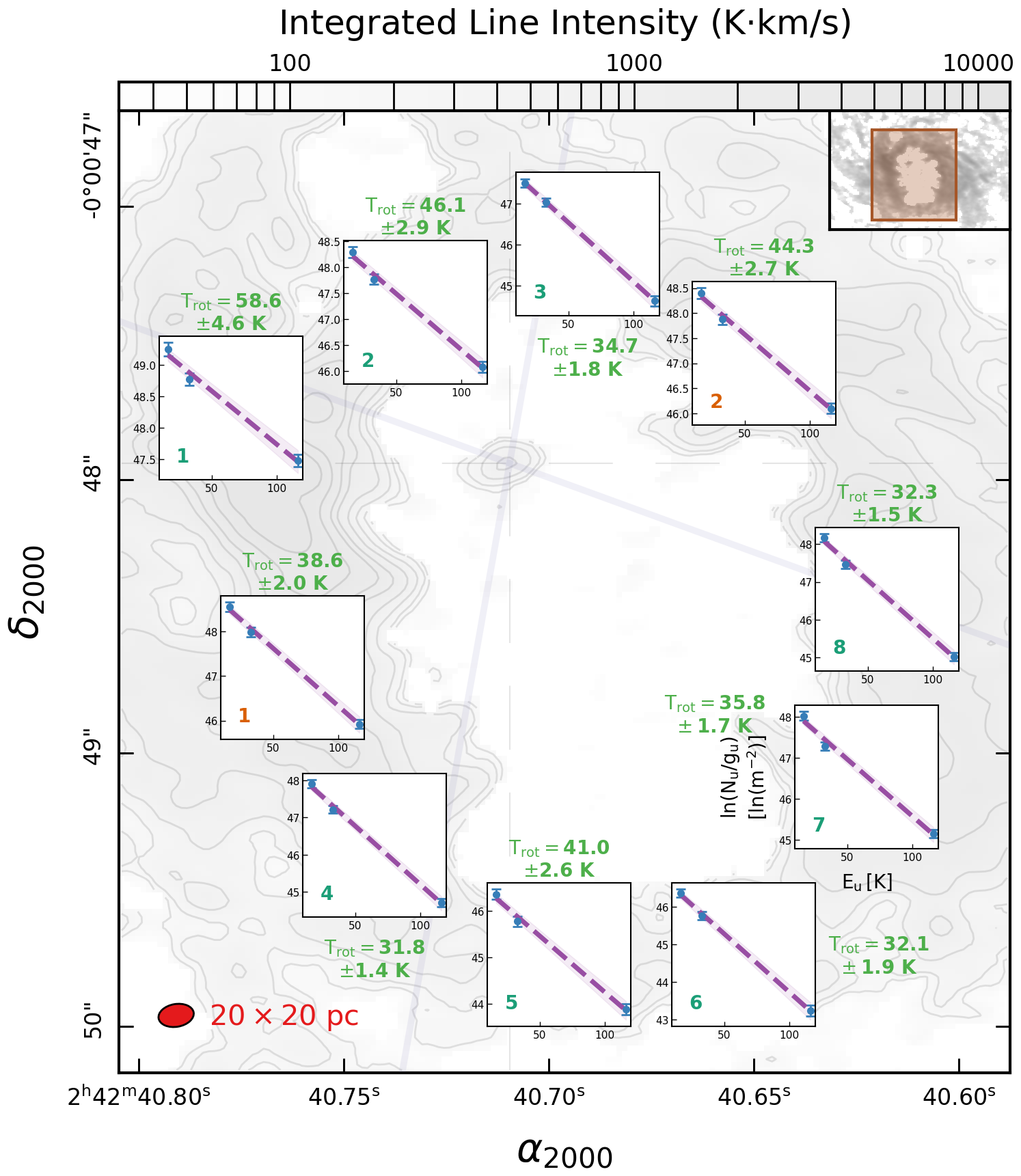}}
\caption{Rotational diagrams of all zoomed-in $20\times20$ regions defined based on the outflow occupying the same locations on the background plot as their corresponding original $40\times40$ pc regions assigned in the region definition in Sect. \ref{rsoutflow} (positions of the inset plots only represent the rough locations of the sampled $20\times20$ pc regions). Region numbers, including their colors, are indicated at the lower left corner of each diagram, corresponding to those appearing in the region codes in Table \ref{tabb12} and are defined in Sect. \ref{zsr}. The sampled region size is indicated in the lower-left corner. All the other symbols and markers of the background plot are the same as Fig. \ref{fig:8}.}

\label{fig:a20}       % Give a unique label to the figure.
\end{figure*}

\begin{figure*}[!htbp]
\resizebox{1\textwidth}{!}{\includegraphics{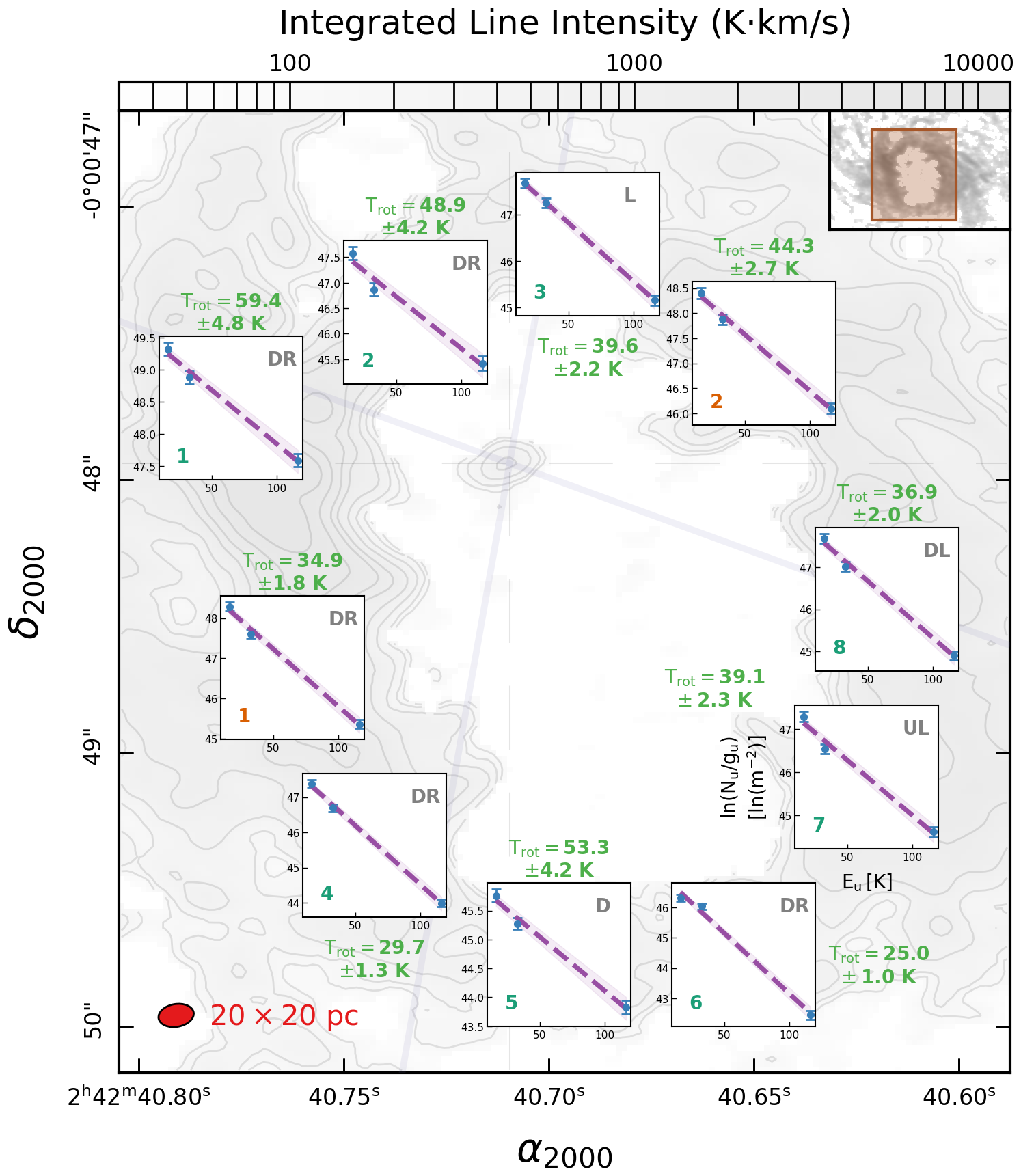}}
\caption{Rotational diagrams of all zoomed-in and shifted regions defined based on the outflow occupying the same locations on the background plot as their corresponding original $40\times40$ pc regions assigned in the region definition in Sect. \ref{rsoutflow} (positions of the inset plots only represent the rough locations of the sampled zoomed-in and shifted regions). Region numbers, including their colors, are indicated at the lower left corner of each diagram, corresponding to those appearing in the region codes in Table \ref{tabb13} and are defined in Sect. \ref{rszs}. The sampled region definitions are indicated at the upper-right corners of all rotational diagrams. All the other symbols and markers of the background plot are the same as Fig. \ref{fig:8}.}

\label{fig:a21}       % Give a unique label to the figure.
\end{figure*}

\begin{table*}[!htbp]
\centering
\begin{tabular}{c c c c c}
\hline
\hline
Region & Rotational Temperature ($T_{\rm{rot}}$) & CO(2-1) & CO(3-2) & CO(6-5) \\ % Repeat this pattern 31 times
\hline
G1\_20 (N) & $\rm{58.6\pm4.6}$ & $69.4\pm9.2$ & $57.3\pm7.9$ & $64\pm13$ \\
G2\_20 (N) & $\rm{46.1\pm2.9}$ & $22.6\pm2.8$ & $19.3\pm2.5$ & $21.4\pm4.3$ \\
G3\_20 (N) & $\rm{34.7\pm1.8}$ & $8.7\pm1.1$ & $8.8\pm1.2$ & $8.8\pm1.9$ \\
G4\_20 (S) & $\rm{31.8\pm1.4}$ & $12.5\pm1.5$ & $10.5\pm1.3$ & $11.8\pm2.4$ \\
G5\_20 (S) & $\rm{41.0\pm2.6}$ & $3.08\pm0.40$ & $2.56\pm0.35$ & $2.89\pm0.66$ \\
G6\_20 (S) & $\rm{32.1\pm1.9}$ & $2.69\pm0.35$ & $2.50\pm0.35$ & $2.64\pm0.71$ \\
G7\_20 (S) & $\rm{35.8\pm1.7}$ & $15.1\pm1.8$ & $11.4\pm1.4$ & $13.6\pm2.6$ \\
G8\_20 (S) & $\rm{32.3\pm1.5}$ & $16.4\pm2.0$ & $13.5\pm1.7$ & $15.4\pm3.2$ \\
O1\_20 (E) & $\rm{38.6\pm2.0}$ & $26.3\pm3.1$ & $22.9\pm2.9$ & $25.1\pm4.9$ \\
O2\_20 (W) & $\rm{44.3\pm2.7}$ & $24.6\pm3.1$ & $21.3\pm2.8$ & $23.3\pm4.7$ \\
\hline
\end{tabular}
\caption{LTE rotational temperatures ($T_{\rm{rot}}$) and total CO column densities ($N$) from the zoomed-in $20\times20\,\rm{pc}$ regions defined based on outflow in units of K and $10^{17}\,\rm{cm^{-2}}$ respectively. The total CO column density calculated from each CO transition is indicated by the name of each transition in the header. The location of each region within the CND is also provided in brackets (e.g., ``N'' stands for the northern CND).}
\label{tabb12}
\end{table*}

\begin{table*}[!htbp]
\centering
\begin{tabular}{c c c c c c}
\hline
\hline
Region & Rotational Temperature ($T_{\rm{rot}}$) & CO(2-1) & CO(3-2) & CO(6-5) \\ % Repeat this pattern 31 times
\hline
G1\_20\_DR (N) & $\rm{59.4\pm4.8}$ & $75.2\pm10.0$ & $64.0\pm8.8$ & $71\pm15$ \\
G2\_20\_DR (N) & $\rm{48.9\pm4.2}$ & $11.5\pm1.8$ & $8.0\pm1.3$ & $10.2\pm2.7$ \\
G3\_20\_L (N) & $\rm{39.6\pm2.2}$ & $11.1\pm1.4$ & $11.1\pm1.4$ & $11.1\pm2.3$ \\
G4\_20\_DR (S) & $\rm{29.7\pm1.3}$ & $7.21\pm0.87$ & $6.32\pm0.81$ & $6.9\pm1.5$ \\
G5\_20\_D (S) & $\rm{53.3\pm4.2}$ & $1.99\pm0.27$ & $1.67\pm0.23$ & $1.86\pm0.41$ \\
G6\_20\_DR (S) & $\rm{25.0\pm1.0}$ & $2.33\pm0.29$ & $3.41\pm0.44$ & $2.60\pm0.62$ \\
G7\_20\_UL (S) & $\rm{39.1\pm2.3}$ & $7.52\pm0.99$ & $5.43\pm0.75$ & $6.7\pm1.5$ \\
G8\_20\_DL (S) & $\rm{36.9\pm2.0}$ & $10.7\pm1.4$ & $8.7\pm1.2$ & $9.9\pm2.1$ \\
O1\_20\_DR (E) & $\rm{34.9\pm1.8}$ & $19.2\pm2.4$ & $15.7\pm2.0$ & $17.8\pm3.8$ \\
\hline
\end{tabular}
\caption{LTE rotational temperatures ($T_{\rm{rot}}$) and total CO column densities ($N$) from the further zoomed-in or shifted regions defined based on outflow in units of K and $10^{17}\,\rm{cm^{-2}}$ respectively. The total CO column density calculated from each CO transition is indicated by the name of each transition in the header. The location of each region within the CND is also provided in brackets (e.g., ``N'' stands for the northern CND).}
\label{tabb13}
\end{table*}

\clearpage

\begin{figure*}[!htbp]
\resizebox{1\textwidth}{!}{\includegraphics{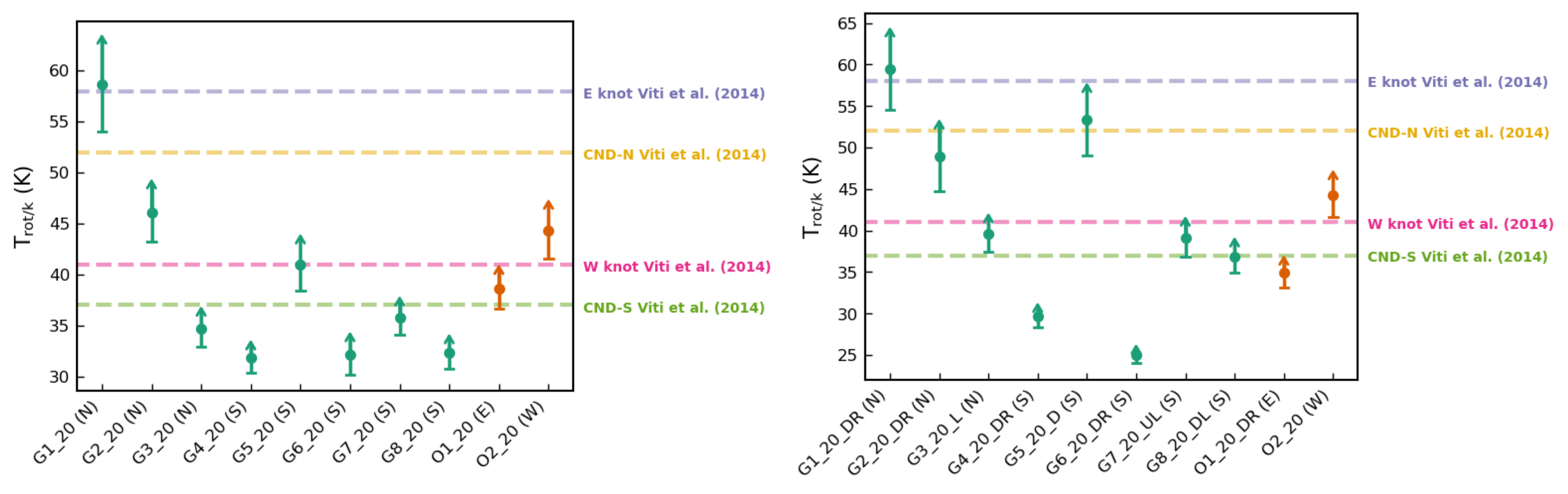}}
\caption{Summary of rotational temperatures for all zoomed-in $20\times20$ pc (\textit{left panel}) or zoomed-in and shifted (\textit{right panel}) regions sampled around the CND defined based on the outflow in Sect. \ref{zsr}. The x-axis marks the region labels along with their locations around the CND (in brackets), while the y-axis indicates the rotational temperatures in the unit of K. The region code (along the x-axis) and color (the color of the data points) of each region match those in Table \ref{tabb12} and are defined in Sect. \ref{zsr}. All other symbols and markers are the same as in Fig. \ref{fig:a23}.}

\label{fig:a30}       % Give a unique label to the figure.
\end{figure*}

\begin{figure*}[!htbp]
\resizebox{1\textwidth}{!}{\includegraphics{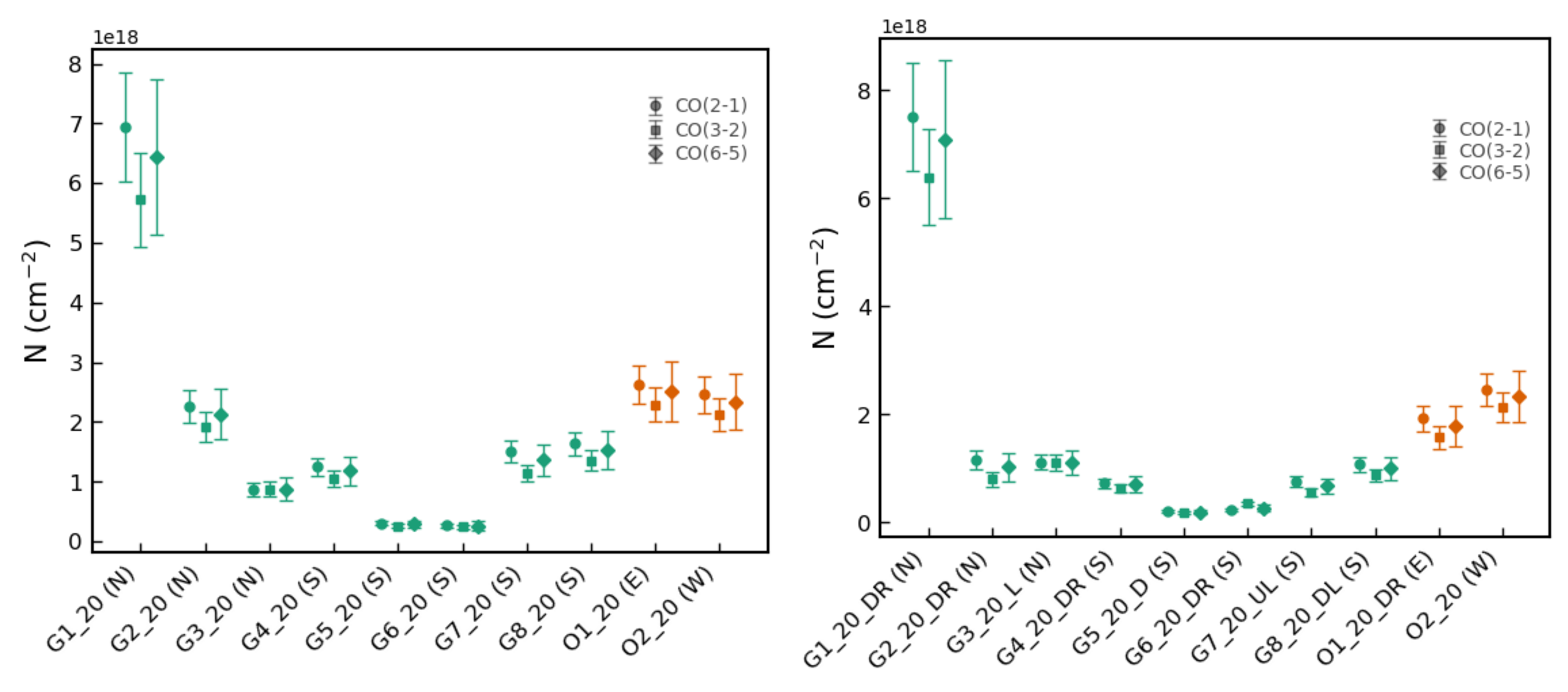}}
\caption{Total CO column densities ($N$) independently calculated from three CO transitions for all zoomed-in $20\times20$ pc (\textit{left panel}) or all zoomed-in and shifted (\textit{right panel}) regions selected based on the outflow. The x-axis marks the region labels along with their locations around the CND (in brackets), while the y-axis indicates the total CO column densities in the unit of $\rm{cm^{-2}}$. The region code (along the x-axis) and color (the color of the data points) of each region match those in Table \ref{tabb12} and are defined in Sect. \ref{zsr}. All other symbols and markers are the same as Fig. \ref{fig:7}.}

\label{fig:a31}       % Give a unique label to the figure.
\end{figure*}

\clearpage

\section{Velocity-integrated Maps of Outflowing CO Gas}
\label{outflow_mom0}

\begin{figure}
\resizebox{0.475\textwidth}{!}{\includegraphics{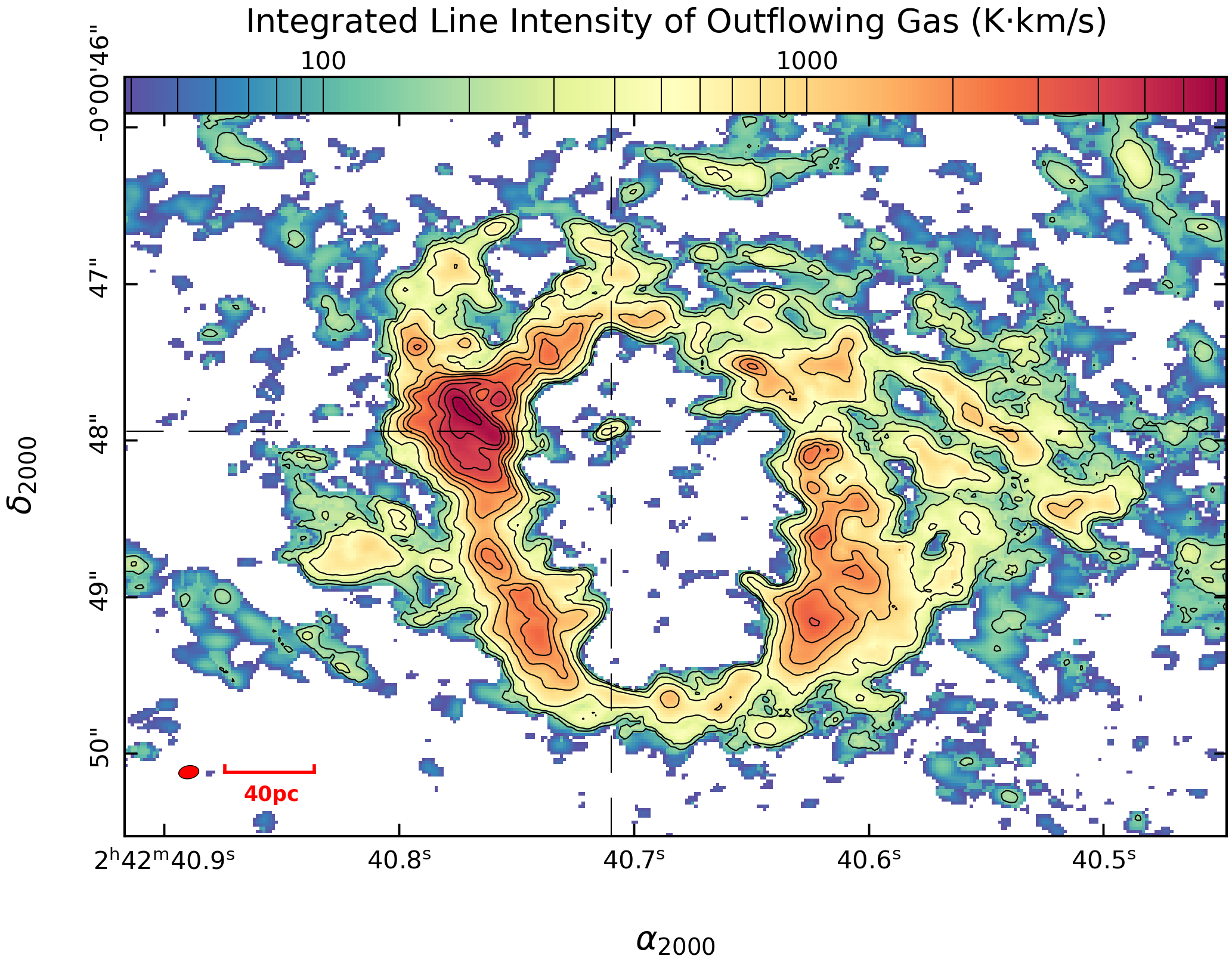}}
\caption{Velocity-integrated map of outflowing CO(2-1) gas in units of $\rm{K\,km\,s^{-1}}$. The color ranges from the minimum to the maximum value in the logarithmic scale, and the contour covers the same extent with levels 3$\sigma$, 5$\sigma$, 10$\sigma$, 20$\sigma$, 30$\sigma$, 40$\sigma$, 60$\sigma$, 80$\sigma$, 100$\sigma$, and 130$\sigma$, where $1\sigma = 50.6\,\rm{K\,km\,s^{-1}}$. All other symbols and markers are the same as in Fig. \ref{fig:1}.}

\label{fig:21outflow}       
\end{figure}

\begin{figure}[!htbp]
\resizebox{0.475\textwidth}{!}{\includegraphics{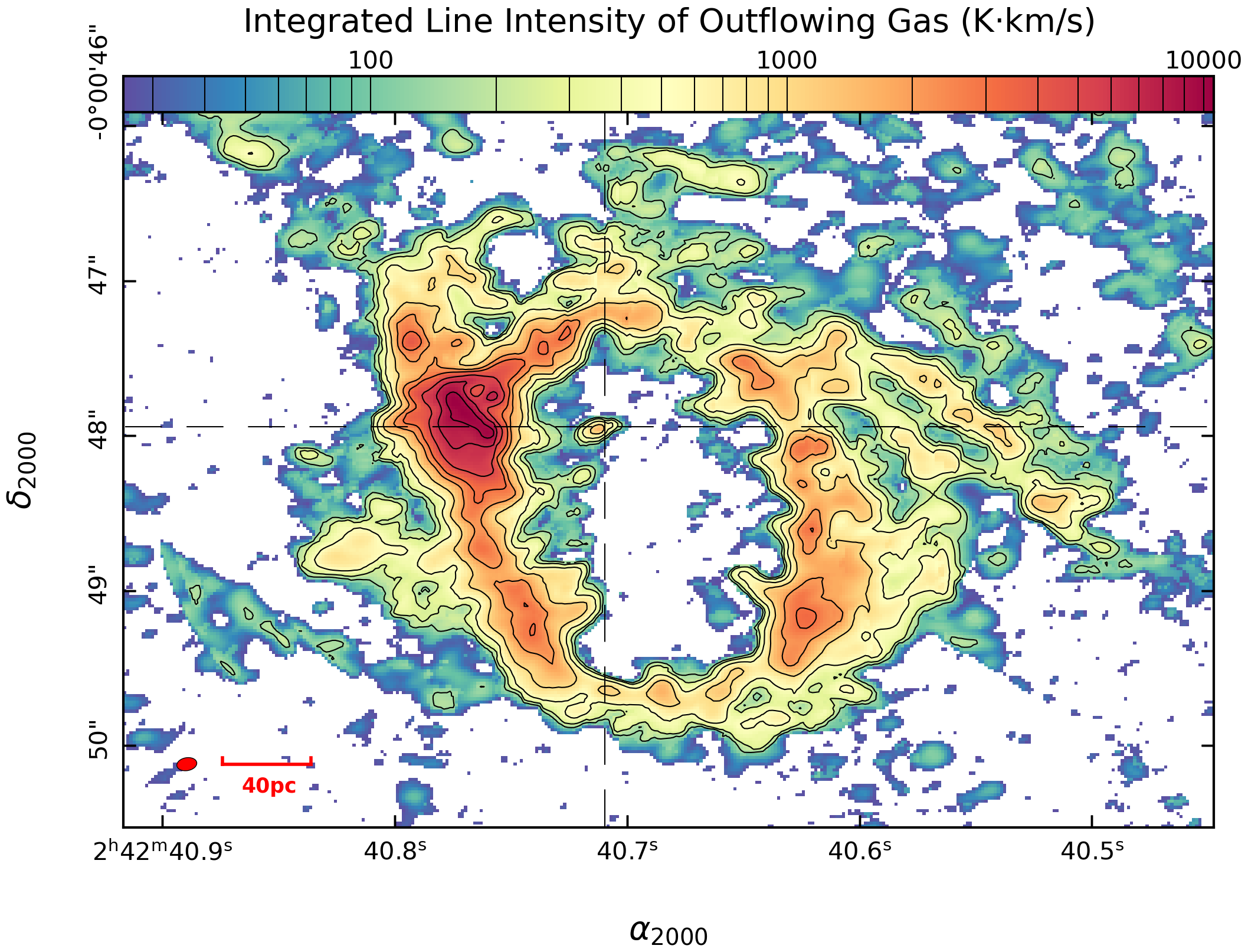}}
\caption{The same as Fig. \ref{fig:21outflow} but for the CO(3-2) transition. The contours cover the same extent as the color bar with levels 3$\sigma$, 5$\sigma$, 10$\sigma$, 20$\sigma$, 40$\sigma$, 60$\sigma$, 100$\sigma$, 140$\sigma$, and 185$\sigma$, where $1\sigma = 49.5\,\rm{K\,km\,s^{-1}}$.}

\label{fig:32outflow}       % Give a unique label to the figure.
\end{figure}

\begin{figure}[!htbp]
\resizebox{0.475\textwidth}{!}{\includegraphics{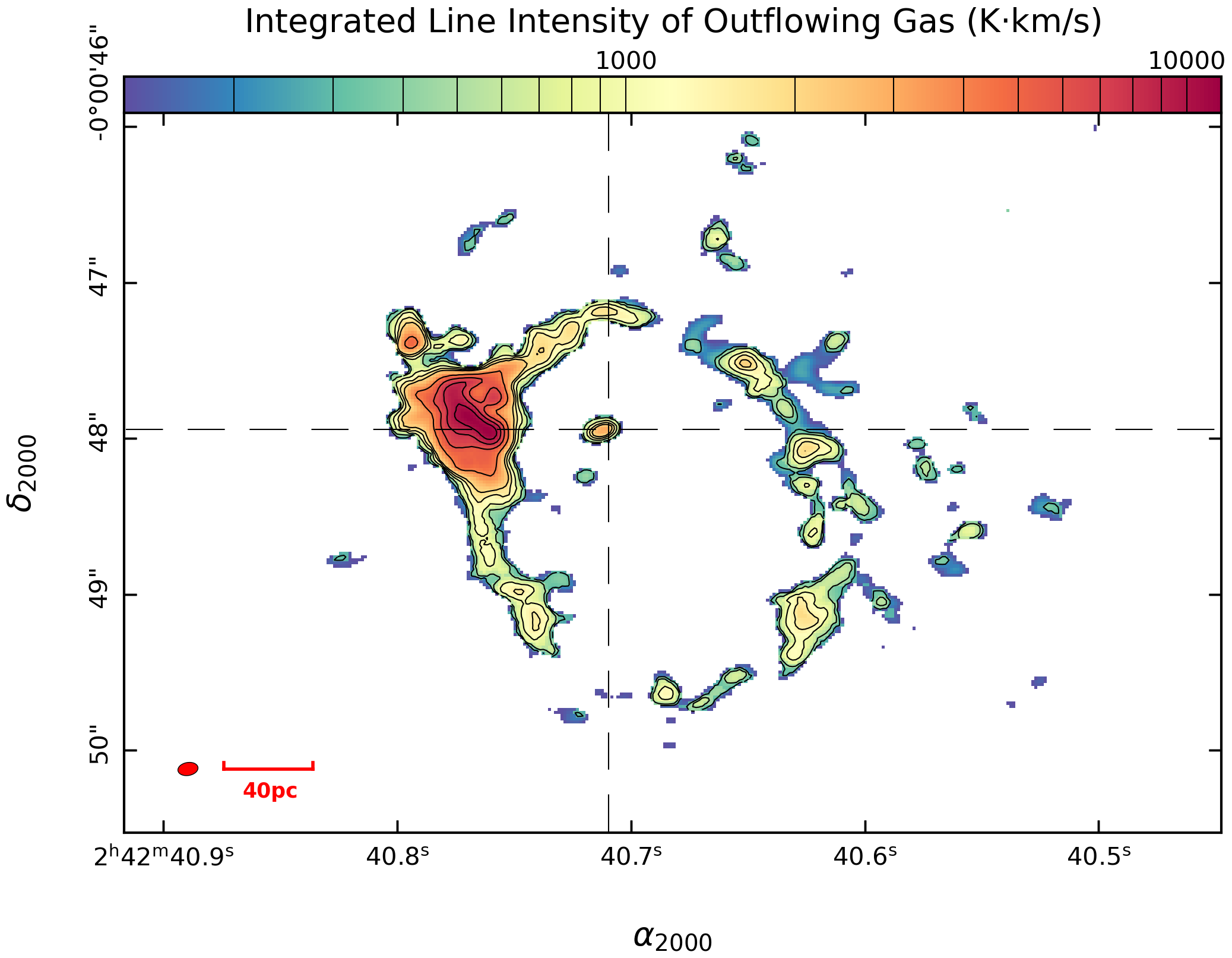}}
\caption{The same as Fig. \ref{fig:21outflow} but for the CO(6-5) transition. The contours cover the same extent as the color bar with levels 3$\sigma$, 5$\sigma$, 10$\sigma$, 15$\sigma$, 20$\sigma$, 40$\sigma$, 60$\sigma$, 80$\sigma$, and 100$\sigma$, where $1\sigma = 99.9\,\rm{K\,km\,s^{-1}}$.}

\label{fig:65outflow}       % Give a unique label to the figure.
\end{figure}

\end{appendix}

%\end{CJK*}
\end{document}